\documentclass{aa}

\usepackage{graphicx}
\usepackage{subfig}
\usepackage{float}
\usepackage{alphalph}
\usepackage{units}
\usepackage{amssymb, amsmath, amsfonts}
\usepackage{natbib}
\bibpunct{(}{)}{;}{a}{}{,}
\usepackage{rotating}
\usepackage{multirow}
\usepackage{url}
\usepackage{txfonts}
\usepackage{hyperref}
\usepackage{lscape}
\usepackage[normalem]{ulem}
\newcommand{\xmm}{{{\it XMM}}}
\newcommand{\xmmn}{{{\it XMM-Newton}}}

\newcommand{\swift}{{\it Swift}}
\newcommand{\rosat}{{\it ROSAT}}

\newcommand{\msun}{M$_{\odot}$}
\newcommand{\cred}{{$C_{red}$ }}

\renewcommand*{\thesubfigure}{\alphalph{\value{subfigure}}}

\begin{document}
   \title{Highly variable AGN from the XMM-Newton Slew Survey}

   \subtitle{}

   \author{N.L. Strotjohann
          \inst{1,2,3}
          \and
          R.D. Saxton\inst{1}
          \and
          R.L.C. Starling\inst{2}
          \and
          P. Esquej\inst{4}
          \and
          A.M. Read\inst{2}
          \and
          P.A. Evans\inst{2}
	\and
	G. Miniutti\inst{5}
          }

   \offprints{nora.linn.strotjohann@desy.de}

   \institute{XMM SOC, ESAC, Apartado 78, 28691 Villanueva de la Ca\~{n}ada, Madrid, Spain
     \and
             Dept. of Physics and Astronomy, University of Leicester, University Road, Leicester LE1 7RH, U.K.
         \and
	Physikalisches Institut, Universit\"at Bonn, Nu\ss allee 12, 53115 Bonn, Germany\\
           Present address: DESY, Platanenallee 6, D-15738 Zeuthen, Germany
         \and
             Centro de Astrobiolog\'ia (CSIC-INTA), E-28850 Torrej\'on de Ardoz, Madrid, Spain
	\and
	Centro de Astrobiolog\'ia (CSIC-INTA), Dep de Astrof\'isica; LAEFF, PO Box 78, E-28691 Villanueva de la Ca\~{n}ada, Madrid, Spain
        }

   \date{Received; accepted}

  \abstract
   {}
   {We investigate the properties of a variability-selected complete sample of AGN in order to identify the mechanisms which cause large amplitude X-ray variability on time scales of years.}
{A complete sample of 24 sources was constructed, from AGN which changed 
their soft X-ray luminosity by more than one order of magnitude 
over 5--20 years between 
\rosat\ observations and the \xmmn\ Slew Survey. Follow-up observations were obtained with the \swift\ satellite. We analyse the spectra of these AGN at the \swift\ and \xmm\ observation epochs, where 6 sources had continued to display extreme variability. Multiwavelength data are used to calculate black hole masses and the relative X-ray brightness $\alpha_{\rm OX}$.}
{After removal of two probable spurious sources, we find that the sample has global properties which differ little from a non-varying control sample drawn from
the wider XMM-Slew/\rosat/Veron sample of all secure AGN detections. A wide range of AGN types are represented in the varying sample. The black hole mass distributions for the varying and non-varying sample are not significantly different. This suggests that long timescale variability is not strongly affected by black hole mass. There is marginal 
evidence that the variable sources have a lower redshift (2$\sigma$) and
X-ray luminosity (1.7$\sigma$). Apart from two radio-loud sources, the sample have normal optical-X-ray ratios
($\alpha_{\rm OX}$) when at their peak but are X-ray weak during their lowest flux 
measurements.}
{Drawing on our results and other studies, we are able to identify a variety of variability mechanisms at play: tidal
disruption events, jet activity, changes in absorption, thermal emission from the inner accretion disc, and variable accretion disc reflection. 
Little evidence for strong absorption is seen in the majority of the sample and single-component absorption can be excluded as the mechanism for most sources.}


   \maketitle
%


\section{Introduction}
The X-ray emission of active galactic nuclei (AGN) likely arises from close to the central engine, and can display large amplitude variability on time scales of hours down to minutes.
This variability is thought to be related to instabilities in the corona where UV photons are scattered to X-ray energies \citep{nandra2001}. Luminosity changes on longer time scales can be caused by perturbations in the accretion flow. This idea is supported by the correlation between luminosity, and thereby black hole mass and accretion rate, and variability time scale \citep[e.g.][]{mchardy2004,mchardy2006}, and by the observation of similar, lower amplitude variability observed in the optical and UV bands \citep[e.g.][]{McCleod10}. A number of components can contribute to the X-ray spectrum of an AGN, and methods such as principle component analysis have identified a number of components contributing to variability across AGN samples on kilosecond (ks) time scales \citep{parker2015}. These have been suggested to include relativistic reflection and changes in partial covering neutral absorption.

On longer time scales of months to years, less is known about the X-ray variability mechanisms of AGN. A number of studies have probed the long-term light curves of samples at energies $\ge$\,2\,keV \citep[e.g.][using data from {\it RXTE} and {\it Swift} BAT]{SobPap09,winter2009,Soldi2014}. They successfully model this with changes in the flux and shape of the intrinsic power law emission. Extending this analysis down to soft X-ray energies, where disc, Compton and absorber contributions may play a significant role, requires data collection over multiple missions to cover long baselines in both time and energy range. This can be achieved with the AGN samples probed by the {\it ROSAT} All Sky Survey \citep[RASS:][]{voges1999} and the {\it XMM-Newton} Slew Survey \citep[XMMSL1:][]{saxton2008}. These X-ray surveys, taken about a decade apart, reach similar depths in the 0.2--2\,keV band.

Detailed studies have been carried out for a small number of individual objects demonstrating extreme X-ray variability. By way of example, WPVS\,007 showed a factor 400 decrease in its soft X-ray flux between 1990 and 1993 \citep{grupe1995}, the X-ray flux of PHL\,1094 fell by a factor 260 over 5 years \citep{miniutti2012} and IRAS\,13224-3809 has shown variations of a factor 50 on time scales of days \citep{boller1997}. For individual objects significant advances in our understanding of variability mechanisms has been possible, but there remains a need both to characterise the highly X-ray variable AGN population and to identify the origins of these extreme flux changes.

Among the proposed variability mechanisms at these energies and time scales, is a change in absorbing column. We know that cold and warm material is present in the broad- and narrow-line regions, and movement of clouds or outflowing material across the line-of-sight can dramatically alter the observed soft X-ray spectral shape \citep[e.g.][]{risaliti05,komossafink,winter2012,starling2014}. The absorption seen in AGN is, in many cases, well described by clumpy, partial covering material which could feasibly provide year time scale orbital variability. In Seyfert 2 galaxies in particular variability of clumpy X-ray absorbing material may be ubiquitous \citep{Risaliti10}, while ionised absorbers may be common among luminous Seyfert 1s \citep{winter2010}. A statistical search for absorption events has been carried out on the long term {\it RXTE} X-ray light curves of a sample of Seyfert galaxies, resulting in probability estimates for observing a source during an absorption event that echo the greater variability expectation for Seyfert 2s over Seyfert 1s \citep{Markowitz}.

A steep flux increase which then decays may be indicative of a more 
catastrophic event such as a tidal disruption event 
\citep[TDE, e.g.][]{rees88}. Candidate TDE have been found in \rosat\ 
\citep{KomossaBade,greiner} and \xmm\ observations
\citep{esquej2007,Saxton2012TDE,Maksym} as well as
at UV \citep[e.g.][]{Gezari2008,VanVelzen2011} and optical wavelengths 
(e.g. \citealt{Komossa08}; \citealt{Arcavi14}; see \citealt{komossa2015} for a review).

Similarly to the likely origins of short time scale variability, intrinsic changes in accretion onto the black hole \citep{miniutti2013,shappee14,saxton14}, as well as jet power and changes in the Comptonising media may be responsible. By observing both the soft and hard X-ray bands simultaneously, the interplay between the underlying emission and any absorption components may be derived.

In this paper we examine a sample of candidate AGN drawn from the {\it XMM-Newton} Slew Survey which have shown large amplitude long time scale soft 
X-ray variability when compared with earlier {\it ROSAT} data. We re-observed these sources in a dedicated {\it Swift} programme, and combined these data with archival multiwavelength data in order both to identify the soft X-ray variability mechanisms in each individual source, and to characterise the highly variable AGN population.

In Section \ref{sample} we detail the sample selection and compare our sample with a wider sample of AGN drawn from the {\it XMM} Slew Survey in Section \ref{comparison}. In Section \ref{observations} we present the {\it Swift} X-ray Telescope (XRT) observations from our targetted programme and show the long-term light curves in Section \ref{curves}. Optical to X-ray flux ratios are calculated in Section \ref{sec:alphaox}. We go on to look at the spectra obtained with \swift\ and \xmm\ for our sample in Section \ref{spectra}, in order to understand the variability mechanisms which may be at play. Finally, we discuss our results in the context of long-term X-ray variability and highly variable AGN populations in Section \ref{discuss}. The results for each individual source are detailed in Appendix \ref{sec:individual}.

A $\lambda$CDM cosmology with ($\Omega_{M},\Omega_{\lambda}$) = (0.3,0.7) and $H_{0}$=70 km$^{-1}$ s$^{-1}$ Mpc$^{-1}$ has been assumed throughout.

\begin{table*}
{\tiny
\hfill{}
\caption{High variability candidate AGN sample.}
\label{tab:highvarsrc}      
\begin{center}
\begin{tabular}{l c c c c c c c c}
\hline\hline                 
XMMSL1 name			& Type		& $z$ 		& $L_{\rm X,0.2-2keV}$		 & XMM/RASS & 0.2-2\,keV CR & 2-10\,keV CR &  M$_\text{bh}$ & Common Name\\   
				& 		&  		& ($10^{42}$ erg s$^{-1}$) & ratio & ct s$^{-1}$ & ct s$^{-1}$ & log(M$_\odot$)& \\   
\hline                        
J005953.1+314934 &  Sy 1.2 	& 0.0149 	& 5.5 		& $12.4\pm{1.4}$ & $8.14 \pm 0.79$ & $1.76 \pm 0.39$ & 6.5&  Mrk\,352 \\
J015510.9-140028 			& - 		& - 		& - 		& $>45.0$ 	& $0.63 \pm 0.27$ & - & - \\
J020303.1-074154 & 		 	 Sy 1 		& 0.0615 	& 25.6 		& $>38.4$ & $2.15 \pm 0.59$ & - & 6.8 & 2MASX~J02030314-0741514\\
J024916.6-041244 & 			 Sy 1.9 	& 0.0186 	& 1.9		& $>24.6$ & $1.84 \pm 0.80$ & - & 5.7 & 2MASX\,J02491731-0412521\\
J034555.1-355959\tablefootmark{a}  	& - 		& - 		& - 		        & $>15.3$ & $1.06 \pm 0.29$ & $1.463 \pm 0.384$ & - & MRSS\,358-033707\\
J044347.0+285822\tablefootmark{a}  	& Sy 1 		& 0.0217 	& 2.6		& $15.7\pm4.2$ & $1.62 \pm 0.46$ & $1.60 \pm 0.58$ & 7.3 & UGC\,3142\\
J045740.0-503053 			& - 		& - 		& -		& $>18.1$ & $2.25 \pm 0.48$ & - & - & 2MASX\,J04574068-5030583\\
J051935.5-323928\tablefootmark{a}  	& Sy 1.5 	& 0.0125 	& 0.76 		& $18.0\pm 6.4$ & $7.57 \pm 0.97$ & $1.021 \pm 0.396$  & 6.6 & ESO\,362-G018\\
J064541.1-590851\tablefootmark{a} 	& - 		& - 		& - 		& $>29.9$ & $1.09 \pm 0.35$ & $0.785 \pm 0.430$ & - & 2MASX\,J06454155-5908456\\
J070841.3-493305\tablefootmark{a}  	& NLS 1 	& 0.0406 	& 14.5 		& $13.4\pm 2.2$ & $6.63 \pm 1.04$ & - & 7.1 & 1H\,0707-495\\
J082753.7+521800\tablefootmark{a} 	& QSO 		& 0.3378 	& 578	 	& $>16.3$  & $1.14 \pm 0.24$ & $0.629 \pm 0.277$  & 7.8 & 87GB\,082409.1+522804\\
J090421.2+170927  			& QSO 		& 0.0733 	& 15.9 		& $>32.5$ & $0.91 \pm 0.25$ & - & 7.4 & SDSS\,J090421.39+170933.2\\
J093922.5+370945  		 	& NLS 1 	& 0.1861 	& 241 		& $>33.9$ & $1.90 \pm 0.39$ & - & 7.9 & 2MASS\,J09392289+3709438\\
J100534.8+392856  		 	& Sy 1 		& 0.1409 	& 106	 	& $>15.4$ & $1.55 \pm 0.22$ & - & 7.8 & 2MASX\,J10053467+3928530\\
J104745.6-375932  		 	& Sy 1 		& 0.0755 	& 47		& $>13.3$ & $2.51 \pm 0.39$ & - & 7.2 & 6dFGS\,gJ104745.7-375932\\
J111527.3+180638  		& liner	& 0.00278 	& 0.073		& $>54.4$ & $4.95 \pm 0.65$ & - & - & NGC\,3599\\
J112841.5+575017\tablefootmark{a} 	& Sy 2 		& 0.0509 	& 10.9 		& $>17.8$ & $1.37 \pm 0.34$ & $0.43 \pm 0.18$ & 7.6 & MCG+10-17-004 \\
J113001.8+020007  			& - 		& - 		& -		& $>21.0$ & $1.47 \pm 0.43$ & - & - & \\
J121335.0+325609  		 	& QSO 		& 0.222 	& 154 		& $21.0\pm 7.3$ & $ 0.82 \pm 0.19$ & - & 7.9 & SDSS\,J121334.67+325615.2\\
J132342.3+482701  		 	& inactive	& 0.0875 	& 39		& $>38.1$  & $ 1.60 \pm 0.37$ & - & - & SDSS\,J132341.97+482701.2\\
J162553.4+562735\tablefootmark{a}  		 	& QSO 		& 0.307 	& 442		& $>26.4$ & $ 1.11 \pm 0.28$ & - & 7.9 & SBS\,1624+565 \\
J173739.3-595625\tablefootmark{a}  	& Sy 2 		& 0.0170 	& 3.09 		& $>19.4$  & $ 4.20 \pm 0.82$ & $1.47 \pm 0.40$ & 7.3 & ESO\,139-G012 \\
J183521.4+611942\tablefootmark{a} 	& blazar	& 2.274 	& 61 449	& $16.9\pm 6.7$  & $ 1.16\pm 0.26$ & - & 9.7 & QSO\,J1835+6119 \\
J193439.3+490922\tablefootmark{a} 	& - 		& - 		& - 		& $>26.1$  & $ 0.95 \pm 0.32$ & - & - & 2MASX\,J19343950+4909211\\
\hline                                   
\end{tabular}
\end{center}
}
\tablefoot{ The information on the AGN type, redshift and common name were obtained from NED and references therein. $L_{X}$ is 
the luminosity as seen in the XMM Slew Survey in the 0.2--2.0\,keV band, calculated using a model of a 
power of index $\Gamma = 1.7$ and Galactic absorption. 
The XMM/RASS ratio is the ratio of the {\it XMM} Slew and \rosat\ fluxes,
or upper limits, for the 0.2--2\,keV energy range, using the same spectral model. Errors are 1$\sigma$. The black hole masses have been estimated using the k-band luminosity as described in Section \ref{comparison} and have a typical uncertainty of 0.3\,dex.\\
\tablefoottext{a}{Some sources have several Slew Survey observations. In those cases we list the first observation in which a significant flux change by at least a factor of ten was observed compared to the \rosat\ flux, or upper limit, 
in the 0.2--2\,keV band. The soft, 0.2--2\,keV, and hard,
2--10\,keV, count rates correspond to the values seen in this slew observation.
The hard band count rate is only quoted if the source was actually detected
in this band.}
} \\
\end{table*}

\section{Sample selection} \label{sample}
The {\it XMM-Newton} Slew Survey \citep[XMMSL1;][]{saxton2008}, with positional 
accuracy of 8$''$ and soft flux sensitivity of
F$_\text{0.2-2.0 keV}=6 \times 10^{-13} \unit[]{ergs\ cm^{-2}s^{-1}}$,
has characteristics which are well matched with
those of the {\it ROSAT} all-sky survey \citep[RASS;][]{voges1999} performed
at the beginning of the 1990s. This allows meaningful flux comparisons to be 
made of sources observed with both satellites.

We have selected a complete sample of XMMSL1 sources, with counterparts identified as
AGN or galaxies\footnote{The identification procedure is outlined in
\cite{saxton2008}. AGN have principally been identified from SIMBAD, NED, SDSS, 
2MASS extended sources and the Veron catalogue \citep{veron2006}},
which have varied by more than a factor of 10 when comparing the 0.2--2\,keV flux in the XMMSL1 observations to fluxes (or 2$\sigma$ flux limits) seen in the RASS or in {\it ROSAT} pointed observations from 5--19 years 
earlier.\footnote{Variability in the harder part of the spectrum is not 
explored here, as no wide area 2--10\,keV survey exists with comparable
sensitivity to XMMSL1. We note that we selected only sources which brightened by a factor of $>10$.} Since some flux measurements have large associated uncertainties we calculate the 1$\sigma$ error on the flux ratio and only select the source if the ratio is still larger than ten when considering the error.
The sample contains all objects that meet these criteria and that are included in the XMMSL1-Delta-3 catalogue which contains
data from slews made between 2001-08-26 and 2009-01-15.

To convert {\it ROSAT} and \xmm\ count rates to fluxes, we have adopted
a spectral model of an absorbed power law with a typical AGN photon 
index of  $\Gamma = 1.7$ \citep{TurnerPounds} and Galactic absorption per source according to \citet{willingale2013}.
Inaccuracies introduced in the
flux ratios by the use of this fixed spectral slope are discussed
in \citet{saxton2011}. The flux ratio will also be affected if the source spectrum
changes between the {\it ROSAT} and \xmm\ slew observations; a point which is
quantitatively addressed in Section \ref{curves}.
This resulted in 24 XMMSL1 sources, listed in Table \ref{tab:highvarsrc}, which form a complete sample selected on the basis of large amplitude soft X-ray variability.

The Slew Survey is sensitive to relatively bright sources and many of 
the detected objects are nearby AGN. Two of our sample sources,
ESO\,362-G018 and 1H\,0707-495, have been extensively discussed in the literature; here we summarise only the most relevant results. A further two sources, NGC\,3599 and SDSS\,J132341.97+482701.2, have been identified as tidal disruption candidates in non-active, or weakly active, 
galaxies \citep[see][]{esquej2007, esquej2008} and will only be peripherally discussed in this paper.

\section{Comparison with the wider XMM Slew Survey AGN sample} \label{comparison}
In order to place our 24 candidate highly variable AGN in context, we examine the count rates of all secure AGN detected in the \xmm\ Slew Survey. We have made a new selection consisting of all sources observed within XMMSL1-Delta-3 and {\it ROSAT} which
are contained in the Veron catalogue of AGN \citep{veron2006}. This produces a sample of 1038 AGN, discussed previously in \citet{saxton2011}, which we call here the Veron sample.

\begin{figure}[htbp]
\centering
\resizebox{\hsize}{!}{\includegraphics{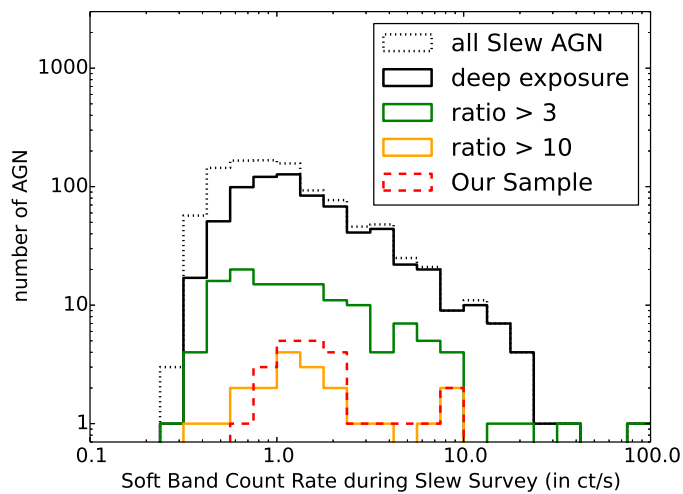}}
\caption[Count rates during Slew Survey.]
{ \label{fig:var_with_countrate} Histogram of XMMSL1 soft band count rates. The black dashed line shows all detected sources from the Veron sample (see text) while the black solid line corresponds to the fraction of the Veron sample
with deep enough {\it ROSAT} observations, such that a factor 10 variability in flux
between XMMSL1 and {\it ROSAT} would have been detected. The green and yellow 
lines represent
sources that fulfil this criterion and that were in addition observed to vary by a factor of more than 3 or 10 respectively. Our so called \emph{constant sample} consists of all sources in between the black and the green line, while the \emph{variable sources} are the ones below the red line. Some low-count sources
contained within the yellow line have errors on their flux ratios which exclude them
from the variable sample and the variable sample includes several AGN which are not in the Veron catalog.}
\end{figure}

For a source to enter our sample of highly variable candidate AGN it must have deep enough observations with both {\it ROSAT} and \xmm, such that strong variability, between the \xmm\ and {\it ROSAT} flux or upper limit, can be detected. To be able to compare the Veron sample to the highly variable AGN candidates we remove all sources from the Veron sample for which such a high variability is not detectable. From the {\it ROSAT} catalogues for bright sources, faint sources and pointed observations, we obtain the exposure times, background rates and extraction radii. With this information we calculate the 90\% noise level during the {\it ROSAT} observation. A source fainter by a factor of 10 compared to the Slew Survey has a count rate reduced by a factor of 70, if observed with {\it ROSAT} \citep[see][]{saxton2011} due to the smaller effective area. We only select the sources from the Veron sample for which this count rate would be above the noise level. This condition is fulfilled by 728 or 70\% of the sources. Figure \ref{fig:var_with_countrate} shows that mainly sources with low count rates during the Slew survey are removed.

In Figure \ref{fig:var_with_countrate} we subdivide the remaining AGN of the Veron sample by their flux ratio. The yellow line shows the population of 35 sources in the Veron sample which are at least 10 times brighter during the Slew Survey. The sample discussed in this paper (shown as a red dashed line) has a large overlap with the highly variable Veron AGN and the count rate distributions are similar. Some of our sources are however not contained in the Veron catalog. In addition we take into account the statistical errors on the count rates and only select sources with a flux ratio significantly above the threshold of 10. We then selected sources which were observed to vary only by a factor of $\leqslant$3 and defined those as the \emph{constant sample}.

\begin{figure}[bt]
{\includegraphics[width=90mm]{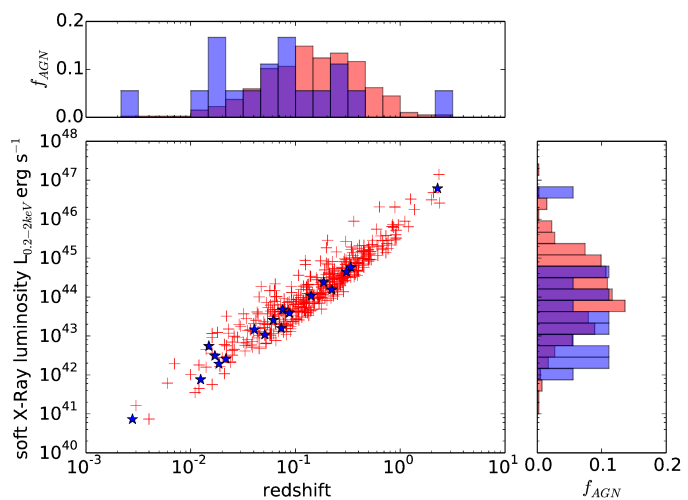}} 
\caption{Comparison between the constant and highly variable samples. In the main panel we plot the 0.2--2\,keV luminosity during the \xmm\ Slew Survey against redshift. Red crosses represent constant sources while the blue stars correspond to the highly variable sources. The subplots show the fraction of sources, $f_{AGN}$, in the samples at each luminosity and redshift, where red bars show constant sources and blue bars variable sources.}
\label{fig:srctypes} 
\end{figure}

In Figure \ref{fig:srctypes} we show the redshift and X-ray luminosity of our highly variable sample overlaid on that of the constant sample. Projecting the distributions on the axes, one sees that our sample is slightly biased towards lower values for both quantities. An exception is XMMSL1\,J183521.4+611942, known to be a bright blazar at $z=2.2$. Using the Kolmogorov Smirnov test we examine whether the variable sources are drawn randomly from the same sample as the constant sources. For the redshift this hypothesis has a probability of 2.2\%, while for the X-ray luminosity it is 2.6\%. Among our highly variable sample we know we have two candidate TDEs, and if we omit these we obtain probabilities of 2.5\% (redshift, deviation at 2$\sigma$ level) and 4.6\% (luminosity, deviation at 1.7$\sigma$ level).
In addition, neglecting the blazar, which is atypical of our candidate AGN sample, would increase the significance by 0.3 $\sigma$ for both distributions. We thus conclude that there are indications that highly variable sources tend to have lower X-ray luminosities than other AGN and are therefore only observed at lower redshifts. However this result is not statistically significant and larger samples would be necessary to confirm this.

One other fundamental AGN property is the black hole mass. Unfortunately precise mass estimates obtained for example from stellar dynamics or reverberation mapping are not available for most AGN in our sample. We therefore have to rely on a more indirect and less precise approach. We here use the empiric correlation between the k-band luminosity and the black hole mass as described in \citet{marconi2003}. The scatter of this method is 0.3 dex. Since all our sources are contained in the 2MASS catalogues, the black hole mass can be calculated in this way for all sources except for the six galaxies lacking a redshift measurement (Table \ref{tab:highvarsrc}). We also omit the two likely tidal disruption candidates. The k-band magnitudes are obtained from the NASA/IPAC Infrared Science Archive. If there is in addition an entry in the extended source catalogue, we use this value, which is only significantly larger for the closest sources.
The k-magnitude has to be corrected for the contribution of the host galaxy 
to obtain the luminosity of the bulge only \citep{simien1986}. However in most cases the type of the host galaxy is unknown, so for those galaxies we use a correction of 0.8 mag corresponding to a lenticular host galaxy, which is in between the correction for an elliptical and a spiral galaxy. The resulting mass estimates are given in Table \ref{tab:highvarsrc}.

The black hole mass has also been estimated using other methods, including the optical H$\beta$ line width or reverberation measurements, for seven of our sample sources as reported in the literature \citep{greene2007,ho2008,shen2008,parisi2009,caramete2010,fabian2012,ponti2012,winter2012,zhou2013}. For some sources we obtain an additional mass estimate using the excess variance method \citep{oneill2005, ponti2012}. For most sources the mass estimates are consistent within an order of magnitude, highlighting that the masses presented here are not very precise and should be considered as order of magnitude estimates.

\begin{figure}[tb]
\centering
\includegraphics[width=88mm]{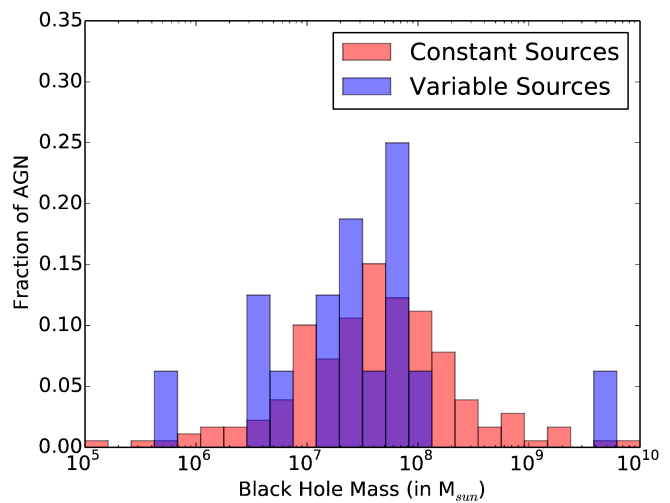}
\caption[Masses of variable and constant sources]
{\label{fig:hist_masses}The mass distribution from the k-band luminosity for the constant sample and for our variable AGN.}
\end{figure}

In Figure \ref{fig:hist_masses} we show the mass distribution of the highly variable sources compared to the masses of the constant sample. Since the variable sources are on average at lower redshifts as shown in 
Figure~\ref{fig:srctypes}, we would expect this to translate into lower black hole masses. There is a small difference between the two samples, however the probability that both distributions are drawn from the same sample is 32\%. Thus we cannot reject this hypothesis. 
Maybe the effect may be diluted by the intrinsic scatter in the relation and the additional uncertainty introduced by not being able to correct for the individual host galaxies. Moreover we know that there are several variability mechanisms and some of them might correlate with the host mass, while others might not or might even be anticorrelated. With our small sample we are not able to distinguish between individual populations of highly variable AGN, for example high mass blazars and low mass intrinsically variable AGN.

We conclude that even though we might expect to see a bias towards lower black hole masses due to the lower redshifts and X-ray luminosities, we do not see any significant deviation.

\section{\swift\ observations} \label{observations}
We observed our full sample of all 24 candidate highly variable AGN with {\it Swift} \citep{gehrels2004} for $\sim$2\,ks each, between 2010 and 2014 as part of a fill-in programme. All XRT \citep{burrows2005} observations were made in photon counting mode with exposure times ranging from 1.6--3.7\,ks. The \swift-XRT data were obtained from the UK {\it Swift} Science Data Centre\footnote{\url{www.swift.ac.uk/user_objects}} and reduced following the procedures of \citet{evans2009} using the {\it Swift} software and calibration database available within HEASOFT v.6.12.
Simultaneous observations were made with the \swift\ BAT \citep{barthelmy2005} at 14--195\,keV and the UV Optical Telescope \citep[UVOT;][]{roming2005} with the u filter applied.

For ten sources, additional archival \swift\ observations were available at the time of writing which we have included and analysed in an identical manner.
Details of all the observations used in this paper are given in the
appendix (see Table~\ref{tab:data_table}).

With the XRT we detected 16 (or two-thirds) of the sample sources in our fill-in observations. Widening our search, we looked at data stacks in the {\it Swift} XRT Point Source Catalogue \citep[1SXPS;][]{evans2014} and other pointed XRT observations and found that a further 5 sources were detected.

Two of the three XRT non-detected sources have only ever been detected in the \xmm\ Slew Survey (Figures \ref{fig:J015} and \ref{fig:J113}), and cannot be identified with any source detected in other wavelength surveys such as 2MASS, WISE, SDSS or 6dF in our searches. One of these, XMMSL1\,J015510.9-140028, lies at the detection threshold of the Slew Survey. The other, XMMSL1\,J113001.8+020007, has a higher significance, however the photons are aligned along a row which indicates that this might not be an astrophysical point source. 
We conclude therefore that those two detections in the Slew Survey are highly likely to be 
spurious. We discuss the spurious fraction further in Section \ref{discuss}.

The remaining XRT-undetected source is XMMSL1\,J193439.3+490922, which has 
three detections in \xmm\ slews and is hence likely real. 
All XRT-detected sources are also detected with UVOT. Three sources (Mrk 352, ESO 362-G018, ESO 139-G012) can be found in the \swift\ BAT 70-month All-Sky Hard X-ray Survey Source Catalog \citep{baumgartner2013}.

\section{Long-term X-ray light curves} \label{curves}

\begin{figure*}
\centering
\subfloat[XMMSL1\,J005953.1+314934]{\label{fig:J005} \includegraphics[width=60mm]{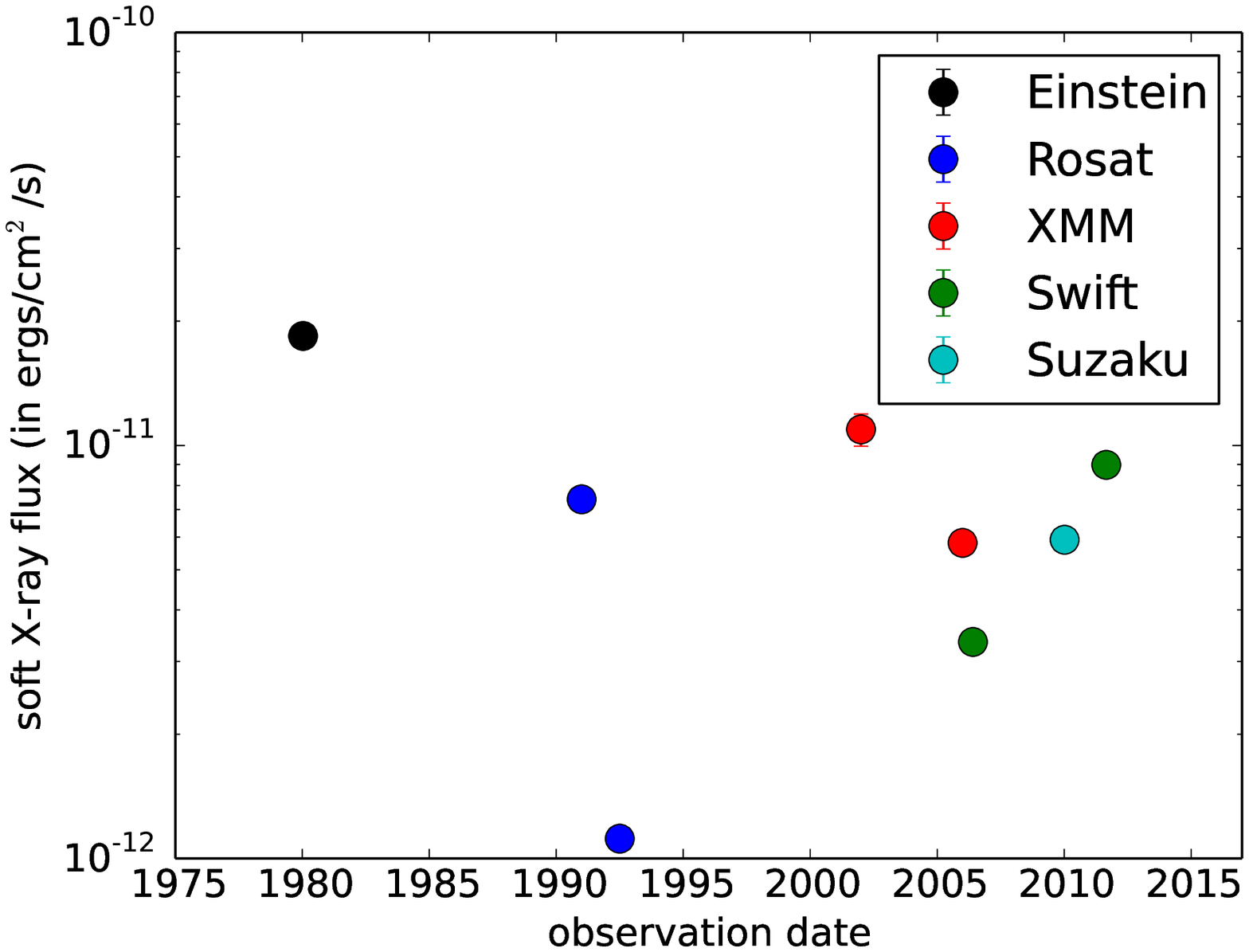}}
\hfill
\subfloat[XMMSL1\,J015510.9-140028]{\label{fig:J015} \includegraphics[width=60mm]{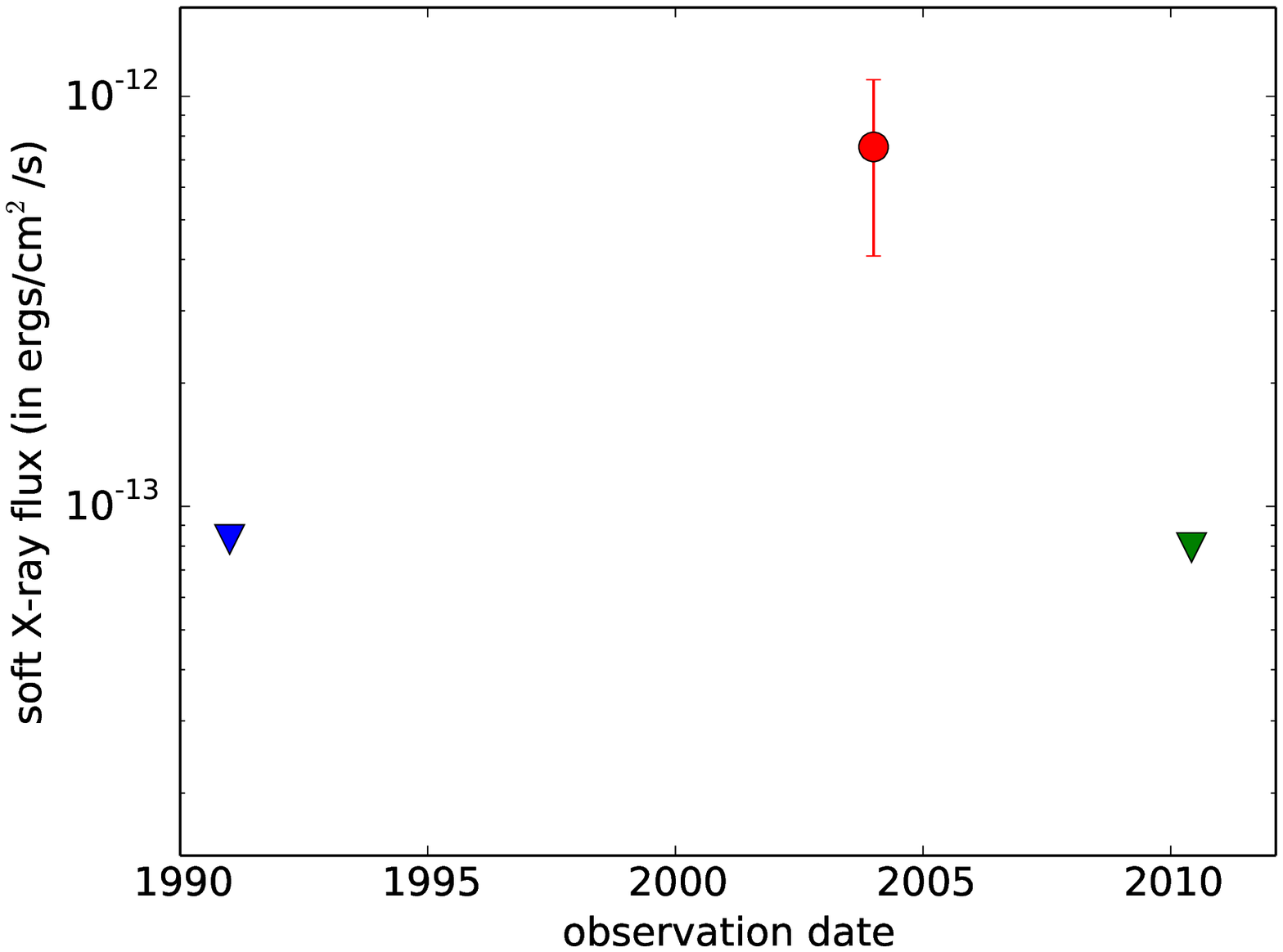}}
\hfill
\subfloat[XMMSL1\,J020303.1-074154]{\label{fig:J020} \includegraphics[width=60mm]{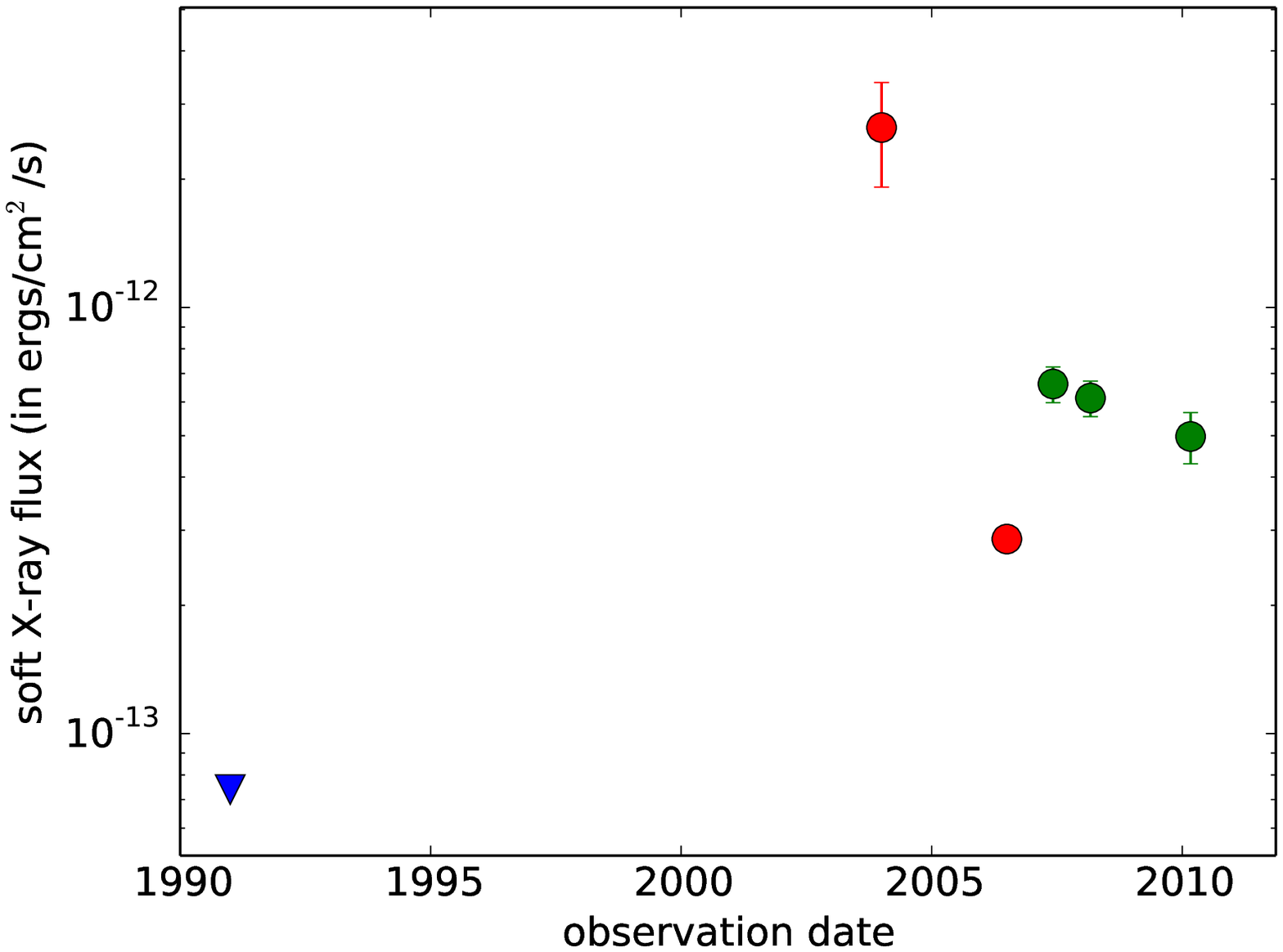}}

\subfloat[XMMSL1\,J024916.6-041244]{\label{fig:J024} \includegraphics[width=60mm]{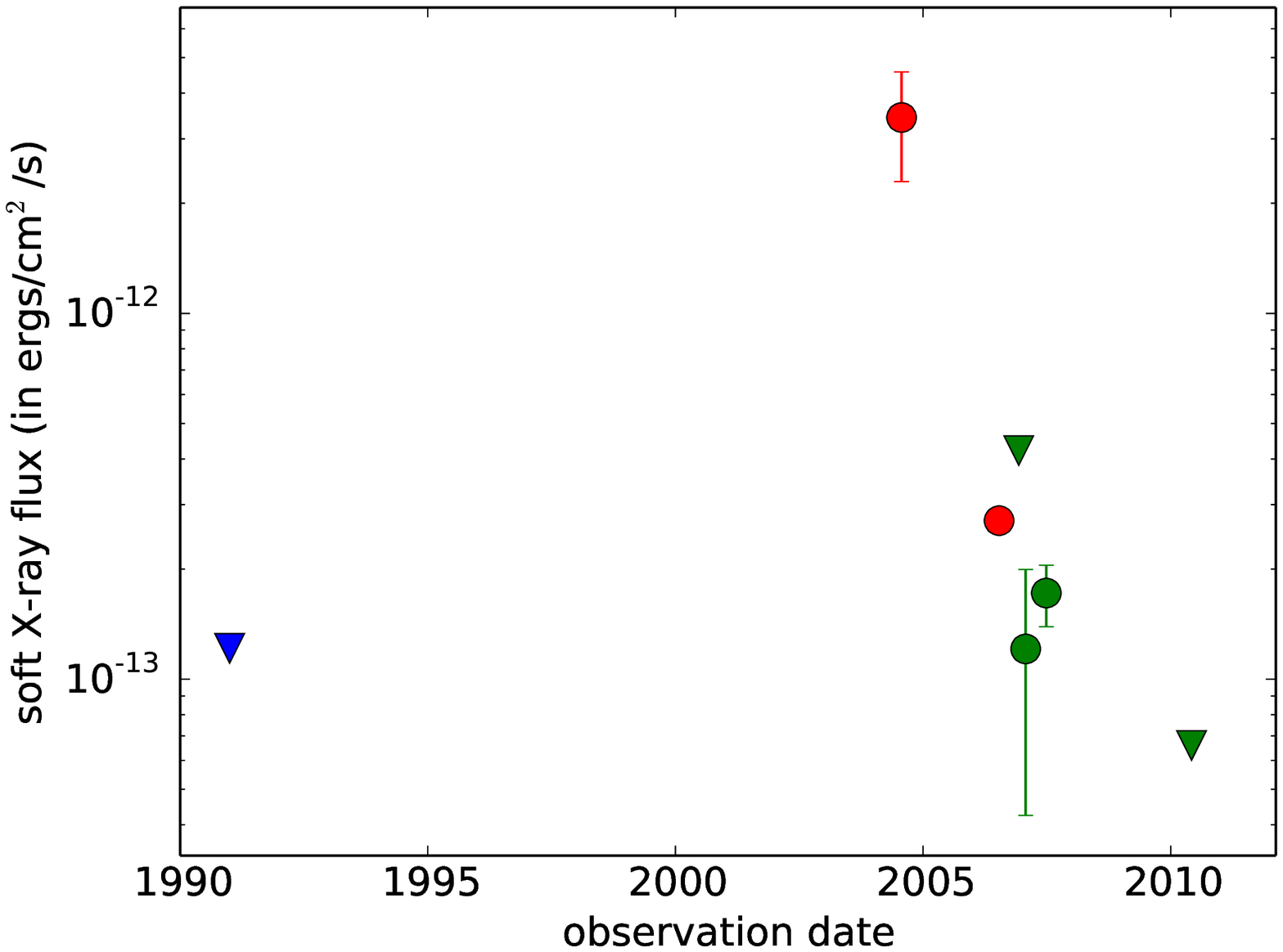}}
\hfill
\subfloat[XMMSL1\,J034555.1-355959]{\label{fig:J034} \includegraphics[width=60mm]{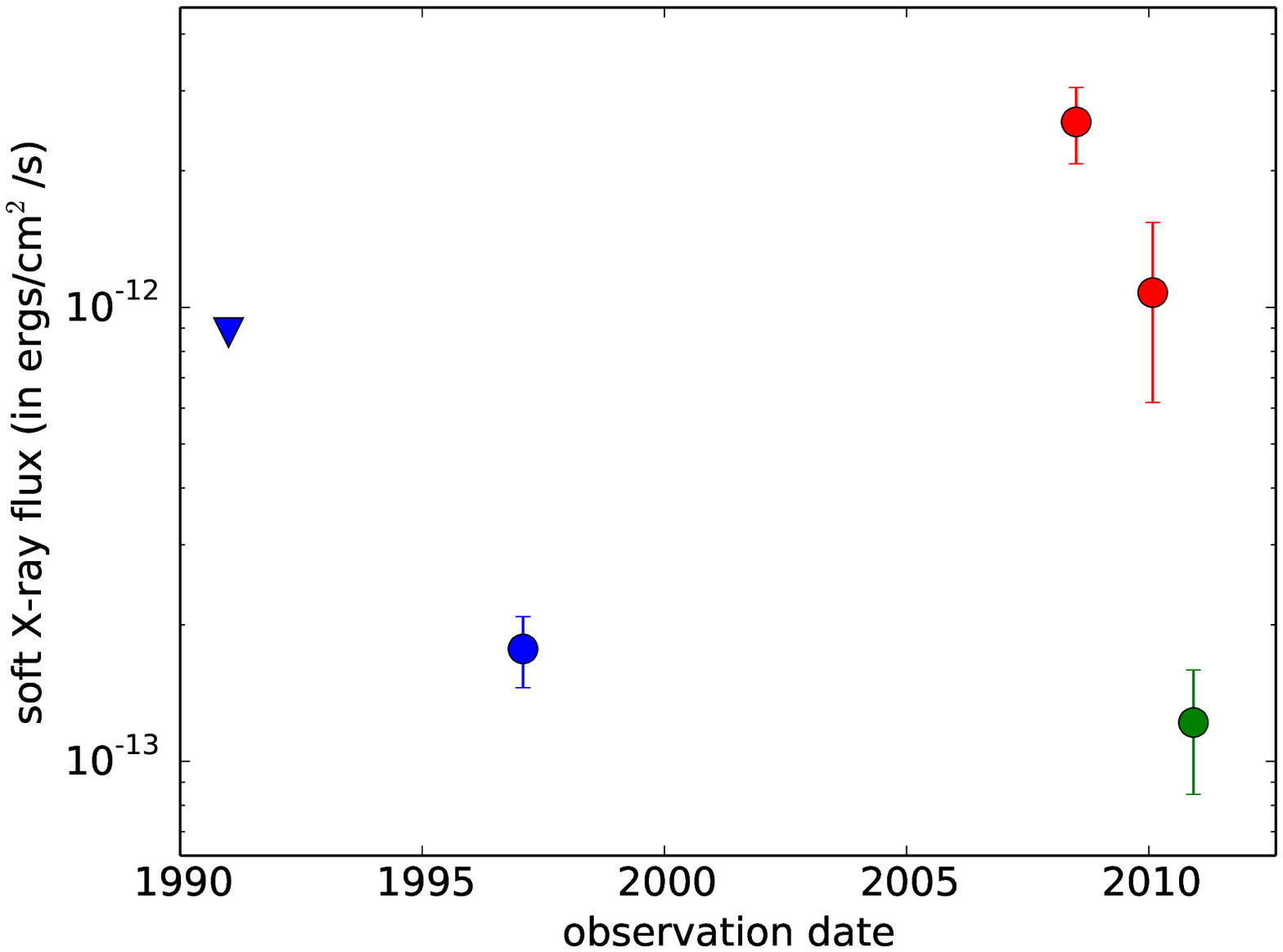}}
\hfill
\subfloat[XMMSL1\,J044347.0+285822] {\label{fig:J044} \includegraphics[width=60mm]{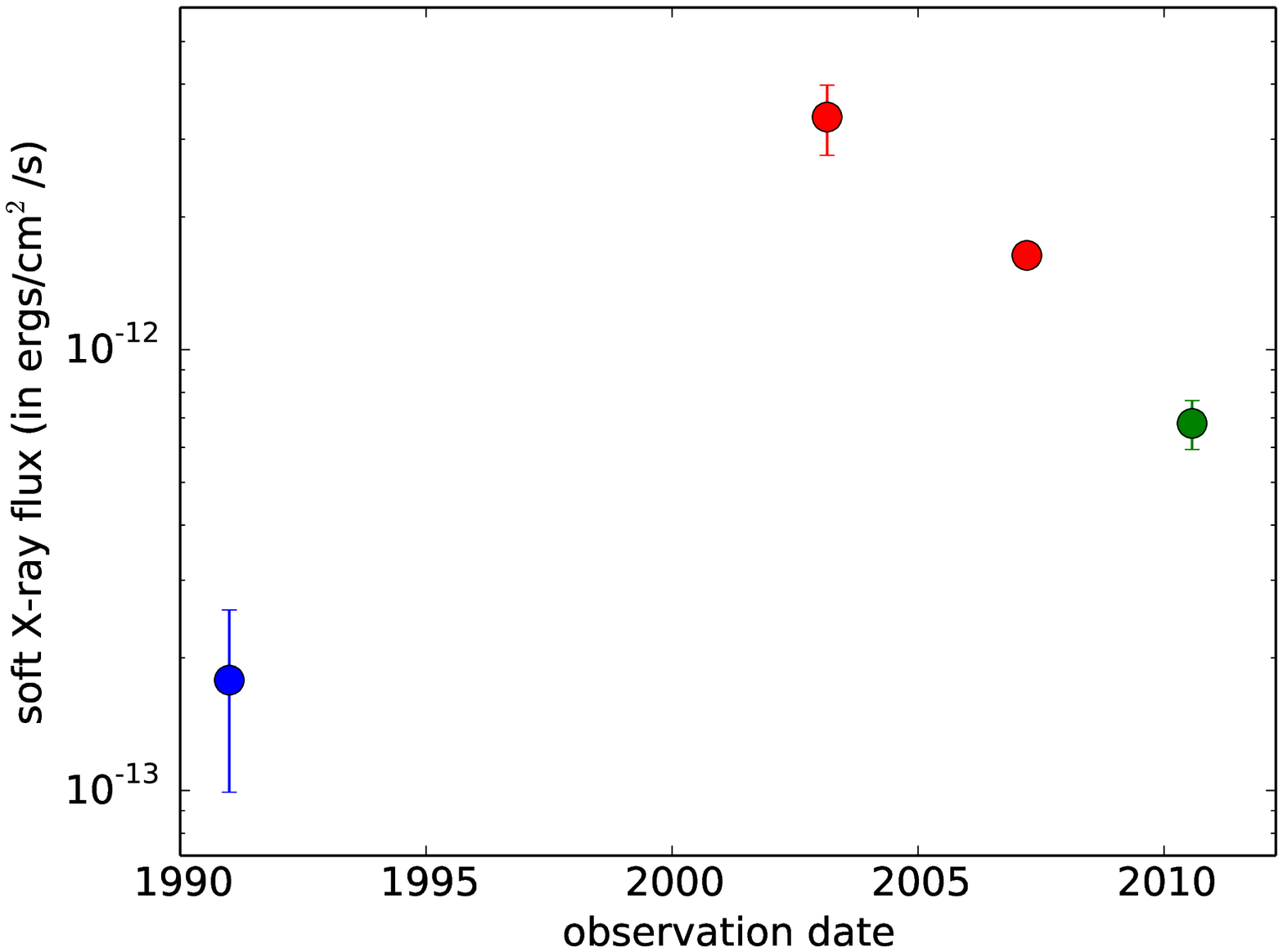}}

\subfloat[XMMSL1\,J045740.0-503053]{\label{fig:J045} \includegraphics[width=60mm]{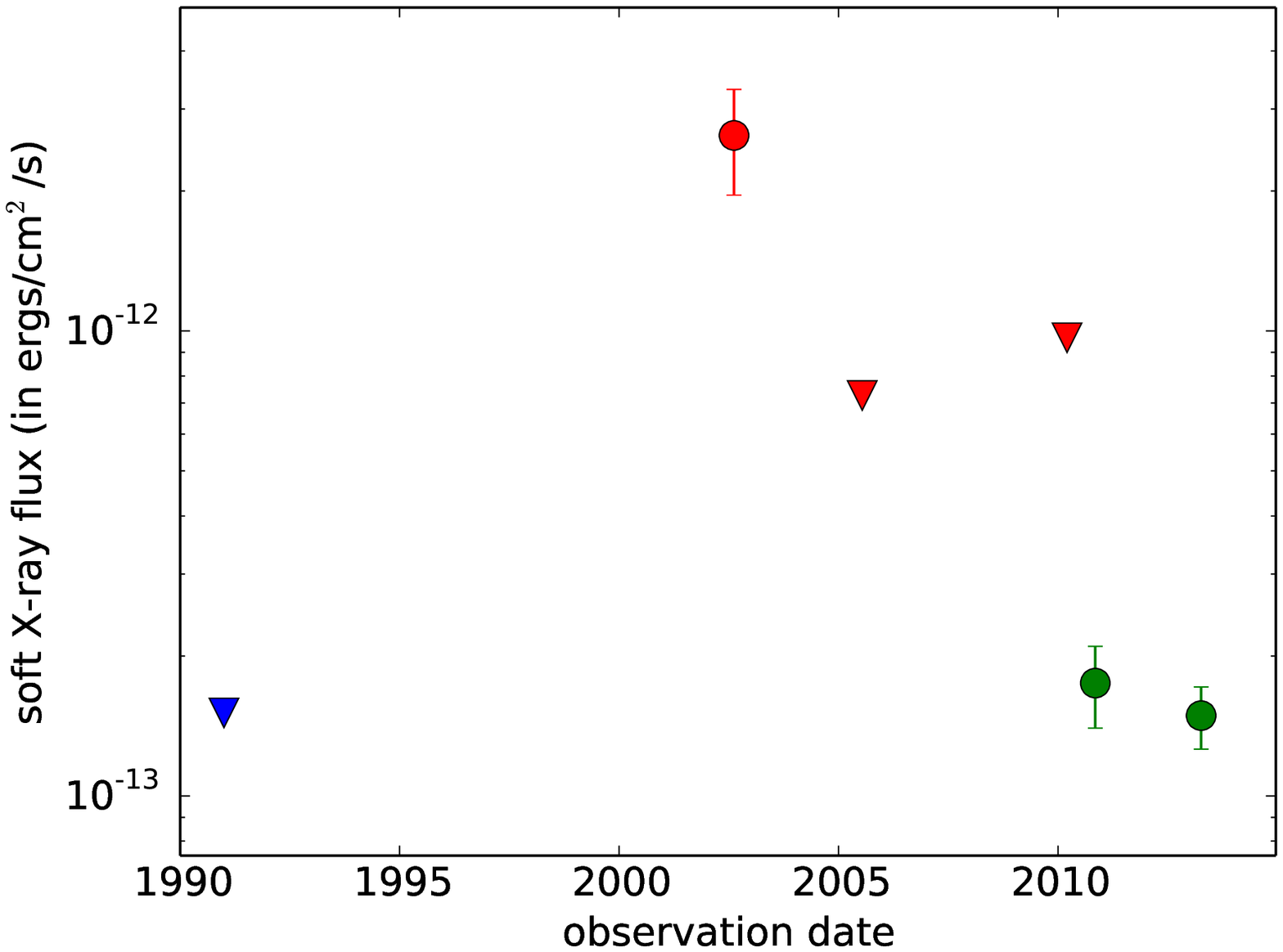}}
\hfill
\subfloat[XMMSL1\,J051935.5-323928]{\label{fig:J051} \includegraphics[width=60mm]{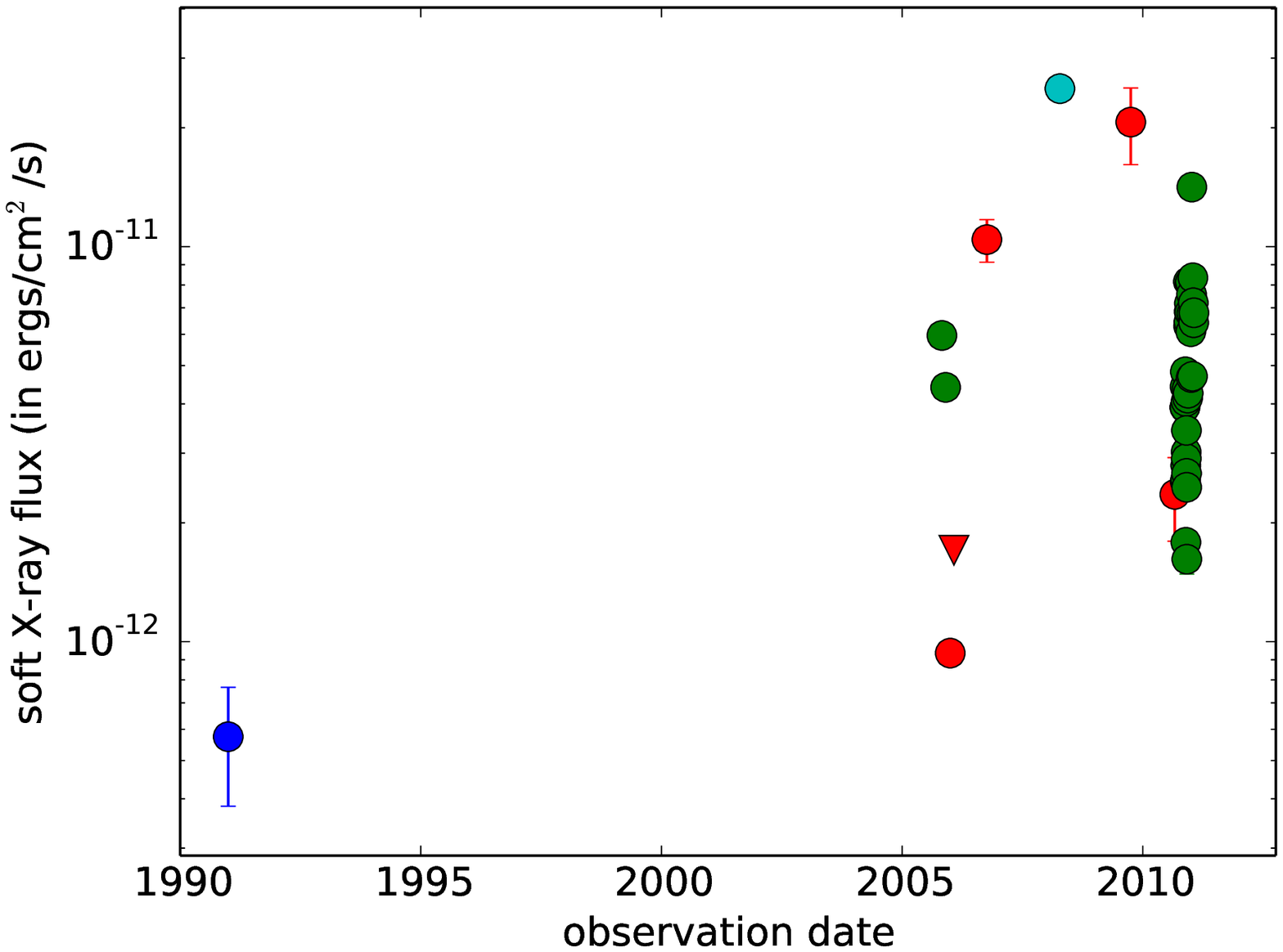}}
\hfill
\subfloat[XMMSL1\,J064541.1-590851]{\label{fig:J064} \includegraphics[width=60mm]{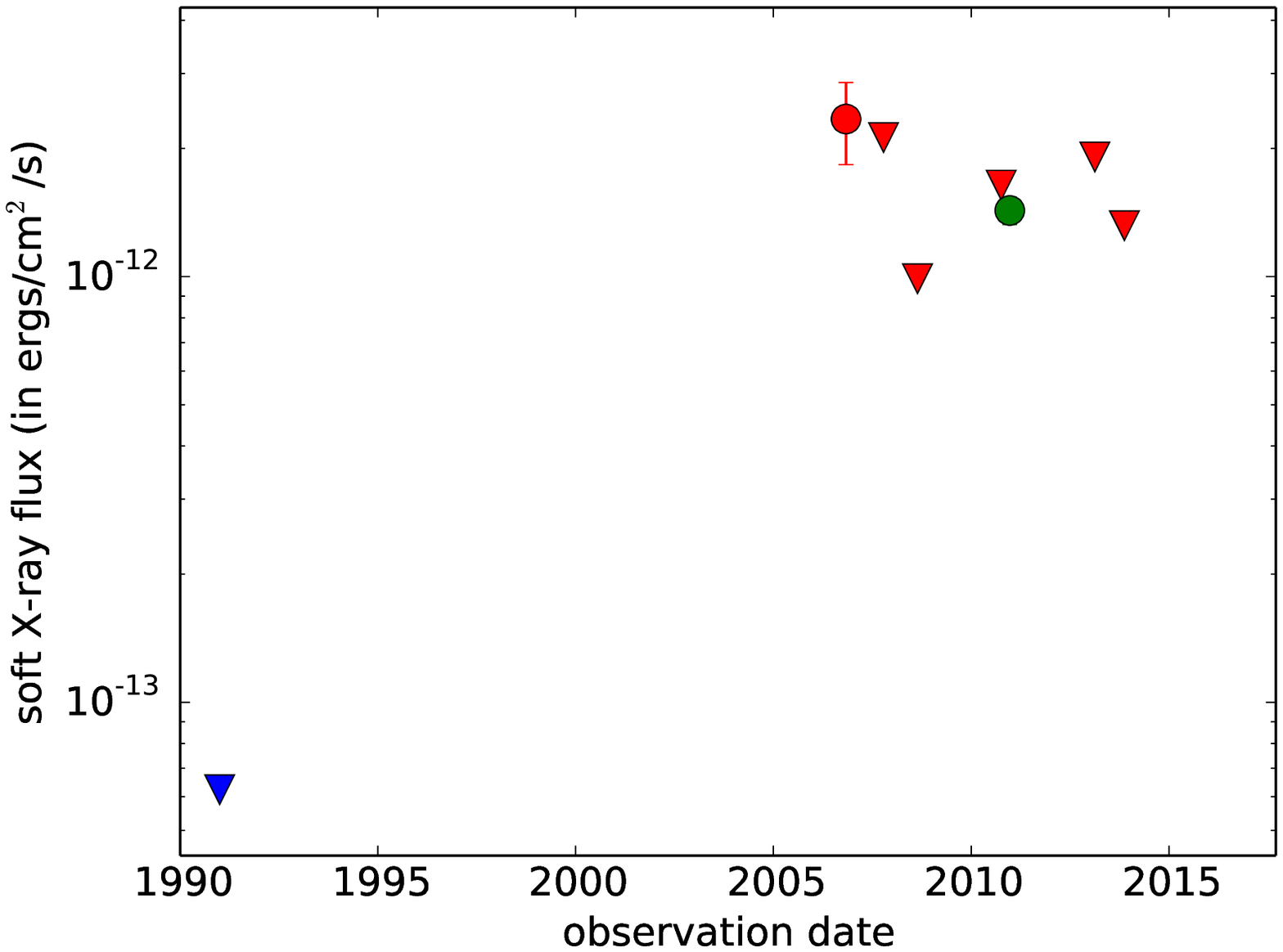}}

\subfloat[XMMSL1\,J070841.3-493305]{\label{fig:J070} \includegraphics[width=60mm]{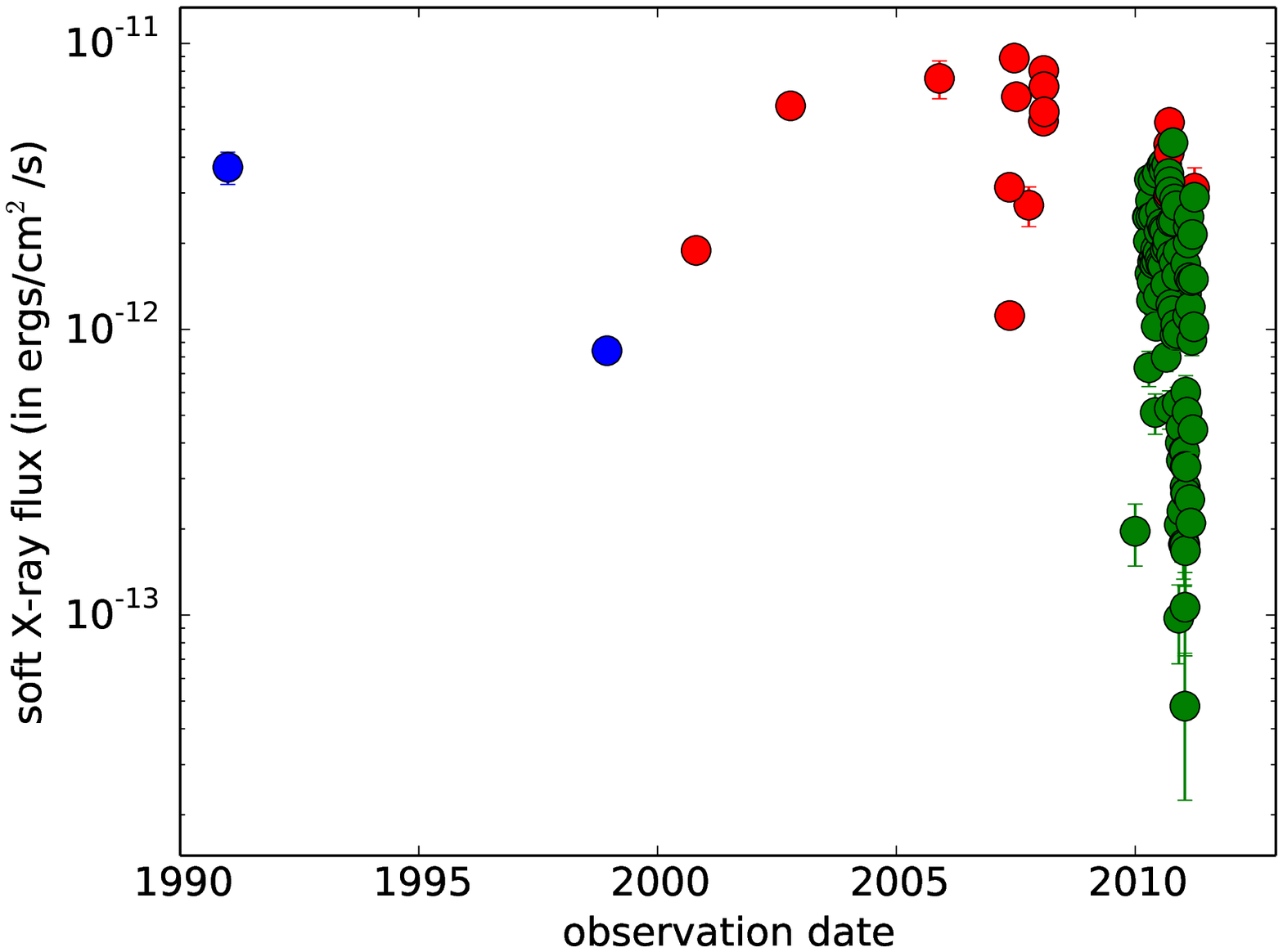}} 
\hfill
\subfloat[XMMSL1\,J082753.7+521800]{\label{fig:J082} \includegraphics[width=60mm]{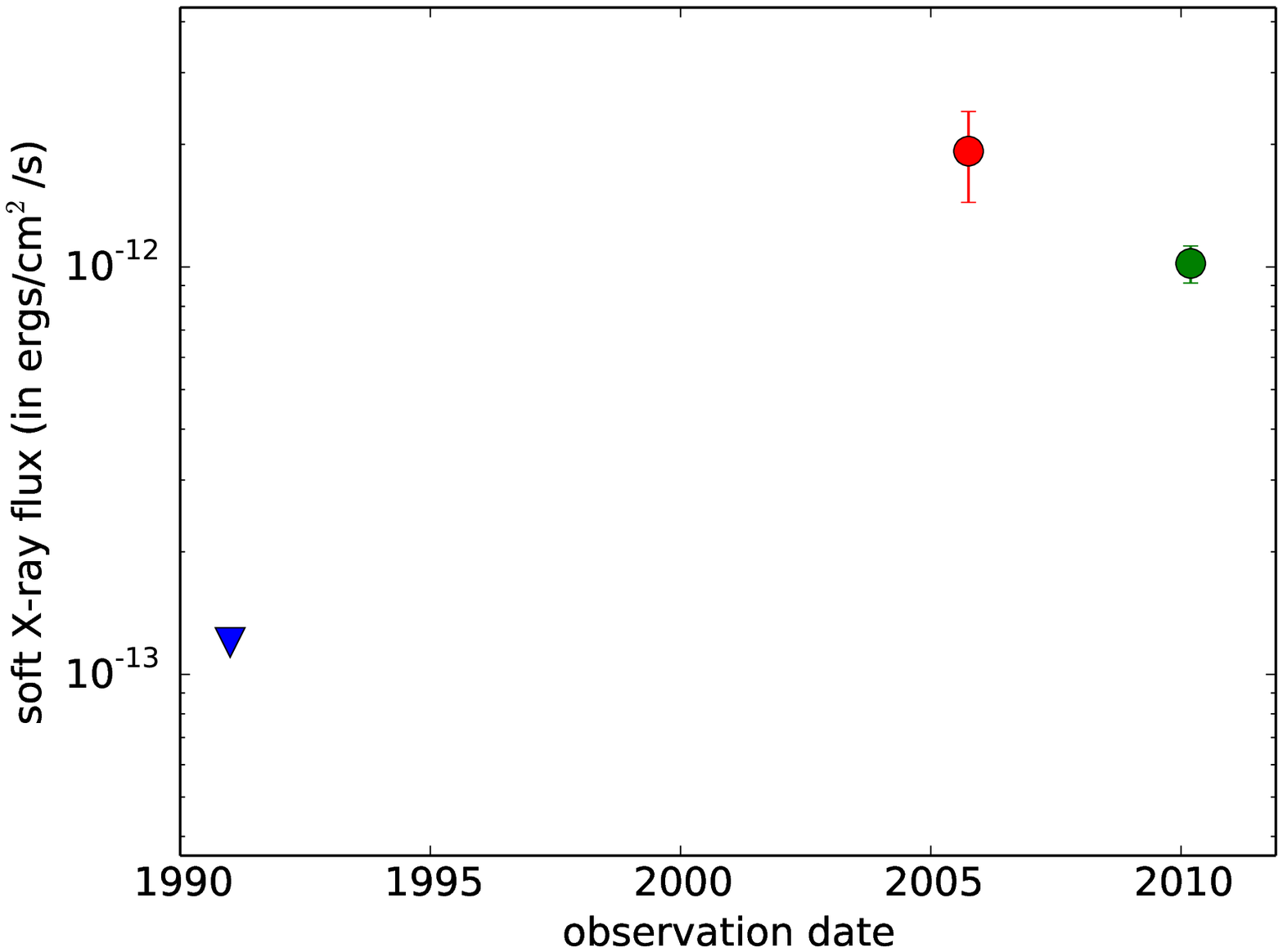}}
\hfill
\subfloat[XMMSL1\,J090421.2+170927]{\label{fig:J090} \includegraphics[width=60mm]{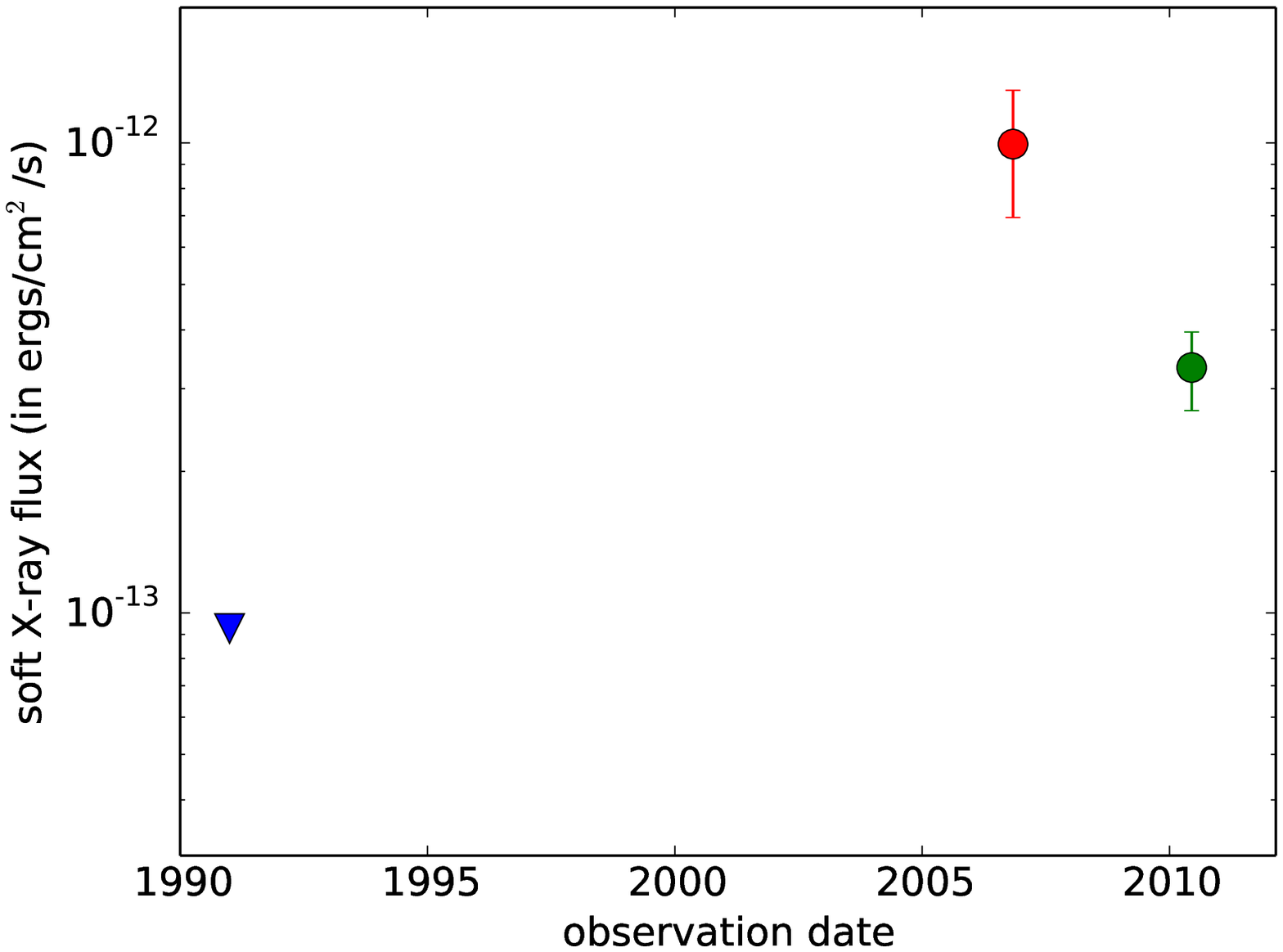}}

\caption[X-ray light curves]{Soft X-ray light curves.}
\end{figure*}
\begin{figure*}[ht]
\centering
\ContinuedFloat
\subfloat[XMMSL1\,J093922.5+370945]{\label{fig:J093} \includegraphics[width=60mm]{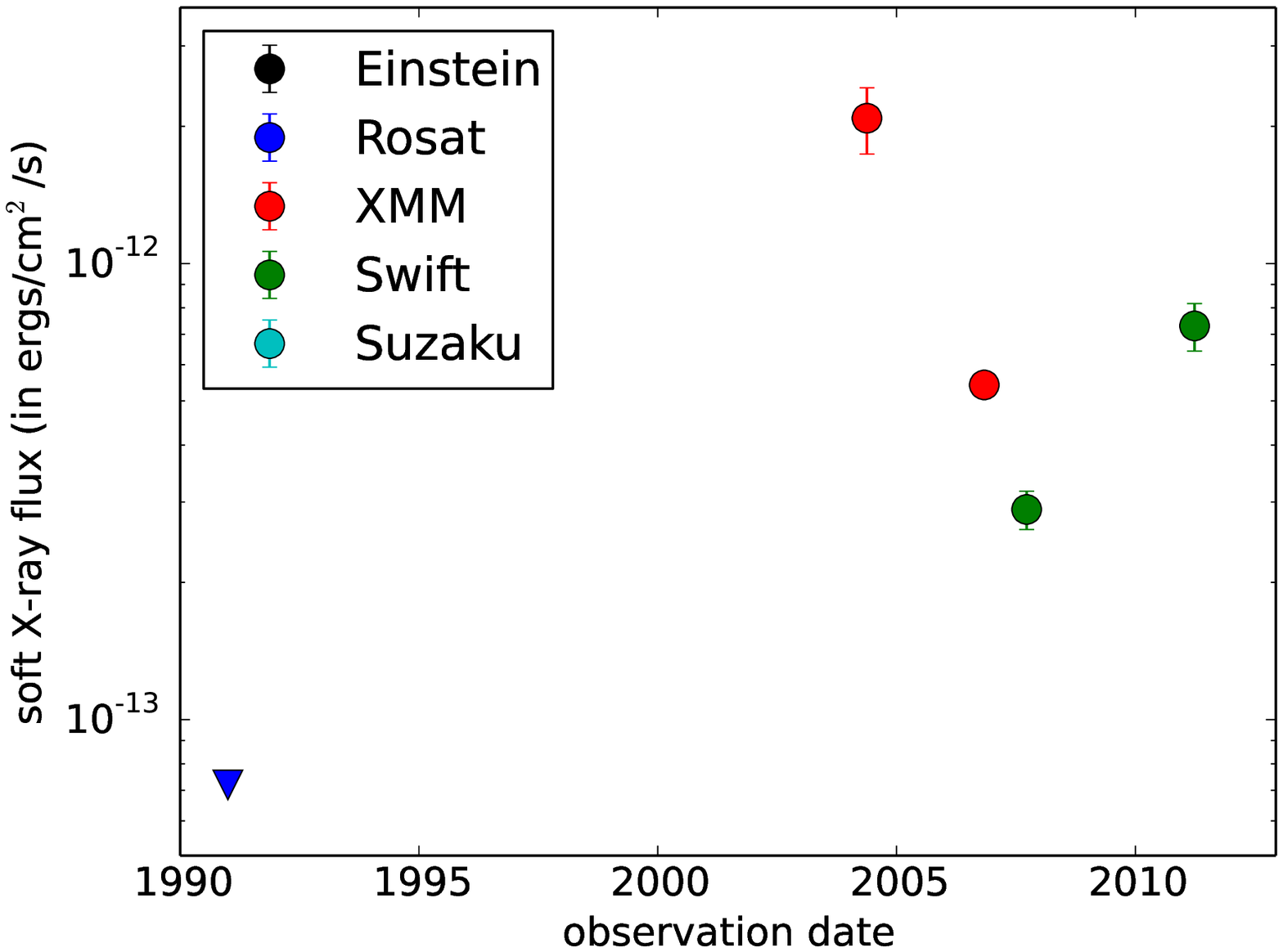}}
\hfill
\subfloat[XMMSL1\,J100534.8+392856]{\label{fig:J100} \includegraphics[width=60mm]{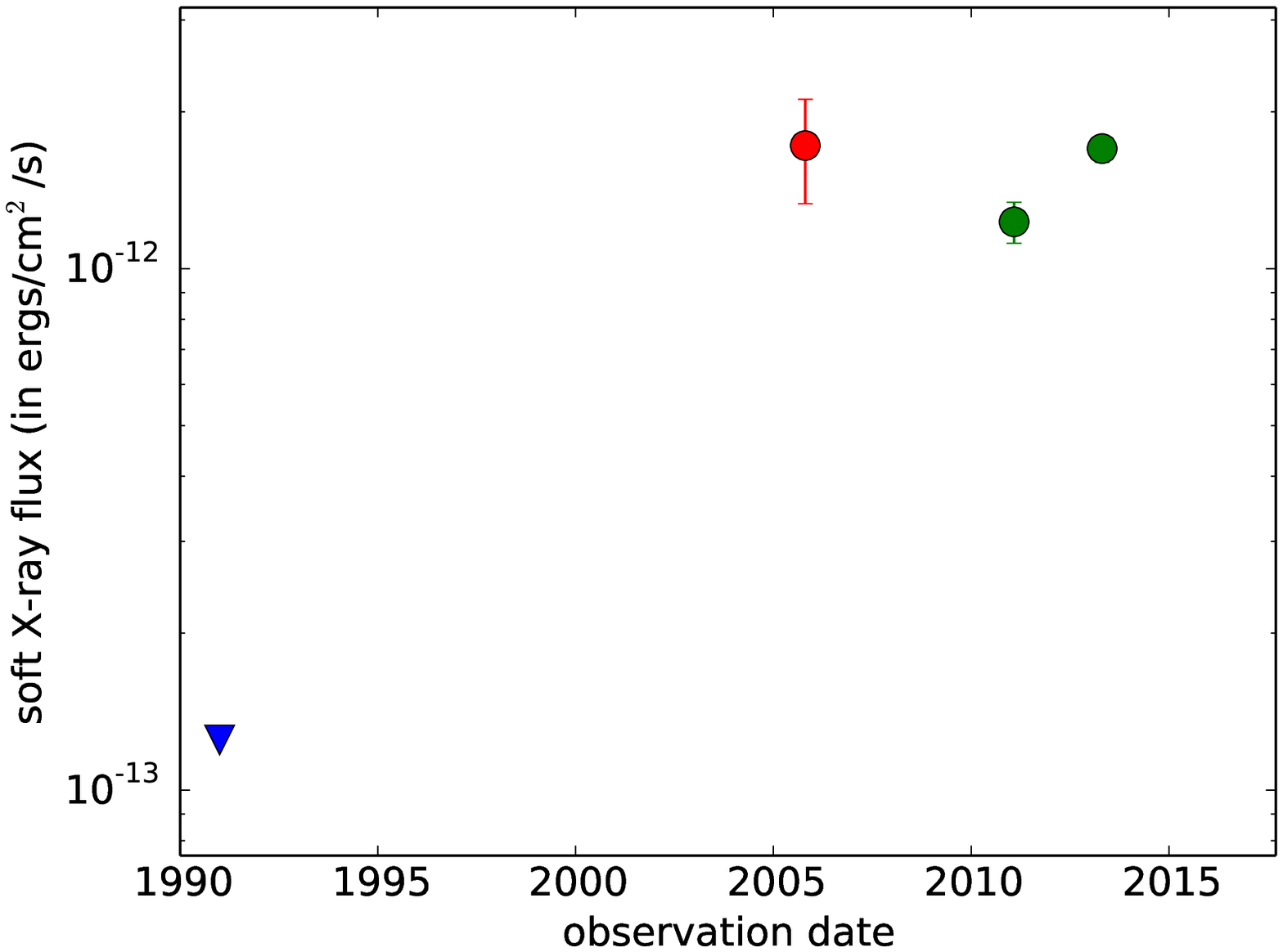}}
\hfill
\subfloat[XMMSL1\,J104745.6-375932]{\label{fig:J104} \includegraphics[width=60mm]{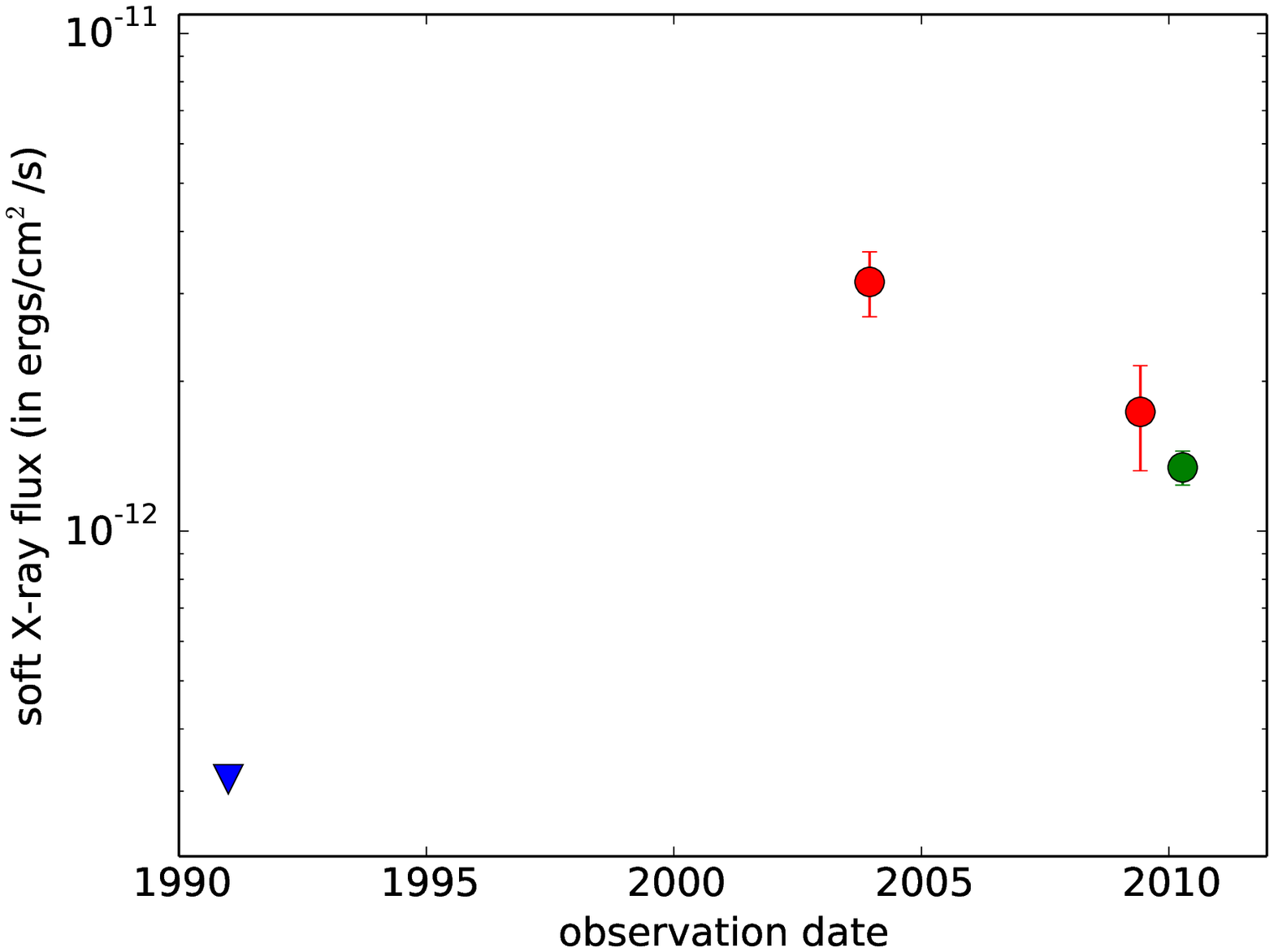}}

\subfloat[XMMSL1\,J111527.3+180638]{\label{fig:J111} \includegraphics[width=60mm]{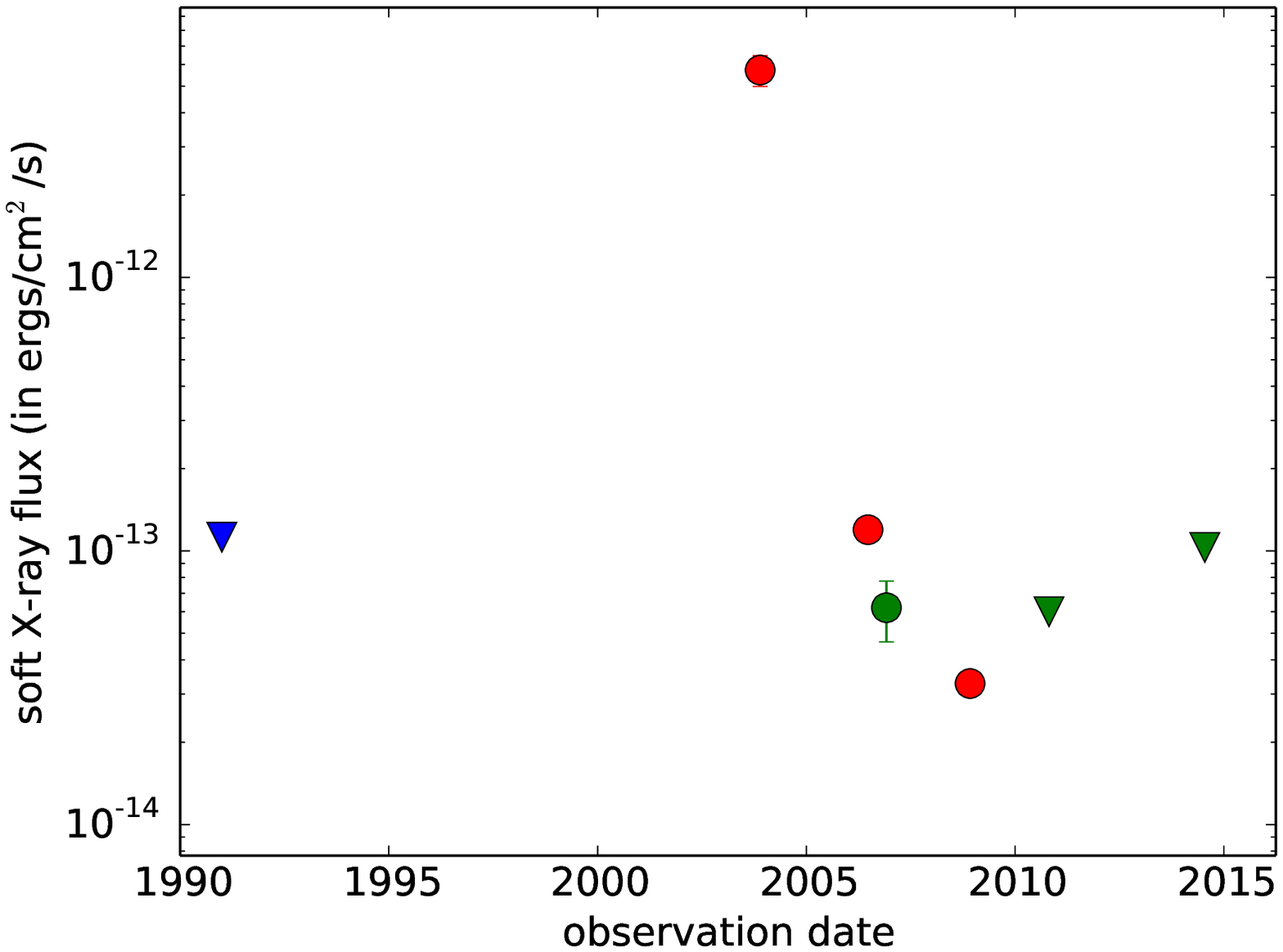}}
\hfill
  \subfloat[XMMSL1\,J112841.5+575017]{\label{fig:J112} \includegraphics[width=60mm]{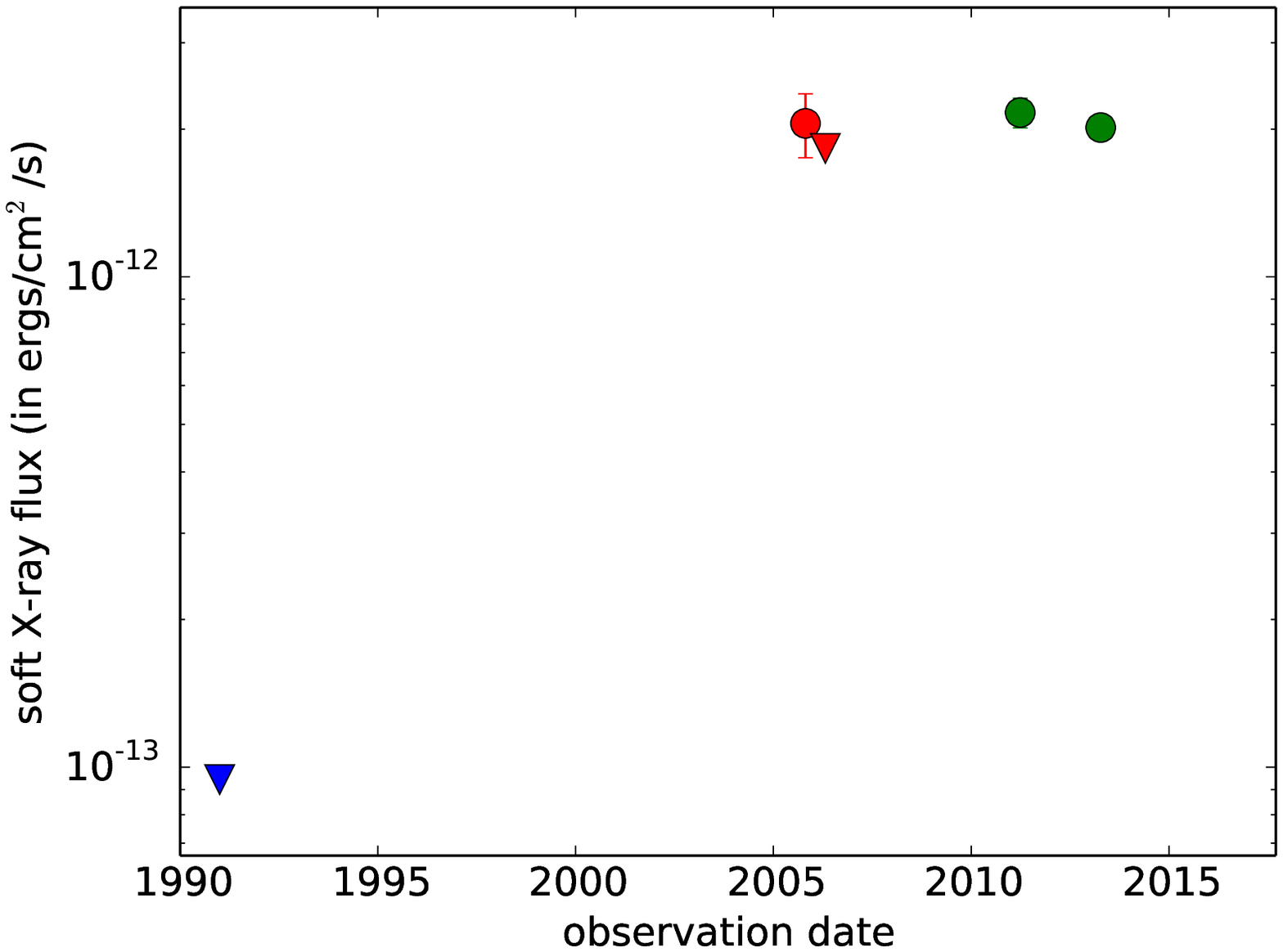}}
\hfill
  \subfloat[XMMSL1\,J113001.8+020007]{\label{fig:J113} \includegraphics[width=60mm]{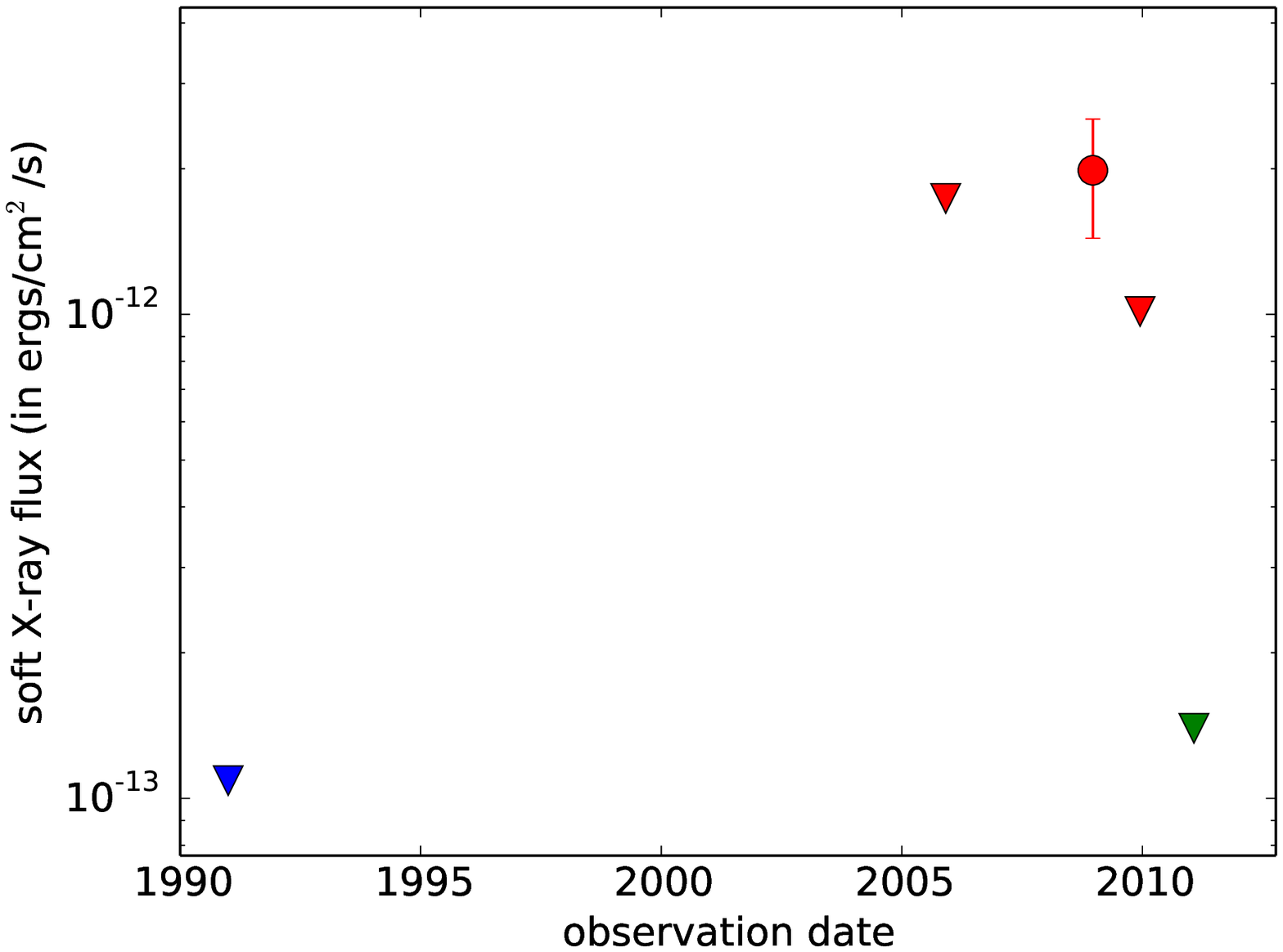}}

  \subfloat[XMMSL1\,J121335.0+325609]{\label{fig:J121} \includegraphics[width=60mm]{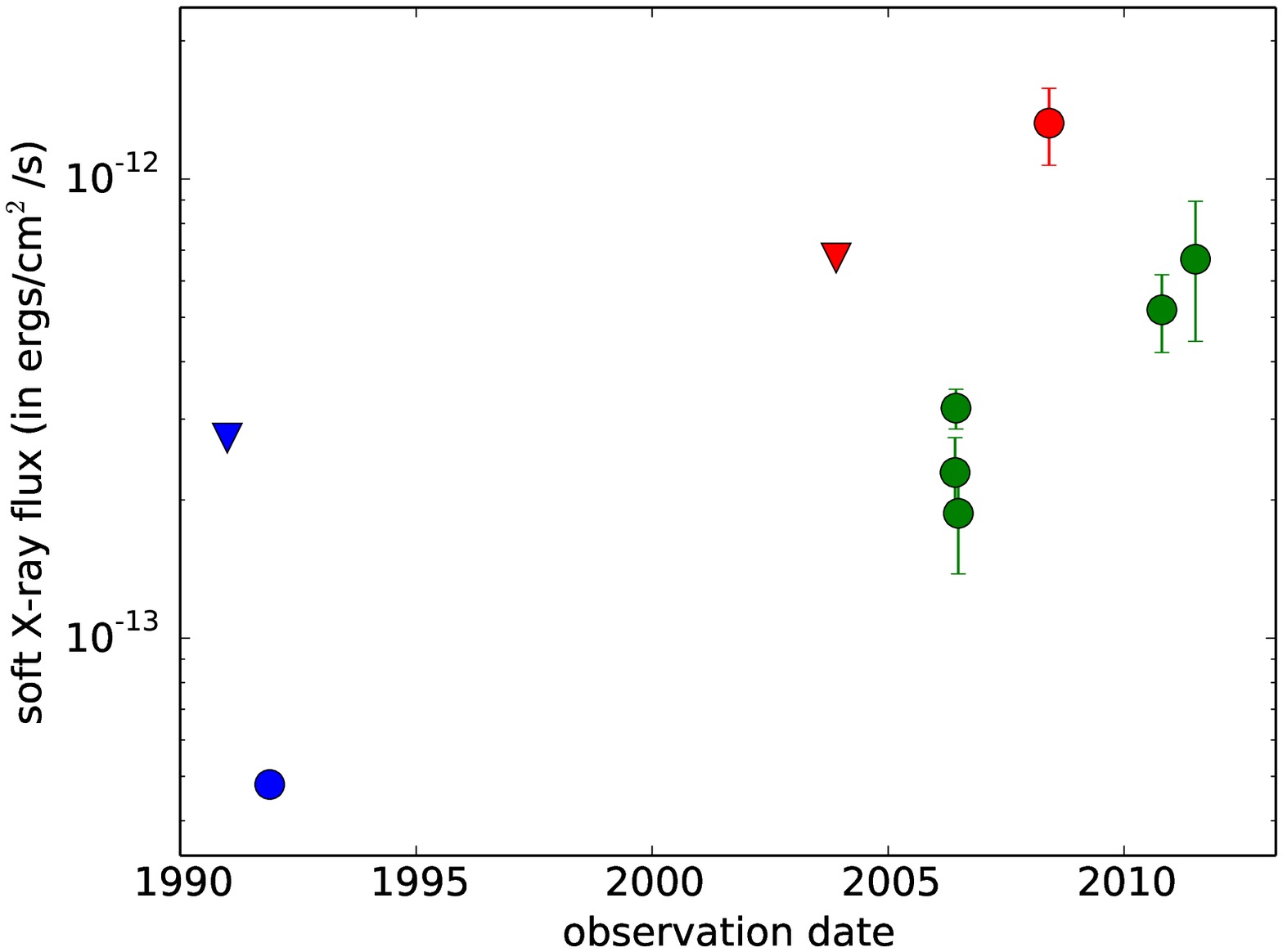}}
\hfill
 \subfloat[XMMSL1\,J132342.3+482701]{\label{fig:J132} \includegraphics[width=60mm]{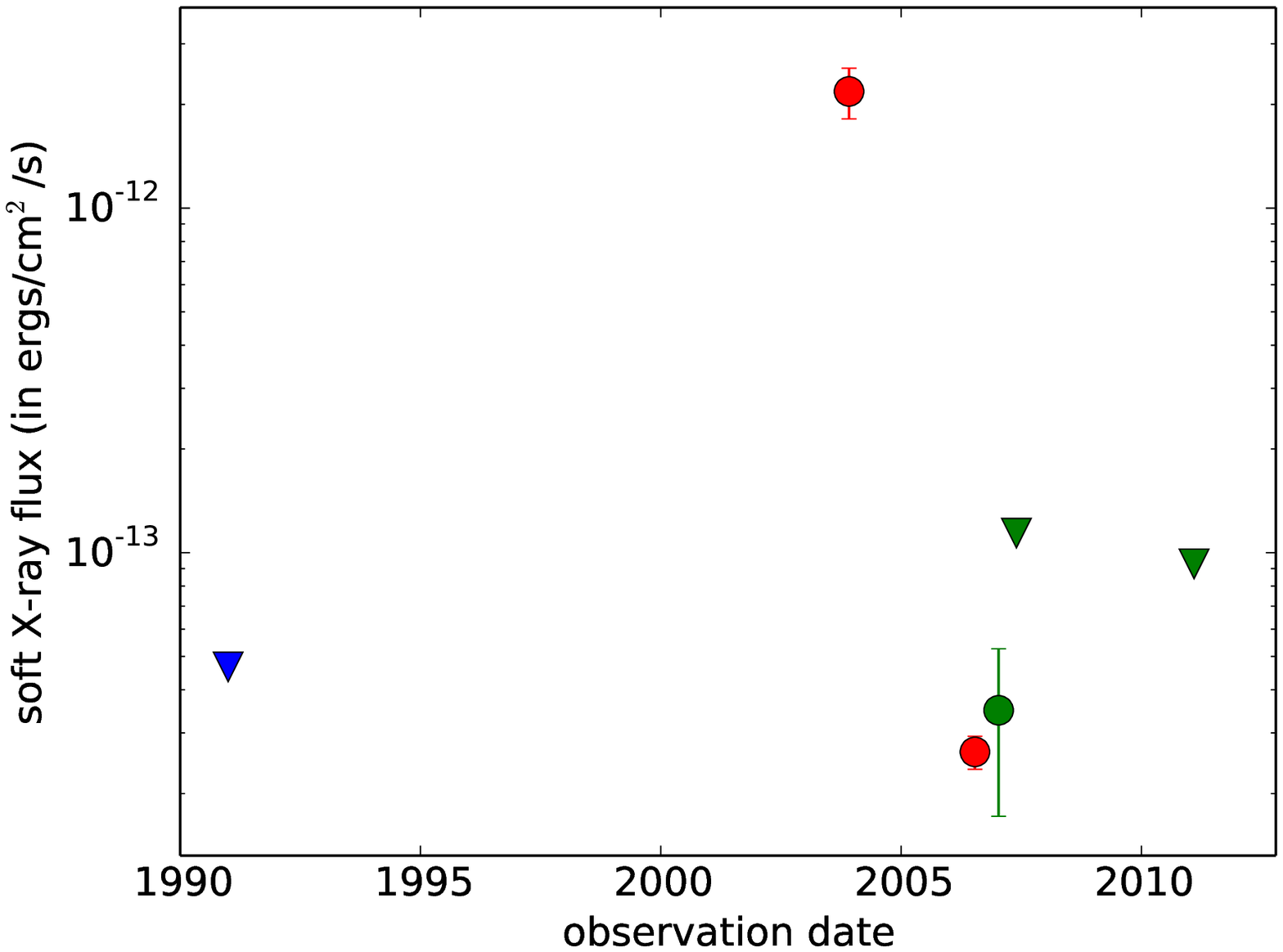}}
\hfill
 \subfloat[XMMSL1\,J162553.4+562735]{\label{fig:J162} \includegraphics[width=60mm]{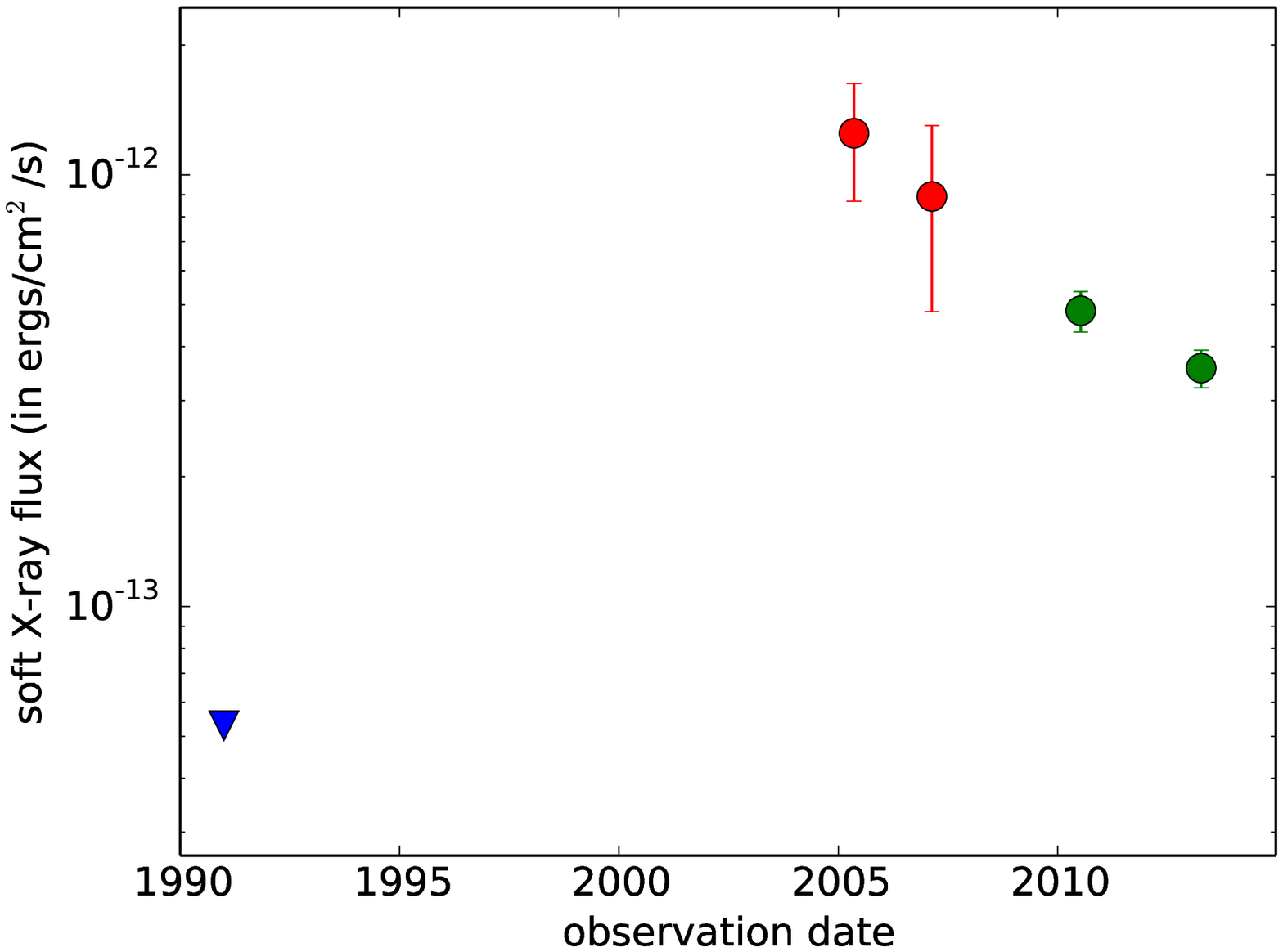}}

  \subfloat[XMMSL1\,J173738.2-595625]{\label{fig:J173} \includegraphics[width=60mm]{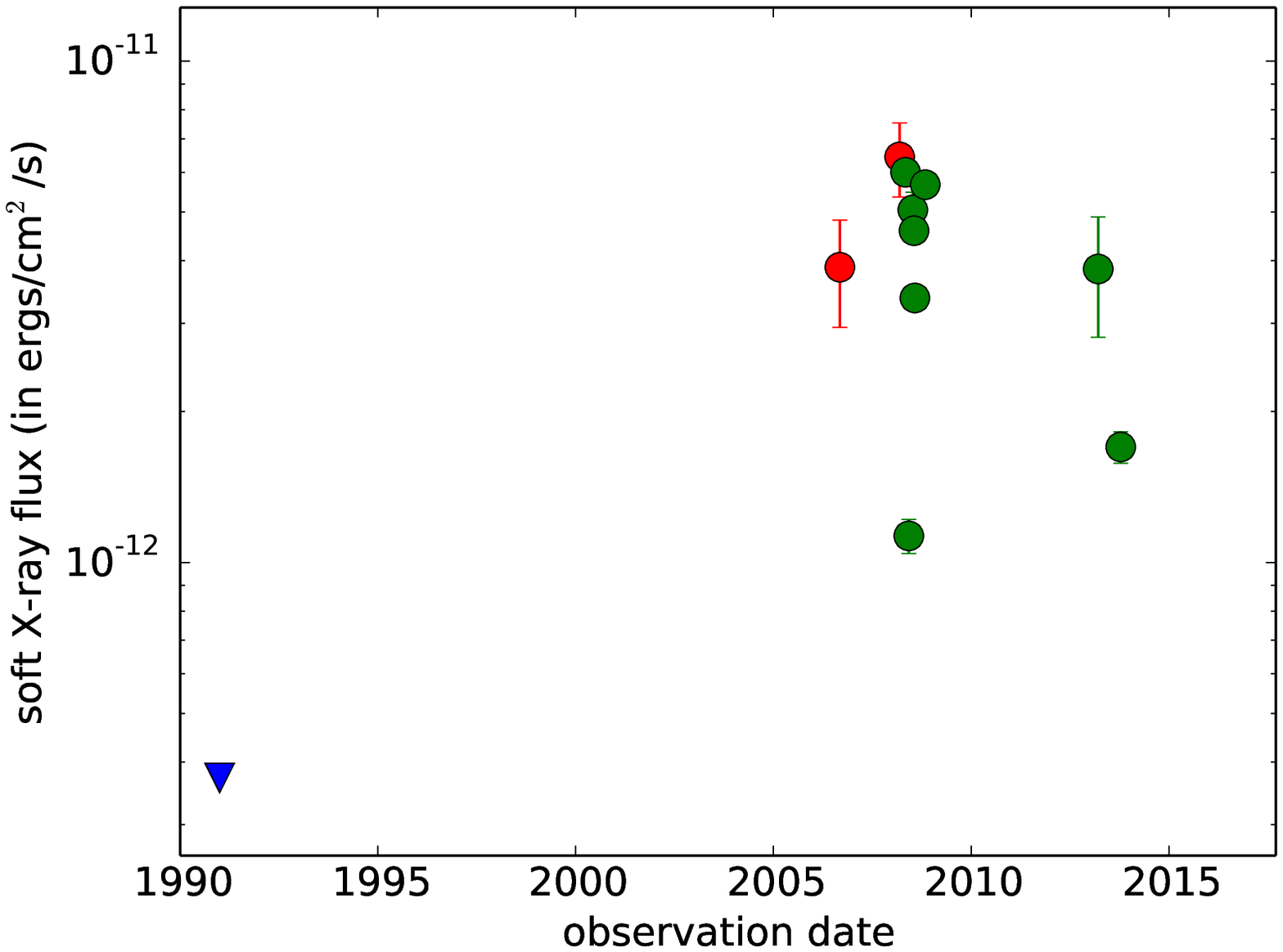}}
\hfill
  \subfloat[XMMSL1\,J183521.4+611942]{\label{fig:J183} \includegraphics[width=60mm]{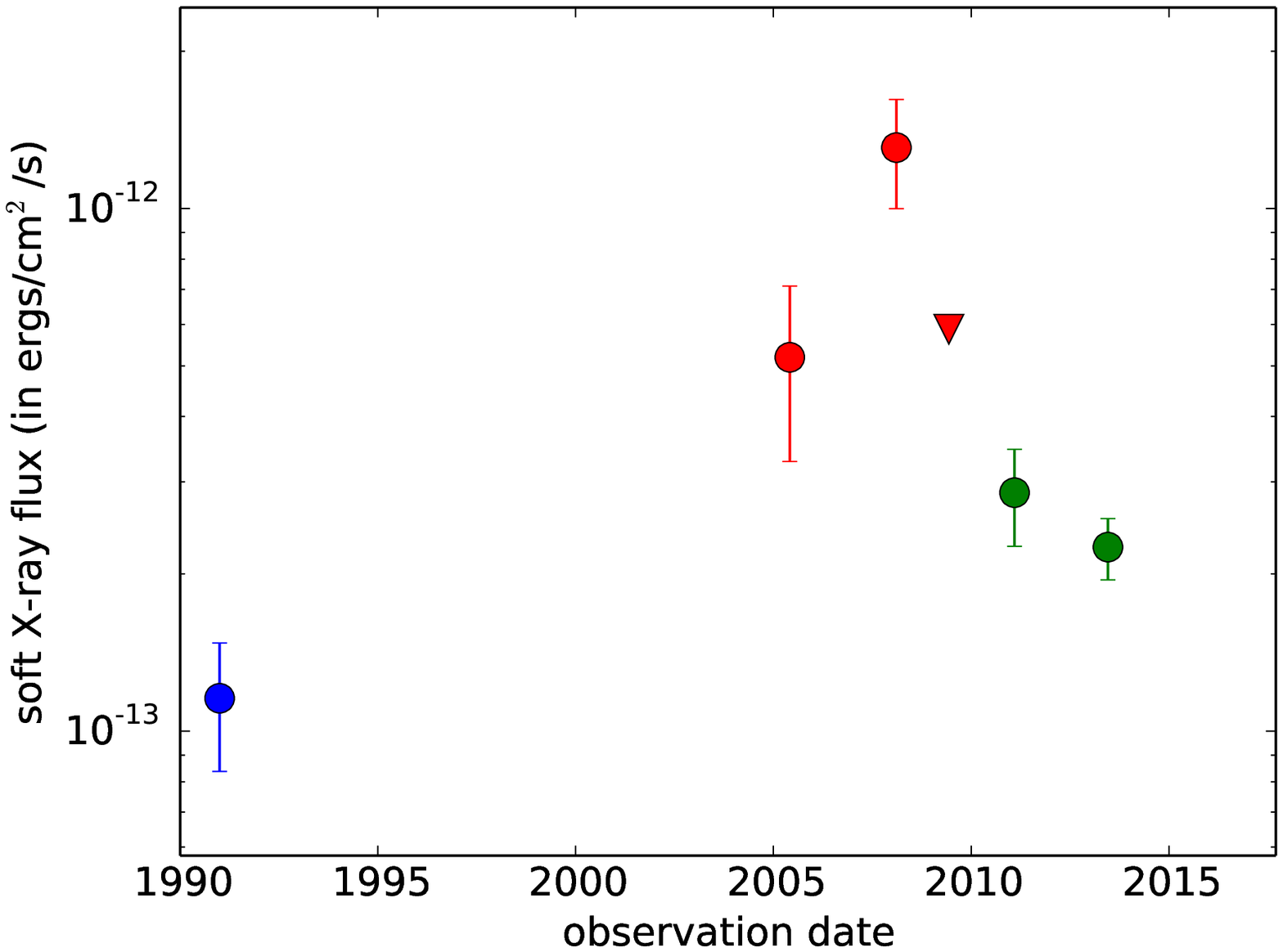}}
\hfill
  \subfloat[XMMSL1\,J193439.3+490922]{\label{fig:J193} \includegraphics[width=60mm]{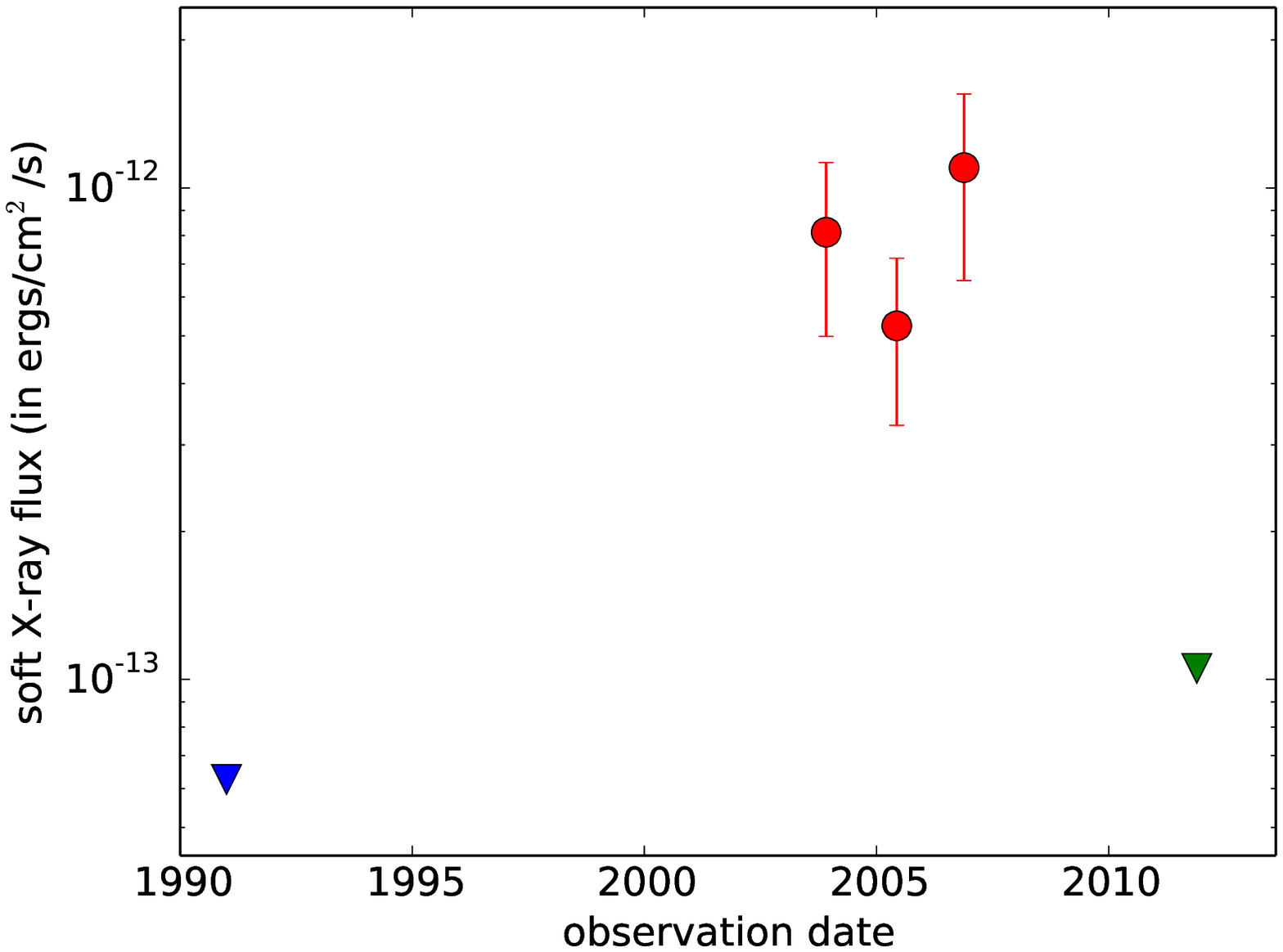}}
\caption[X-ray light curves]{\label{fig:lightcurves} Soft X-ray (0.2--2\,keV) light curves for all the sources in our sample. Count rates of the different telescopes were converted to the 0.2--2.0\,keV range using PIMMS under the assumption that the spectral shape can be described by a power law with a photon index of 1.7 and absorption in our Galaxy only. Circles mark detections, while triangles symbolise 95\% upper limits on the flux. All error bars are 1$\sigma$.}
\end{figure*}

In Figure \ref{fig:lightcurves} we plot the soft X-ray light curves for our candidate highly variable AGN using available X-ray data taken by the satellite missions {\it Einstein, ROSAT, XMM, Suzaku} and {\it Swift}. The count rates were obtained from different archives including HEASARC, the \xmm\ Science Archive, the \swift\ UKSSDC and from our own {\it Swift} XRT data analysis, and for upper limits the 1SXPS catalogue \citep{evans2014} and the \xmm\ upper limit server\footnote{\url{http://xmm.esac.esa.int/external/xmm\_products/}\\\url{slew\_survey/upper\_limit/uls.shtml}} were queried.

The count rates of the different satellites were converted to fluxes between 0.2--2.0\,keV using PIMMS\footnote{\url{http://heasarc.gsfc.nasa.gov/Tools/w3pimms.html}} assuming a power law with a photon index of 1.7 as a spectral shape taking into account Galactic extinction as given by \citet{willingale2013}. 

\citet{SobPap09} found a positive correlation between flux and spectral slope for a sample
of bright RXTE AGN in the 2--10\,keV band. This could affect the relative fluxes seen
in our sample and plotted in Figure \ref{fig:lightcurves}. We have attempted to quantify this for the
different detectors used in the creation of our light curves. The sample of \citet{SobPap09}
showed spectral changes with observed power-law slope varying between 1.0
and 2.0 (see their Figure 7). For a typical Galactic absorption of $3\times10^{20}\text{cm}^{-2}$ the change from slope of 1.0 to 2.0 would alter
our estimated fluxes by $-$14\% (ROSAT), $-$13\% (XMM-Newton), +7\% (Swift-XRT), +76\% (Suzaku), +25\% (Einstein-IPC). The change is large for Suzaku observations since in this case we use the count rate between 2.0--10.0\,keV and extrapolate it to the soft band. All other satellites are sensitive in the soft band and the fluxes are hence less dependent upon the assumed spectral index.

Six sources within our sample (XMMSL1\,J024916.6-041244, J034555.1-355959, J045740.0-503053, 
J051935.5-323928, J070841.3-493305 and J193439.3+490922) display factor 10 or greater 
variation in flux between at least one pair of \xmm\ and \swift\ observations, 
on timescales of months to years. 
The ratio between the soft X-ray flux observed with \swift\ and that 
observed with \xmm\ for the remaining sources is typically a factor 
of a few. 
We observed the two TDE candidates with XRT, and found that both had faded 
significantly, following expectations from previous and later fluxes and upper limits (Figures \ref{fig:J111} and \ref{fig:J132}).

\section{Relative X-ray luminosity} \label{sec:alphaox}
\begin{figure}[htb]
\centering
\includegraphics[width=88mm]{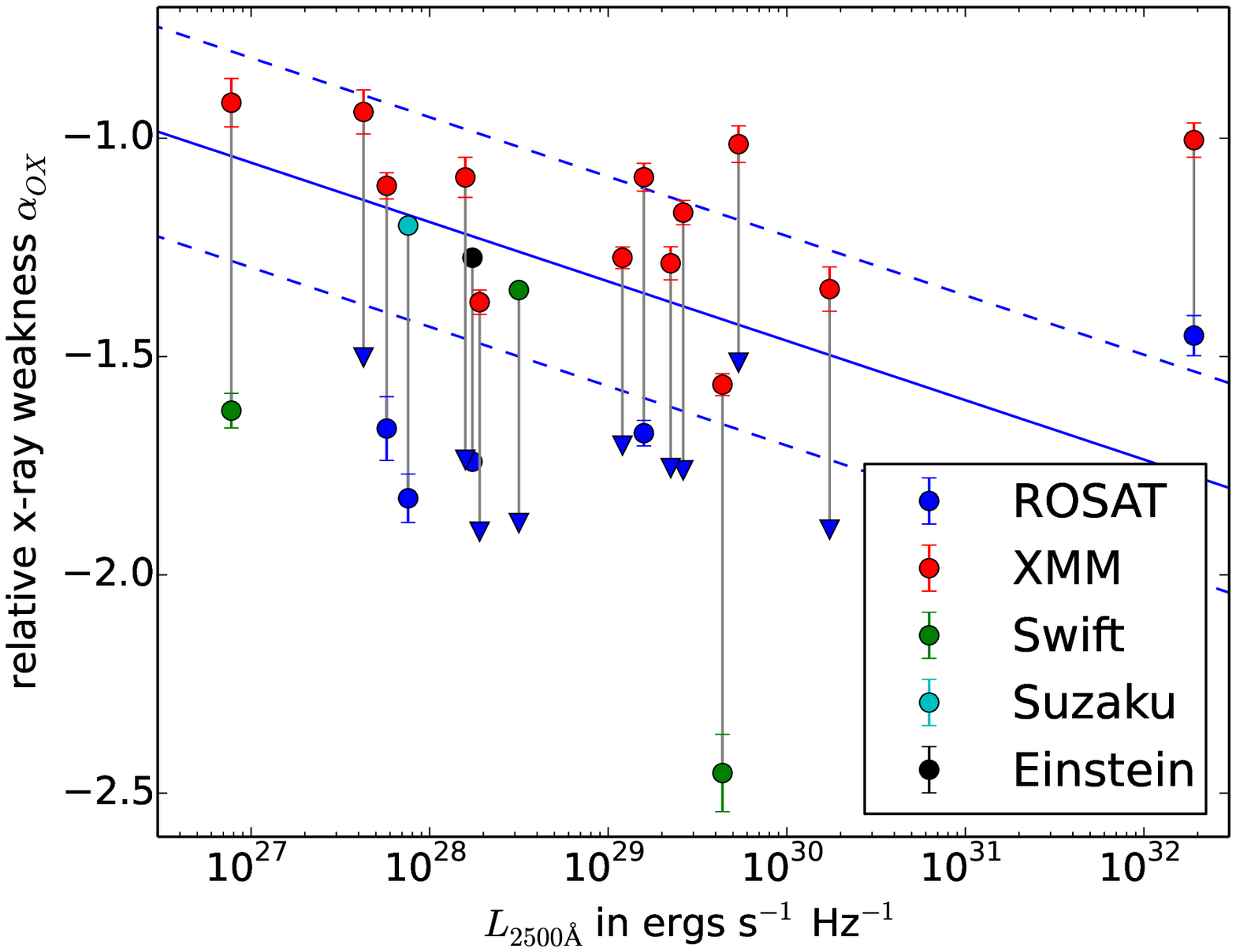}
\caption[$\alpha_{OX}$ vs redshift]{ \label{fig:alpha_ox} $\alpha_{OX}$
plotted against $L_{2500\AA}$ for the high-variability sample. 
For clarity only the brightest and faintest X-ray observations are shown.
$L_{2500\AA}$ is measured from the SED of each source and is assumed to 
be constant between observations.}
\end{figure}

Since the X-ray flux of AGN mainly consists of UV photons which gain energy in inverse Compton scattering processes the UV and the X-ray flux are closely correlated as described in \citet{just2007}. This allows us to estimate whether the X-ray luminosities of our sources are relatively bright or faint compared to AGN with the same UV luminosity. In Figure \ref{fig:alpha_ox} we compare the highest and lowest observed X-ray flux to the expected value using the relation given in \cite{just2007}, now omitting the two probable TDEs and the two probable spurious detections.

The relative X-ray brightness $\alpha_\text{OX}$ is defined as
\begin{equation}
\alpha_\text{OX}=0.3838 \cdot \log \left( \frac{F_\text{2keV}}{F_\text{2500\AA}}\right),
\end{equation}
where $F_\text{2keV}$ is the monochromatic flux at 2\,keV and $F_\text{2500\AA}$ the one at 2500\,$\AA$. Since the highest observed fluxes normally are \xmm\ slew observations and the lowest ones either upper limits or observations with few counts, we can not use the spectra to determine the monochromatic flux at 2\,keV. Instead we use PIMMS to obtain the flux $F_\text{broad}$ between the energies $E_1$=1.9\,keV and $E_2$=2.1\,keV, assuming as before an absorbed power law with a photon index of $\Gamma=1.7$. This result can be converted to the monochromatic flux at 2\,keV using the relation
\begin{equation}
F_\text{2 keV}=\frac{(2-\Gamma) E^{(1-\Gamma)}}{E_2^{(2-\Gamma)}-E_1^{(2-\Gamma)}}\cdot F_\text{broad}.
\end{equation}
The flux at 2500\,$\AA$ is approximated using measurements by the \xmm\ Optical Monitor, by the \swift\ UVOT or by Galex, depending on which is closest to the wavelength after considering the redshift of the AGN. For most sources, 
several measurements are available and we linearly interpolate the two data points which bracket the required wavelength, in double logarithmic space. If only one measurement exist we use this value.
For two sources, XMMSL1\,J044347.0+285822 and XMMSL1\,J173738.2-595625, there are no reasonably close measurements available and we estimate a value by
extrapolating the SED by eye.
For those two cases 
the numbers have to be treated as an order of magnitude estimate, 
while for the other AGN we estimate the uncertainties on the UV flux,
introduced by the interpolation, to be $\leq0.1$ dex.

Since the UV light is heavily affected by extinction, we need to correct for the Galactic hydrogen column density \citep{schlegel1998}. It would be preferable to consider intrinsic absorption as well, however as mentioned above for most of those data points no spectra are available, such that we do not have any information about intrinsic absorbers. The X-ray fluxes are also corrected for Galactic absorption.

We note that the UV and X-ray data are typically not simultaneous, 
and as these are variable sources this could, in principle, introduce 
errors on our measured values of $\alpha_{\rm OX}$. In practise, X-ray
variability tends to be much greater than UV variability 
\citep[e.g.][]{Grupe12mkn335,saxton14} and so changes
in $\alpha_{\rm OX}$ will be dominated by the X-ray luminosity.
Our calculated values are given in Table \ref{tab:alphaox}.


\begin{table}
{\small
\hfill{}
\caption{2\,keV:2500\,\AA\, flux ratio $\alpha_\text{OX}$.}
\label{tab:alphaox}      
\begin{center}
\begin{tabular}{l c c c}        
\hline\hline                 
XMMSL1 name & $\alpha_{\rm OX,exp}$ & $\alpha_{\rm OX,min}$ & $\alpha_{\rm OX,max}$ \\ \hline
J005953.1+314934 &$-$1.2& $-$1.741 $\pm$ 0.008  & $-$1.274 $\pm$ 0.005 \\
J020303.1-074154 &$-$1.2& $\le-$1.7  &$-$1.09 $\pm$ 0.05 \\
J024916.6-041244 &$-$1.0& $-$1.62 $\pm$ 0.04 &$-$0.92 $\pm$ 0.06 \\
J034555.1-355959 & -&- &- \\
J044347.0+285822 &$-$1.2& $-$1.67 $\pm$ 0.07 &$-$1.11 $\pm$ 0.03 \\
J045740.0-503053 &-&- &- \\
J051935.5-323928 &$-$1.2&$-$1.82 $\pm$ 0.05 &$-$1.20 $\pm$ 0.01 \\
J064541.1-590851 &-&- &- \\
J070841.3-493305 &$-$1.4& $-$2.45 $\pm$ 0.09 &$-$1.56 $\pm$ 0.03\\
J082753.7+521800 &$-$1.4& $\le-$1.5  &$-$1.01 $\pm$ 0.04\\
J090421.2+170927 &$-$1.1& $\le-$1.5  &$-$0.94 $\pm$ 0.05\\
J093922.5+370945 &$-$1.4& $\le-$1.8  &$-$1.17 $\pm$ 0.03\\
J100534.8+392856 &$-$1.4& $\le-$1.8  &$-$1.29 $\pm$ 0.04\\
J104745.6-375932 &$-$1.3& $\le-$1.7  &$-$1.27 $\pm$ 0.02\\
J112841.5+575017 &$-$1.3& $\le-$1.9  &$-$1.35 $\pm$ 0.01\\
J121335.0+325609 &$-$1.4& $-$1.68 $\pm$ 0.03 &$-$1.09 $\pm$ 0.03\\
J162553.4+562735 &$-$1.5& $\le-$1.9  &$-$1.35 $\pm$ 0.05\\
J173738.2-595625 &$-$1.2& $\le-$1.9  &$-$1.38 $\pm$ 0.03\\
J183521.4+611942 &$-$1.8& $-$1.45 $\pm$ 0.05 &$-$1.00 $\pm$ 0.04\\
J193439.3+490922 &-&- &- \\
\hline                                   
\end{tabular}
\end{center}
}
\tablefoot{
Columns give the source name, the expected $\alpha_\text{OX}$
for a source with this $L_{2500\AA}$ from the correlation of
\citet{steffen2006}, $\alpha_\text{OX}$ calculated from the minimum  
X-ray flux and $\alpha_\text{OX}$ calculated from the maximum X-ray flux
that we have recorded for that particular source.
}
\hfill{}
\end{table}

We are now able to say how the different X-ray observations of our sources compare to other AGN which have the same UV luminosity. In Figure \ref{fig:alpha_ox} we show the relative X-ray luminosity for the brightest and the faintest X-ray observation and the values for the individual sources are also given 
in Table \ref{tab:alphaox}. The straight line is the relation between 
$\alpha_{\rm OX}$ and UV luminosity, found for an optically-selected sample of 
333 AGN by \citet{steffen2006}, with the $1 \sigma$ deviations indicated. 
Radio-loud sources are relatively X-ray bright \citep{gibson2008}, since additional X-rays arise from the jet activity,
and are expected to deviate from this relation. Indeed the two radio loud sources of our sample lie clearly above the relation. Those are XMMSL1\,J082753.7+521800 with a UV luminosity of $10^{29.6}\text{ergs s}^{-1}\text{Hz}^{-1}$ and XMMSL1\,J183521.4+611942, the blazar at the highest UV luminosity.

For the other sources we summarise that in nearly all cases their most 
luminous observation corresponds to the expected X-ray luminosity, while 
in their faintest states they are X-ray weak. We note that 
none of our non-radio-loud sources have been observed to be X-ray bright.

One source, 1H\,0707-495,
with $\log L_\text{2500\AA}=29.5$ erg, reaches very low values of $\alpha_{\rm OX}$. This low state
has been attributed to a collapse of the X-ray corona to a region so close to the black hole that only a few hard X-rays escape \citep{fabian2012}.

\citet{dong2012} analysed a sample of 49 optically selected, low-mass, 
narrow-line AGN with UV luminosities in the range 
$10^{27-28} \unit[]{erg\ s^{-1} Hz^{-1}}$. They found that nearly
50\% of their sources lie below the 1$\sigma$-region of the relation 
from \citet{steffen2006}. Their result may well be explained 
by X-ray variability.

\section{Spectral analysis} \label{spectra}
The high-variability sample has been selected from \rosat - \xmm\
variability. To attempt to understand the variability mechanism in 
each source, it is 
essential to have spectral information at high and low fluxes, covering 
an energy band broad enough to constrain absorbing column densities.
This is possible for the pointed \xmm\ and \swift\ spectra of our sources but not
for the earlier \rosat\ low-energy spectra or upper limits and for most of the Slew observations. 
While all of our sources have undergone extreme variability in the past, all except six sources show less extreme variations between the Slew and \swift\ observations, common among AGN on months--year time scales.

We analyse the X-ray spectra of \swift\ observations from our programme using the full spectral range from 0.3--10.0\,keV and using models available in Xspec. Many of the spectra suffer from low statistics, such that complex models are poorly constrained. A simple power law model with Galactic 
absorption is a reasonable fit for most of the spectra (using Cash statistics, see Table \ref{tab:models}). Below we explore whether a cold or an ionised absorber can explain the observed variability.

Some of our sources have been observed by \xmm\ in pointed observations. We consider the additional spectra as well as published results in comparison to the observations of this programme. By applying the more complex models of the high quality spectra to our observations, we can observe how the parameters of the model change. 

The results of the spectral fits for individual sources are discussed in Appendix \ref{sec:individual}. The parameters of the best fitting models are given in Table \ref{tab:models}.
The spectra of XMMSL1\,J024916.6-041244 are of particular interest being
very soft and lacking the usual power law emission seen 
above 2\,keV. This source is discussed further in \ref{sec:J024}.

\subsection{Neutral absorber}
One obvious reason for variability is a change in absorption along the line-of-sight. In order to quantify whether this is a possible explanation for the 
change in brightness of our sources, we check whether the spectra are compatible with thick enough absorbers to explain the observed variability. 

\begin{figure}[htb]
\centering
\includegraphics[width=88mm]{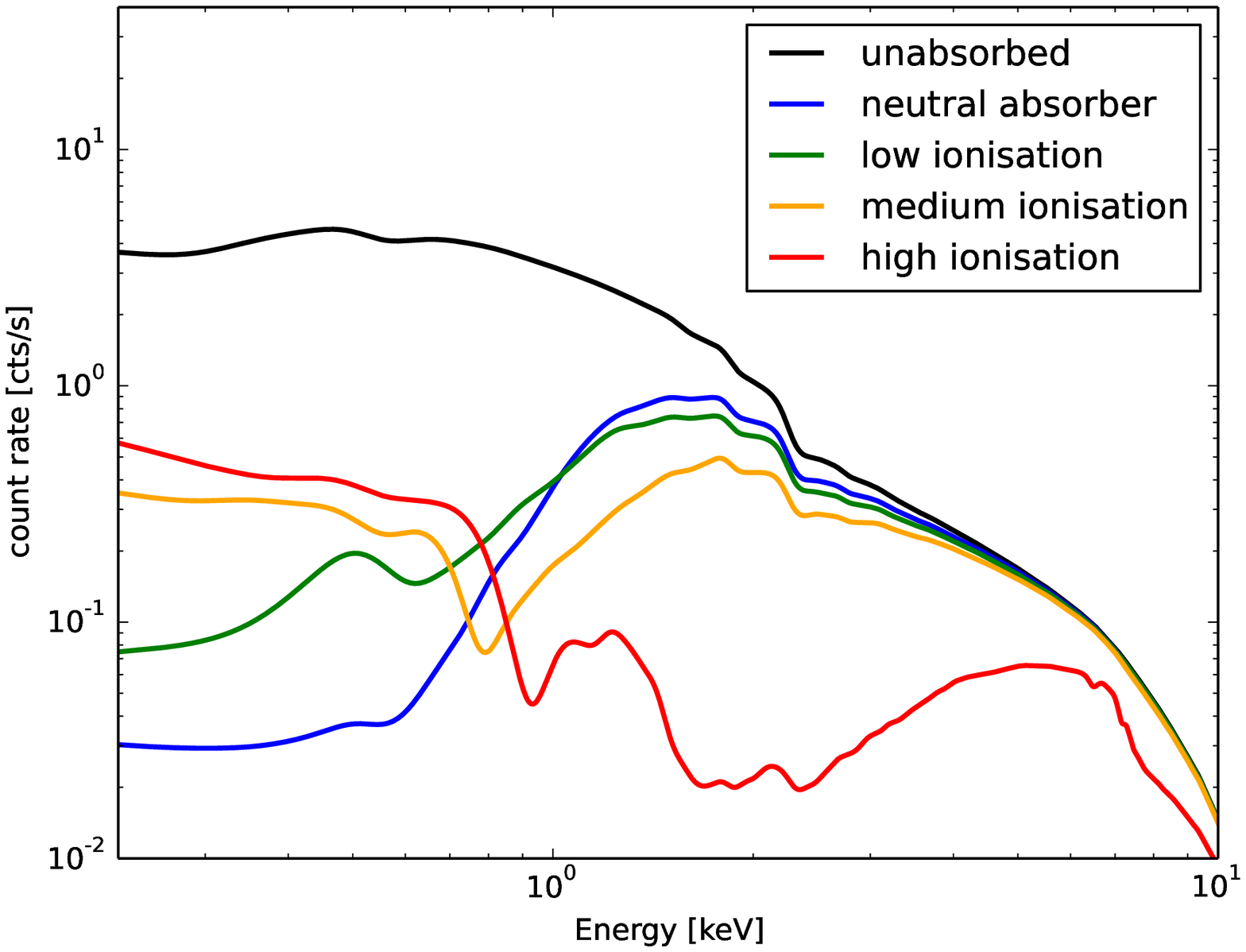}
\caption[Spectra absorbed by an ionised absorber.]
{ \label{fig:absorber} The effect of absorption on a power law spectrum
of $\Gamma=1.7$ and Galactic absorption of 
$N_H=3 \times 10^{20}\unit[]{cm^{-2}}$ here adopting the effective area of \xmm. Absorbers with various ionisation
parameters have been tuned such that the soft (0.2--2\,keV) X-ray flux is 
reduced by a factor ten. The necessary column densities are 
$0.89 \times 10^{22} \unit[]{\,cm^{-2}}$ for the neutral absorber, 
$0.67 \times 10^{22} \unit[]{\,cm^{-2}}$ for low ionisation ($\log(\xi)=-1$), 
$2.26 \times 10^{22} \unit[]{\,cm^{-2}}$ for medium ionisation ($\log(\xi)=0.5$) 
and $36.1 \times 10^{22} \unit[]{\,cm^{-2}}$ for high ionisation ($\log(\xi)=2$).
}
\end{figure}

We add a further, intrinsic neutral absorber to account for 
absorption in excess of that of our Galaxy using the xspec model {\it zpcfabs} for a partially covering neutral absorber at the source distance. We fix the absorbers covering fraction to one to reduce the number of free parameters. 
The effect of a neutral or ionised absorber on the spectrum can be seen in Figure \ref{fig:absorber}. The redshift is set to the source redshift or to $z=0.05$ if the redshift is unkown.

The spectra of all sources except XMMSL1\,J044347.0+285822 are fit best without any additional neutral absorption. We apply the largest column density allowed at the 90\% level and compare the soft flux of the absorbed spectrum to the soft flux without the additional absorber. Assuming that the brightest observed flux corresponds to the unobscured source, we check whether the obtained column densities are high enough to account for the difference between the brightest point and the flux measured in the \swift\ follow-up observation.
In general the photon index is a free parameter in this fit but for several poorly constrained spectra we fix it to 1.7 to avoid unphysical results. The results of the calculations are shown in Table \ref{tab:neutralh}. 

In most cases, absorption by neutral hydrogen does not offer an explanation for the observed variability. Only one source, XMMSL1\,J044347.0+285822, is consistent with a high column density of neutral hydrogen, that can account for the observed flux change. Indeed for this source, the detailed analysis of the \xmm\ pointed observation spectrum indicates that there are several neutral partially covering absorbers \citep{ricci2010} in the line-of-sight.

In two cases, XMMSL1\,J064541.1-590851 and XMMSL1\,J112841.5+575017, the \swift\ observation 
is of similar strength to the brightest observed point such that small 
column densities are sufficient to explain the rather small variability 
between those two points. For those cases absorption can not be excluded, as the \swift\ spectrum might correspond to a rather unobscured state.

Two sources, 1H\,0707-495 and XMMSL1\,J024916.6-041244, have a spectrum 
which deviates significantly from an absorbed power law.
We here neglect those two sources since we can not obtain a good fit with an absorbed powerlaw model. 1H\,0707-495 is likely variable due to its changing corona (see Section \ref{sec:J070} for a summary). The emission of XMMSL1\,J024916.6-041244 seems to be purely thermal (see Section \ref{sec:J024}).

Please note that the observed flux ratios in Table \ref{tab:neutralh} correspond to the numbers shown in Figure \ref{fig:lightcurves}. We hence assumed that the spectral shape between the observations does not change, which of course is not true for variable absorption. Variable absorbers can change the effective spectral index in the soft band. The resulting error on the flux rates are quantified in Section \ref{curves}.

\subsection{Ionised Absorber}

An alternative to intrinsic neutral absorption is to consider absorption by 
warm, ionised gas, which leads to a different spectral signature as shown in Figure \ref{fig:absorber}. 
To test this possibility, we use the Xspec {\it zxipcf} model, where we as before fix the covering fraction and redshift.

The absorption in addition depends on the degree of ionisation $\log(\xi)$ as can be seen in Figure \ref{fig:absorber}. Many of our spectra are however not good enough to fit the ionisation fraction reliably. We therefore chose three different ionisation levels which we use in the fits. Highly ionised gas absorbs less efficiently at soft energies; for $\log(\xi)=3$ even a huge column density of $N_{\rm H}=5\cdot10^{24}\unit[]{\,cm^{-2}}$ 
only reduces the 0.2--2\,keV flux by a factor of four. Since in general we 
see higher variations, those are rather unlikely to be caused by such highly ionised gas clouds. We therefore only use lower ionisations of $\log(\xi)=2$, $\log(\xi)=0.5$ and $\log(\xi)=-1$. For even lower values the effect of the absorber would be similar to 
that of a neutral absorber.

In Table \ref{tab:ionised_absorber}, we quote the results for the ionisation 
degree that leads to the highest absorption.
Five of the spectra require an ionised absorber when fitted with a power law model. These are: XMMSL1\,J020303.1-074154, XMMSL1\,J024916.6-041244, XMMSL1\,J044347+285822, where a neutral absorber is preferred (see Appendix \ref{sec:J044}), 
XMMSL1\,J090421.2+170927 and XMMSL1\,J121335.0+325609.

The analysed \swift\ spectrum of XMMSL1\,J020303.1-074154 is best described with a highly ionised absorber which however can not explain the observed flux change (see Appendix \ref{sec:J020}). When using a power law model, the fit to the \swift\ spectrum of XMMSL1\,J024916.6-041244 is improved by adding an ionised absorber. The pointed \xmm\ observation of the source is however best described by a pure black body model which implies thermal emission instead (compare Appendix \ref{sec:J024}). The absorber required by the spectrum of XMMSL1\,J090421.2+170927 can explain variability by a large factor up to 7 and there are no further spectra which could indicate a different explanation. For this source absorption by ionised material might hence explain the observed variability. For XMMSL1\,J121335.0+325609 the situation is comparable to XMMSL1\,J020303.1-074154: Some ionised absorption improves the fit, however the spectrum is not consistent with a large enough column density to explain the observed variability.

In addition there are four sources (XMMSL1\,J064541.1-590851, XMMSL1\,J082753.7+521800, XMMSL1\,J100534.8+392856 and J104745.6-375932) where an additional ionised absorber does not improve the fit, however the absorption allowed at the 90\% level is sufficient to explain the relatively small differences between the analysed spectra and the brightest observations.

Hence, we found one source, XMMSL1\,J090421.2+170927, where absorption by ionised gas is a likely explanation for the observed variability. Two further sources might exhibit ionised absorbers which are however not massive enough to be the only reason for the variability. Moreover there are four sources where we can not rule out variable ionised absorption as the reason for small flux changes between the analysed spectra and the brightest observation.

In summary there is little direct evidence that variability is caused
by absorption effects, a conclusion also reached by \citet{SobPap09} in an
analysis of {\it RXTE} AGN observations.

\section{Discussion} \label{discuss}
We have defined a complete sample of candidate highly variable AGN within the {\it XMM-Newton} Slew Survey, and followed these up with the {\it Swift} satellite. Together with archival data, we have used the temporal and spectral information to identify potential variability mechanisms, and to better understand the variable AGN population.

\subsection{Highly variable AGN as a population}
The \xmm\ Slew Survey AGN sample \citep{saxton2011} is dominated ($\sim$80\%) by sources which are constant within a factor of 3, and just 5\% can be classed as highly variable (flux changes of a factor of 10 or more). Among the highly variable sample we present here, a small number are drawn from rare types: 
we find one (possibly two) blazars (XMMSL1\,J183521.4+611942 and maybe XMMSL1\,J082753.7+521800), one low-mass, extremely soft source (XMMSL1\,J024916.6-041244) and two 
nearby tidal disruption candidates (XMMSL1\,J111527.3+180638 and XMMSL1\,J132342.3+482701). An overview of the possible variability mechanisms is presented in Table \ref{tab:summary}.
Disregarding the two spurious detections, the remainder appear to be more 
typical AGN, spanning a wide range of types including QSO, Seyfert 1, Seyfert 1.5, Seyfert 2 and NLS1. Therefore, the highly variable sample do not appear to 
be a fundamentally different class, but are drawn from all AGN populations. The spread in black hole masses
for the sample supports this idea (Table \ref{tab:highvarsrc}). They are, however, marginally more likely to be found at lower redshifts.

We do not find any {unusually X-ray} bright AGN among the highly variable sample, and at their faintest, these sources are generally X-ray weak. We confirm that in their brightest states the X-ray fluxes are consistent with other AGN with the same UV luminosity. 

Many of the AGN go through more than one high and low state 
between the {\it ROSAT} and latest {\it Swift} observations, but a small number
(XMMSL1\,J064541.1-590851, J100534.8+392856 and J112841.5+575017) initially observed at faint fluxes, brightened, and remained at the same flux level in subsequent observations. These are what \citet{kanner13} call
state-change objects, which should not be confused with the state-changes 
reported in Galactic black-hole binary systems that are due to disc 
structure changes \citep{esin97}.

Our detection method is well suited for finding AGN which have transited into the high-soft state but we do not find any, 
which suggests that such
transitions are very rare (or very slow) in AGN \citep[but see][]{miniutti2013}. 

\subsection{Variable absorption}
To investigate the cause of the variability we examined absorption in the available spectra. 
Among the 18 sources with spectra, only XMMSL1\,J044347.0+285822 can be variable due to neutral obscuration alone, while XMMSL1\,J090421.2+170927 might feature a moderately ionised absorber thick enough to explain the variation of a factor of $\sim$6 compared to its brightest point. More complicated absorption models can possibly explain the variability of the Sy 1.2, XMMSL1\,J005931.1+314934 (see Appendix \ref{sec:J005}) and of the Sy 1.5, XMMSL1\,J051935.5-323928 \citep{agis14}.
Our conclusion is thus that for a subsample of our sources (one quarter) the variability is likely due to changing absorbers.

The other sources are likely to exhibit absorption to some degree 
(see Tables \ref{tab:neutralh} and \ref{tab:ionised_absorber}), however 
the column densities are not high enough to explain the factors 
of $>$10 variability observed at earlier epochs (between {\it ROSAT} and \xmm), 
implying that for most sources an additional mechanism is required.

\subsection{Variable intrinsic emission}
If not due to absorption then the observed behaviour must be due to 
variable emission. We have approximated the AGN spectra with a simple power law model when comparing
fluxes but actual spectra are more complex, typically showing excess
emission above an extension of the power law below 1--2\,keV 
\citep[e.g.][]{comastri92,saxton93,scott2012}. 
The nature of this soft excess is not
agreed upon and may vary from source to source. Candidate mechanisms include
thermal emission from the inner region of the accretion disc, relevant
for low-mass BH \citep{yuan10,terashima12,miniutti2013}, a low-electron-temperature
Comptonisation zone \citep{done12} or relativistically blurred 
reflection of the primary power law continuum from the inner disc 
\citep{georgeFabian91,fabian89}
where the strong variability may be enhanced by gravitational amplification 
\citep{minfab04}. 

Based on the X-ray spectra, XMMSL1\,J024916.6-041244 is a strong 
candidate for variability 
caused by changes in thermal disc emission. This ties in with it being the
lowest-mass BH in our sample; none of the other sources show evidence
for thermal emission. Variable disc reflection is a more widely-applicable 
mechanism,
previously cited for the spectral and temporal variations of XMMSL1\,J070841.3-493305 
\citep{fabian2009,fabian2012} and 
proposed for strong variability seen in PHL\,1094  
\citep{Miniutti09, miniutti2012}, Mrk\,335 \citep{gallo13} and
PG\,2112+059 \citep{schartel10} among others. Here, the flux and spectral 
variability is attributed to the expansion and contraction of a comptonising
electron cloud, leading to a highly variable disc reflection 
\citep{wilkinsGallo15}.  It is difficult to discriminate
between complex absorption and blurred reflection models 
purely from the medium-energy 
X-ray spectrum. Studies with {\it NuSTAR} have broken the degeneracy by providing
compelling evidence for a Compton hump at higher energies, 
consistent with the reflection model \citep{parker14,WilkinsGallo2}.

\subsection{Spurious sources and unconfirmed AGN}
We conclude that two of our candidate AGN are in fact spurious detections in the Slew Survey because no other X-ray detection was made and no nearby multiwavelength catalogued counterparts could be found.
The detection likelihood of XMMSL1\,J015510.9-140028 was the lowest 
of our sample and close to the lower limit required for the Clean 
catalogue\footnote{The {\em Clean} slew survey catalogue contains sources 
with detection likelihoods$>$10 \citep[see][]{saxton2008} }. 
While on visual inspection, XMMSL1\,J113001.8+020007 does not appear
to have the profile of a typical point source. Furthermore, they both have quite a large offset from
their nearest SDSS galaxy counterpart (0.3 and 0.2 arcmin).
It is possible that these sources are normally very faint,
and are bona fide transients, but the lack of subsequent detections together with the large offset to the counterpart makes it likely that these are 
indeed spurious.

\begin{figure}[htb]
\centering
\includegraphics[width=88mm]{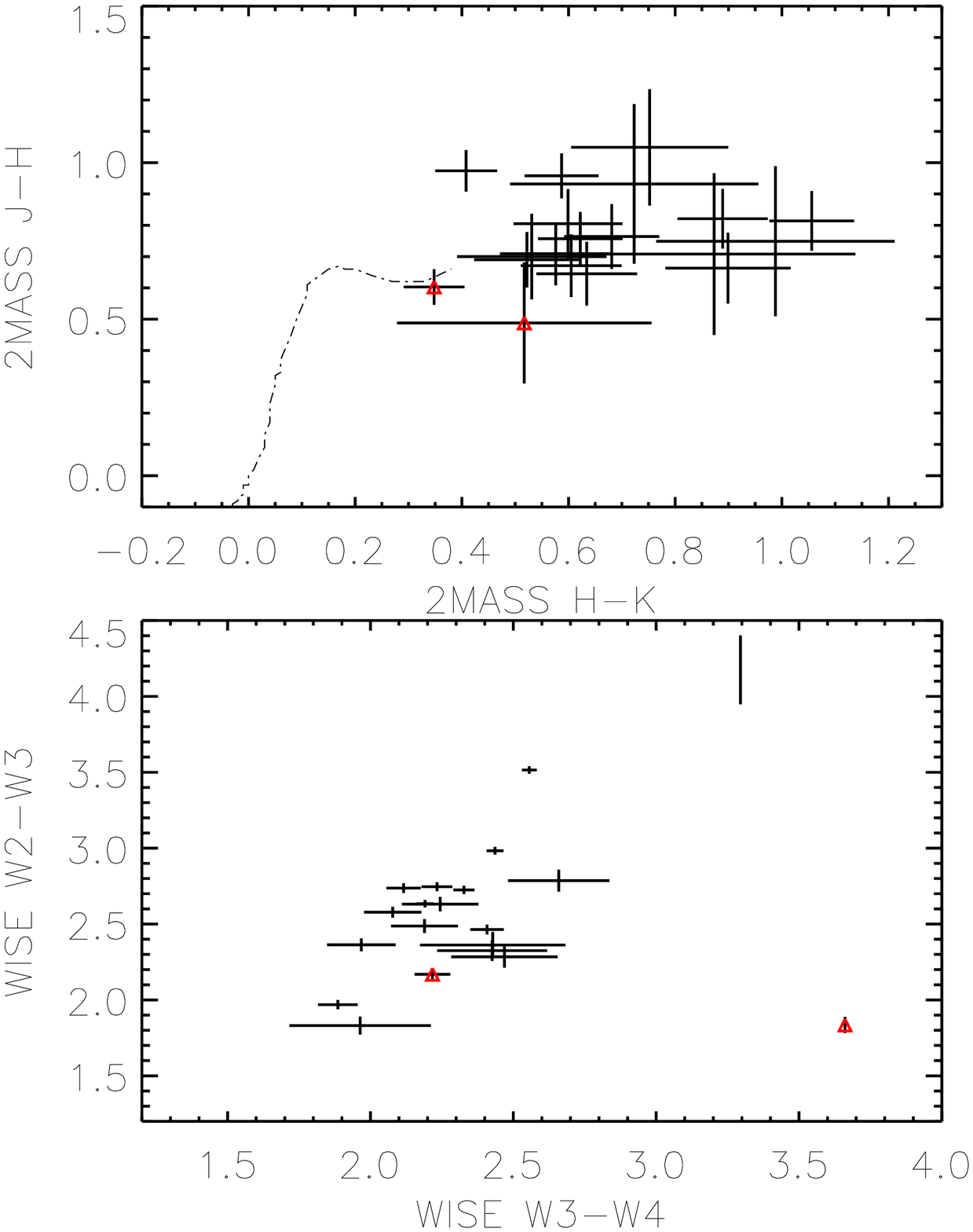}
\caption[Wise and 2mass colours]{ \label{fig:WISE2MASS} 2MASS (upper panel) 
and WISE (lower panel) colours of all sources in our sample detected 
in both infrared surveys. 
Candidate TDEs are marked with red triangles. The black dashed line in the 
upper plot indicates the expected colours for main sequence stars, 
while colours to the upper right of the end of this line are typical of AGN. 
All the WISE colours we find are consistent with an AGN or galaxy nature, and 
stellar colours would lie off the plot to the bottom left.}
\end{figure}

Four of our sources, excluding the spurious detections, are associated with
galaxies with no known redshift. 
All of these have counterparts in 2MASS and WISE with
IR colours which indicate an AGN nature and rule out a stellar classification,
using the technique illustrated in the Slew Hard Band Survey 
\citep[see Figure~7 of][for a comparison]{warwick2012} and a sample 
of unidentified transients in the 
Slew Survey \citep{starling2011}. We show the results in Figure \ref{fig:WISE2MASS}.
In two cases, XMMSL1\,J064541-590851 and XMMSL1\,J121335.0+325609, there are two possible counterparts, within the \xmm\ Slew error circle. 
For XMMSL1\,J121335.0+325609 both of these are consistent with an AGN, while the 2MASS colours for XMMSL1\,J064541-590851 indicate that one of the 
possible counterparts (the fainter one) has colours typical of a star on 
the main sequence. We have used the more 
accurate position of the {\it Swift}-XRT observation to narrow the match down to 
just one candidate, an AGN, in both of these sources. 

\subsection{Conclusions}
We have investigated the properties of a well-determined sample of AGN selected on the basis of their long-term X-ray variability. After removal of two 
spurious detections the sample spans a wide range of optical classifications
and appears to be drawn from the general AGN 
population, albeit with 
a slightly lower average redshift and luminosity than a non-varying control
sample. Of the 22 AGN, two are radio-loud and likely vary due to jet activity,
two are candidate tidal disruption events, four show variable neutral or 
multi-phase absorption, the lowest mass object has variable intrinsic thermal disc 
emission and the well-studied variability in the NLS1 XMMSL1\,J070841.3-493305 has 
been elsewhere attributed to reflection from a changing comptonising, 
power law emission component.

We cannot say with confidence what the variability mechanism is for the
remaining sources. We see that the peak X-ray flux
is consistent with that expected from the UV luminosity but in the lowest flux measurements the sources
are generally X-ray weak. This may indicate absorption although we can exclude
changes in a single component absorber in most cases. Changes in multiple 
absorbers cannot be excluded with our data and provide perhaps the most likely explanation for the two Seyfert 2 galaxies (XMMSL1\,J112841.5+575017 and J173739.3-595625). The NLS1 XMMSL1\,J093922.5+370945 might share the same variability mechanism as J070841.3-493305. A summary of proposed variability mechanisms is given in Table \ref{tab:summary}.

A larger sample, extracted from future releases of the {\it XMM} Slew
Survey or from dedicated survey instruments, such as {\it eRosita} \citep{Predehl2010}, will be useful 
to further investigate differences in the variable AGN population.

\section{Acknowledgments}
NLS was supported by the ESAC Traineeship programme.
RLCS was supported by a Royal Society Dorothy Hodgkin Fellowship. PAE acknowledges support from the
UK Space Agency.
We thank Kim Page for assistance with the {\it Swift} data.
This work made use of data supplied by the UK {\it Swift} Science Data Centre at the University of Leicester. The XMM-Newton project is an ESA science mission with instruments and contributions directly funded by ESA member states and the USA (NASA). This research has made use of the NASA/IPAC Extragalactic Database (NED) which is operated by the Jet Propulsion Laboratory, California Institute of Technology, under contract with the National Aeronautics and Space Administration. This publication makes use of data products from the Wide-field Infrared Survey Explorer, which is a joint project of the University of California, Los Angeles, and the Jet Propulsion Laboratory/California Institute of Technology, funded by the National Aeronautics and Space Administration. This publication makes use of data products from the Two Micron All Sky Survey, which is a joint project of the University of Massachusetts and the Infrared Processing and Analysis Center/California Institute of Technology, funded by the National Aeronautics and Space Administration and the National Science Foundation.

\bibliographystyle{aa}
\bibliography{references}

\begin{sidewaystable*}
 \caption{\label{tab:models}Best fitting models for all spectra analysed here.}
\begin{center}
\begin{tabular}{l c l c c c c c l}       
\hline\hline                
XMMSL1 name		& obs. date 	& instr.	& $C_{red}$	& power law 	& cold abs. 	& \multicolumn{2}{c}{warm abs.} \\
		& 	 	& 		&		& $\Gamma$	& $N_{\rm H}$ (10$^{22}$ cm$^{-2}$)	& $N_{\rm H}$ (10$^{22}$ cm$^{-2}$) & $\log(\xi)$ \\
\hline                        %
J005953.1+314934 & 2002-01-15	& XMMSL& 
1.01 
& 2.21 (F)		& -	& - & -	\\
 		 & 2006-01-24	& XMM P.& 
0.95 	
& 2.21$^{+0.08}_{-0.10}$	& 2.3$^{+0.4}_{-0.5}$, $f_c=0.36^{+0.08}_{-0.10}$	& 117$^{+44}_{-62}$, $f_c=0.81^{+0.07}_{-0.17}$ & 2.05$^{+0.12}_{-0.06}$ \\
 		 & 2006-05-29	& Swift	& 
1.04 	
& 2.21 (F)	& 2.3 (F), $f_c=0.50\pm 0.03$	& - & -	\\
 		 & 2011-08-28	& Swift	&  
0.79 	
& 2.21 (F)	& 2.3 (F), $f_c<0.28$	& -	& - \\
[1ex] \\ [-2.5ex]

J020303.1-074154		 & 2006-07-03 	& XMM P.& 
0.95 	
& 2.31$^{+0.07}_{-0.06}$	& -	& 2.6$^{+1.1}_{-0.5}$	& 1.94$^{+0.17}_{-0.06}$\\
		 & 2007-06-03	& Swift	& 
0.77 
& 2.2$\pm0.2$	& -	& -	& - \\
		 & 2008-03-02	& Swift	& 
1.06 	
& 1.9$\pm0.2$	& -	& -	& - \\
		 & 2010-03-01	& Swift	& 
0.88 	
& 1.5$\pm0.4$	& -	& -	& - \\
[1ex] \\ [-2.5ex]

J024916.6-041244 & 2006-07-14	& XMM P.& 
1.11 	
& $kT=124^{+17}_{-12}$\,eV & 
0.14$^{+0.02}_{-0.03}$	& 65$^{+11}_{-5}$ & 2.95$^{+0.03}_{-0.05}$ \\
& 2007-07-14	& Swift & 
0.55 
& $kT=100^{+24}_{-21}$\,eV	& 0.14 (F)	& 65 (F)	& 2.95 (F) \\
[1ex] \\ [-2.5ex]

\multirow{2}{*}{J044347.0+285822}  & \multirow{2}{*}{2007-03-18}	& \multirow{2}{*}{XMM P.}& \multirow{2}{*}{1.00} & \multirow{2}{*}{1.59 $\pm0.12$}	& 1.3$^{+0.3}_{-0.4}$, $f_c=0.92^{+0.02}_{-0.04}$	& \multirow{2}{*}{-}	& \multirow{2}{*}{-}\\
  & 	& 	& 	& 	& 4.3$^{+1.6}_{-1.2}$, $f_c=0.64\pm0.11$	& 	& 	& \\
 		 & 2010-07-27	& Swift	& 
1.06 	
& 1.59 (F)	& 10$^{+8}_{-4}$, $f_c=0.95^{+0.03}_{-0.04}$	& -	& -\\
[1ex] \\ [-2.5ex]

J045740.0-503053 & 2010-10-03	& Swift	&  
0.99	
& 1.6$\pm0.6$	& -	& -	& -\\
[1ex] \\ [-2.5ex]

J064541.1-590851 & 2010-12-26	& Swift	& 
0.75 	
& 1.9$\pm$ 0.2		& -	& -	& -	\\
[1ex] \\ [-2.5ex]

J082753.7+521800 & 2010-03-11	& Swift	&  
1.11 	
& 2.0$\pm0.3$	& -	& -	& -	\\
[1ex] \\ [-2.5ex]

J090421.2+170927 & 2010-06-12	& Swift	& 
0.86 	
& 1.7 (F) 	& - 
& $3.0^{+4.6}_{-2.3}$	& $1.38^{+0.68}_{-0.88}$ \\
[1ex] \\ [-2.5ex]

J093922.5+370945 & 2006-11-01	& XMM P.& 
1.07 	
& 2.94$\pm0.07$	& -	& -	& - \\
 		 & 2007-09-21	& Swift	& 
0.68 	
& 2.7$\pm0.3$		& -	& -	& \\
 		 & 2011-03-28	& Swift	& 
1.15 	
& 3.0$\pm0.4$		& -	& -	& \\
[1ex] \\ [-2.5ex]

J100534.8+392856 & 2011-01-31	& Swift	& 
1.20 	
& 1.9$\pm0.2$	& -	& -	& -	\\
     & 2013-04-23     & Swift & 
0.81       
& 1.75$\pm0.15$        & -     & -     & -     \\
[1ex] \\ [-2.5ex]

J104745.6-375932 & 2010-04-11	& Swift	& 
0.89 	
& 2.0$\pm0.2$	& -	& -	& -	\\
[1ex] \\ [-2.5ex]

J112841.5+575017 & 2013-04-10       & Swift & 
0.82       
& 1.7$\pm0.1$   & -     & -     & -     \\
[1ex] \\ [-2.5ex]

J121335.0+325609 & 2010-10-19	& Swift	& 
1.03 	
& 2.1$\pm{0.4}$	& -	& -	& -	\\
[1ex] \\ [-2.5ex]

J162553.4+562735 & 2010-07-10	& Swift	& 
0.87 	
& 2.3$\pm0.5$	& -	& -	& -	\\
[1ex] \\ [-2.5ex]

J173739.3-595625 & 2008-05 -- 2008-11	& Swift	& 
1.03 
& 1.85$\pm0.07$	& -	& -	& -	\\
 \\ 

J183521.4+611942 & 2011-02-04	& Swift	&  
1.49 	
& 1.9$\pm0.5$		& -	& -	& -	\\
\hline                                   
\end{tabular}
\end{center}
\tablefoot{
XMMSL = Slew Survey observation, XMM P. = pointed XMM observation. Galactic absorption was included and fixed in all fits. Parameters followed by ``(F)'' were frozen. XMMSL1\,J005953.1+314934 was fit with an absorbed power law plus black body model, J024916.6-041244 was fit with an absorbed black body spectrum only, J044347.0+285822 has two partially covering neutral absorbers in the 2007 observation. 
$f_c$ is the covering fraction of the absorber. Errors are quoted at 90\% confidence.
}
\end{sidewaystable*}

\begin{table*}[hp]
{\small
\hfill{}
\caption{Allowed column densities of neutral hydrogen.}
\label{tab:neutralh}      
\begin{center}

\begin{tabular}{l l c c c c c c}       
\hline\hline                
XMMSL1 name		& obs. date & $C_{red}$ & spectral    & $N_{\text{H, max}}$\tablefootmark{a}  	& implied\tablefootmark{b}	& observed\tablefootmark{c} & can it explain \\   
	& 	&   & index &    ($10^{20} \text{cm}^{-2}$) 	& flux change	& flux change & variability?\\ 
\hline
J005953.1+314934    & 2011-08-28    & 0.87  & 1.9   & 1.6		& 1.1		& 1.9	& no\\
J020303.1-074154	& 2010-03-01    & 0.90  & 1.7\tablefootmark{d}   & 1.3		& 1.2	    & 5.8	& no\\
J024916.6-041244	& 2007-06-27    & 0.95  & 1.7\tablefootmark{d}   & 2.3 		& 1.2	    & 17	& no\\
J044347.0+285822	& 2010-07-27    & 1.07  & 1.7\tablefootmark{d}	& 554		& 117       & 7.3	& yes\\
J045740.0-503053	& 2010-10-03    & 1.00  & 1.7\tablefootmark{d}	& 16.9		& 2.0	    & 17	& no\\
J064541.1-590851    & 2010-12-26    & 0.77  & 1.9	& 5.58		& 1.2	    & 1.4	& maybe\\
J082753.7+521800    & 2010-03-11    & 1.12  & 2.0	& 4.76		& 1.2	    & 2.7	& no\\
J090421.2+170927	& 2010-06-12    & 1.21  & 1.7\tablefootmark{d}	& 75		& 2.7	    & 3.6	& no\\
J093922.5+370945    & 2011-03-28    & 1.17  & 3.0	& 12.8		& 2.5	    & 3.4	& no\\
J100534.8+392856    & 2011-01-01    & 1.22  & 1.9	& 1.55		& 1.1	    & 1.7	& no\\
J104745.6-375932    & 2010-04-11    & 0.90  & 2.0	& 2.97		& 1.2	    & 1.6	& no\\
J112841.5+575017	& 2011-03-01    & 0.86  & 1.7\tablefootmark{d}	& 7.20		& 1.5	    & -	    & Swift point brightest\\
J121335.0+325609    & 2006-06\tablefootmark{e} & 0.97  & 1.4	& 6.6		& 1.3	    & 4.9	& no\\
J162553.4+562735    & 2010-07-10    & 0.88  & 2.3	& 10.2		& 1.6	    & 3.2	& no\\
J173739.3-595625    & 2008-11-02    & 0.91  & 1.9	& 1.4		& 1.1	    & 2.0	& no\\
J183521.4+611942    & 2011-02-01    & 1.51  & 1.7\tablefootmark{d}	& 16.2		& 1.4	    & 5.4	& no\\
\hline                                   
\end{tabular}
\end{center}
}
\tablefoot{
\tablefoottext{a}{Maximum neutral Hydrogen column allowed by power law fit to the 
\swift\ spectrum (90\% confidence).} \\
\tablefoottext{b}{Fractional change in observed 0.2--2\,keV flux due to this $N_{\rm H}$ column.}\\
\tablefoottext{c}{Change in 0.2--2\,keV flux actually observed between the highest flux 
measurement and the \swift\ observation.}\\
\tablefoottext{d}{Power law slope fixed at 1.7 during fit.}\\
\tablefoottext{e}{Merged spectra taken between 2006-06-02 and 2006-06-12.}
}
\hfill{}
\end{table*}

\begin{table*}[hp]
{\small
\hfill{}
\caption{Allowed column densities of ionised gas.}
\label{tab:ionised_absorber}     
\begin{center}

\begin{tabular}{l l c c c c c c c}       
\hline\hline                
XMMSL1 name		& obs. date & $C_{red}$ & spectral    & ionisation\tablefootmark{a}    & $N_{\text{H, max}}$\tablefootmark{b}  	& implied\tablefootmark{c}	& observed\tablefootmark{d} & can it explain \\   
	& 	&   & index &   &    ($10^{20} \text{cm}^{-2}$) 	& flux change	& flux change & variability?\\ 
\hline
J005953.1+314934    & 2011-08-28    & 1.07  & 1.7\tablefootmark{e}   & low       & 5.1		& 1.2		& 1.9	& no\\
J020303.1-074154	& 2010-03-01    & 0.74  & 1.7\tablefootmark{e}   & high      & 69		& 3.3	    & 5.8	& no\\
J024916.6-041244    & 2007-06-27    & 1.03  & 1.7\tablefootmark{e}   & high\tablefootmark{f}      & 150 		& 110	    & 17	& yes\\
J044347.0+285822    & 2010-07-27    & 1.05  & 1.7\tablefootmark{e}	& low       & 210		& 260       & 7.3	& yes\\
J045740.0-503053    & 2010-10-03    & 0.99  & 1.7\tablefootmark{e}	& medium    & 72		& 2.5	    & 17	& no\\
J064541.1-590851    & 2010-12-26    & 0.77  & 2.0	& high      & 190		& 1.6	    & 1.4	& yes\\
J082753.7+521800    & 2010-03-11    & 1.13  & 2.3	& high      & 650		& 3.2	    & 2.7	& yes\\
J090421.2+170927    & 2010-06-12    & 0.83  & 1.7\tablefootmark{e}	& high      & 1400		& 7.3	    & 3.6	& yes\\
J093922.5+370945    & 2011-03-28    & 1.20  & 3.2	& low       & 9.9		& 2.2	    & 3.4	& no\\
J100534.8+392856    & 2011-01-01    & 1.22  & 1.7\tablefootmark{e}	& high      & 180		& 1.5	    & 1.7	& maybe\\
J104745.6-375932    & 2010-04-11    & 0.88  & 2.2	& high      & 700		& 1.7	    & 1.6	& yes\\
J112841.5+575017    & 2011-03-01    & 0.74  & 1.7\tablefootmark{e}	& high      & 7.0		& 3.1	    & -	    & Swift point brightest\\
J121335.0+325609    & 2006-06\tablefootmark{g} & 0.99  & 1.5	& low       & 7.9		& 1.3	    & 4.9	& no\\
J162553.4+562735    & 2010-07-10    & 0.89  & 2.2	& low       & 13		& 1.9	    & 3.2	& no\\
J173739.3-595625    & 2008-11-02    & 0.91  & 2.0	& high      & 140		& 1.4	    & 2.0	& no\\
J183521.4+611942    & 2011-02-01    & 1.51  & 1.7\tablefootmark{e}	& high      & 610		& 1.6	    & 5.4	& no\\
\hline                                   
\end{tabular}
\end{center}
}
\tablefoot{
\tablefoottext{a}{Ionisation level which yields largest flux change: low (log($\xi$)=$-$1); medium (log($\xi$)=0.5); 
high (log($\xi$)=2)}\\
\tablefoottext{b}{Maximum ionised hydrogen column allowed by power law fit to the 
\swift\ spectrum (90\% confidence).}\\
\tablefoottext{c}{Fractional change in observed 0.2--2\,keV flux due to this $N_{\rm H}$ column.}\\
\tablefoottext{d}{Change in 0.2--2\,keV flux actually observed between the highest flux 
measurement and the \swift\ observation.}\\
\tablefoottext{e}{Power law slope fixed at 1.7 during fit.}\\
\tablefoottext{f}{A medium ionisation level yields even stronger absorption, however the fit is worse.}\\
\tablefoottext{g}{Merged spectra taken between 2006-06-02 and 2006-06-12.}\\
}
\hfill{}
\end{table*}

\clearpage
\begin{appendix}

\section{Individual sources}
\label{sec:individual}
We now summarise the results for each source individually, 
providing further detailed information in some cases, and list the
conclusions in Table \ref{tab:summary}.

\subsection{XMMSL1\,J005953.1+314934}
\label{sec:J005}
\begin{figure}[htbp]
\centering
\includegraphics[width=8cm]{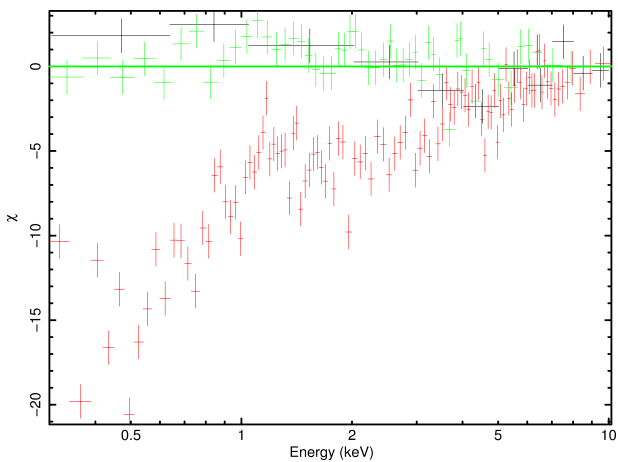}
\caption[Residuals of J005]{ \label{fig:J005_residuals} Residuals of the \xmm\ Slew 2002 (black),
\swift\ 2006 (red) and \swift\ 2011 (green) observations of XMMSL1\,J00534.8+392856 when compared to the best fit to the \xmm\ spectrum of 2006, a simple power law plus a black body model.}
\end{figure}

This Seyfert 1.2 galaxy is well known and several spectra are available. 
Over the last 30 years it has experienced several bright and dim states. No publication dedicated to this source is available but it has been used in a sample to support the
model of multi-cloud absorption \citep{tatum13}.
Apart from the Slew spectrum, three rather good quality spectra are available. 
We fit the best one, the pointed \xmm\ observation from January 2006, 
and compared the residuals of the other spectra relative to this fit. 
The \xmm\ spectrum can be reasonably well fit by a simple model of 
a power law plus a black body ($kT=120$\,eV) and Galactic absorption. 
The \xmm\ spectrum corresponds to a rather bright state, a factor of 7 brighter than the dimmest observation. In Figure \ref{fig:J005_residuals}, the {\it Swift} spectrum from June 2006 looks to be absorbed, while the later {\it Swift} spectrum and earlier \xmm\ Slew Survey spectrum do not deviate much from the \xmm\ pointed observation. We did not reanalyse the {\it Suzaku} spectrum from 2010 here but according to \citet{winter2012} 
it can as well be fit with a power law and a soft black body, so it is probably similar to the XMM spectrum.

This is consistent with the light curve (Figure \ref{fig:J005}), where the June 2006 observation exhibits a flux lower by a factor of 3 compared to the \xmm\ observation from five months earlier, while in all other spectra the AGN had roughly the same luminosity.\\
Motivated by this observation we tried to find an absorption model, that can be applied to all spectra. 
A better fit to the 2006 \xmm\ spectrum is achieved by adding a partially covering neutral as well as an ionised absorber with the parameters given in Table \ref{tab:models}. 

Since the other spectra have fewer counts we freeze as many parameters as possible. The {\it Swift} 2006 observation can be fitted with the model from the \xmm\ 2006 spectrum by adjusting only the covering fractions of the two absorbers and the normalisation of the power law. The covering fraction of the neutral absorber rises from 36\% to 50\%, which explains the dimming, while the ionised absorber is no longer needed. 
The two other spectra are less constrained. They can both be fit by changing the covering fraction of the neutral absorber, but they cannot constrain the ionised one. All fit parameters are given in Table \ref{tab:models}. We conclude that this AGN is likely variable due to absorption by both neutral and ionised partially covering absorbers.  
There are UVOT observations in the u band for both {\it Swift} observations. For the 2006 observation which is absorbed in X-rays the UV flux is lower by 30\%. This is most probably due to the presence of the absorber, which covers 
the UV emitting region partially. It has been variously shown in the 
literature that multi-phase absorption cannot be spectrally 
distinguished from emission due to relativistically-blurred disc reflection  
in the 0.2--10\,keV band. In the case of XMMSL1\,J00534.8+392856 the significant change in UV flux 
may make it more likely that the variability seen here is due to absorption rather
than reflection.

\subsection{XMMSL1\,J015510.9-140028}
This source was only detected in the \xmm\ Slew Survey being just above 
the detection threshold. Since it could not be found before or after that 
and does not have a counterpart in other wavelength bands, other than a
rather distant (offset=20\arcsec) galaxy, APMUKS(BJ)\,B015244.14-141523.0,
we assume that it is a false detection.

\subsection{XMMSL1\,J020303.1-074154}
\label{sec:J020}
\begin{figure}[htbp]
\centering
\rotatebox{-90}{\includegraphics[height=8cm]{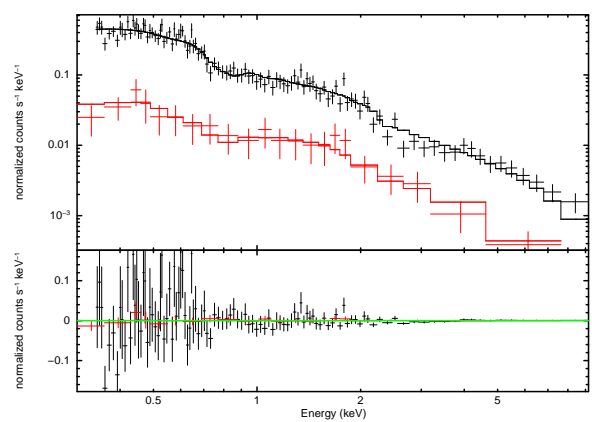}}
\caption[Residuals of J005]{\label{fig:J020_sim_fit} Simultaneous fit of the {\it XMM} 2006 and the merged {\it Swift} spectra of XMMSL1\,J020303.1-074154. The model consists of a power law with $\Gamma=2.29^{+0.07}_{-0.05}$ and a strongly ionised thick absorber with $\log(\xi)=1.97^{+0.14}_{-0.19}$ and $N_\text{H}=2.8^{+1.0}_{-0.5} \times 10^{22} \unit[]{cm^{-2}}$. All parameters are the same for both models, except the normalisation, which is 40\% larger for the merged Swift spectra.}
\end{figure}

After brightening in 2004 the source was found to remain in a rather 
constant luminosity state a factor of five below the \xmm\ slew observation. 
The spectrum of the \xmm\ pointed observations is best fit with
an ionised absorber. However some residuals remain in the spectrum and removing the absorber only increases the soft flux by ~40\%. 
The {\it Swift} spectra are all quite similar and we merge these to 
improve statistics before
comparing with the \xmm\ spectrum, which can be fitted with the same model but
with lower normalisation, in Figure \ref{fig:J020_sim_fit}. 

The soft absorber we found is not thick enough to explain the high variability
and the mechanism remains unclear.

\subsection{XMMSL1\,J024916.6-041244}
\label{sec:J024}

\begin{figure}[htbp]
\centering
\includegraphics[height=8cm,angle=-90]{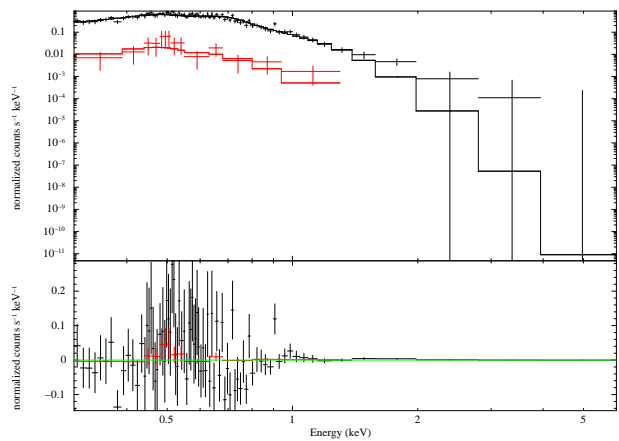}
\caption[Combined spectral fot to J024]
{\label{fig:J024_sim_fit} Simultaneous fit
of an absorbed black body model to \xmm\ pointed (black) and {\it Swift}-XRT (red) observations 
of XMMSL1\,J024916.6-041244. The black body temperature and normalisation 
is independent in each fit while the absorption has been tied.}
\end{figure}

The long-term light curve consists of a single flare which faded over a
number of years.
The underlying spectrum is soft and seems to be predominantly thermal.
An \xmm\ pointed observation may be fit by a black body, of temperature
$kT=110-140$\,eV, absorbed by both cold ($N_{\rm H} \sim 10^{21}$\,cm$^{-2}$) and
ionised ($N_{\rm H} \sim 6 \times 10^{23}$\,cm$^{-2}$; $\log(\xi)\sim$3) gas. A later \swift-XRT spectrum, shows a
consistent spectral shape (Figure~\ref{fig:J024_sim_fit}).
Fixing the absorption to the values found by \xmm\ gives a consistent
black body temperature of
$kT=75-115$\,eV, with a flux reduced by a factor 2--3.

The X-ray light curve and spectra of this source are typical for a tidal disruption event, but the optical spectrum shows clear narrow OIII lines, which indicates ongoing AGN activity. It seems very unlikely that a tidal disruption happens within a rare X-ray weak AGN.

The temperature of the emission is typical of the effective temperature
seen in the ubiquitous
soft excess but more interestingly, is also consistent with
thermal emission from the inner edge of the accretion disc around a
$5\times10^5$ M$_{\odot}$ black hole. In this respect the source resembles
2XMM\,J123103.2+110648 \citep{terashima12,ho2012,lin2013} and GSN069 \citep{miniutti2013}, which also host low mass M$\le 10^6$
M$_{\odot}$ black holes (see Table~\ref{tab:highvarsrc}) and have spectra 
apparently dominated by thermal emission with
little or no contribution from a power law component.

The flux detected with the {\it Swift}-UVOT u filter is constant, within errors,
for observations made between 2006 and 2011.

\subsection{XMMSL1\,J034555.1-355959}
This unclassified AGN at an unknown redshift was detected in a {\it ROSAT} pointed observation and later twice in the Slew Survey. The source was barely detected during the \swift\ follow-up observation, hence no good spectrum is available. The source is variable on relatively short time scales: in 2010 it faded within 10 months by at least a factor of 11.

\subsection{XMMSL1\,J044347.0+285822\label{sec:J044}}
\begin{figure}[htbp]
\centering
\includegraphics[width=80mm]{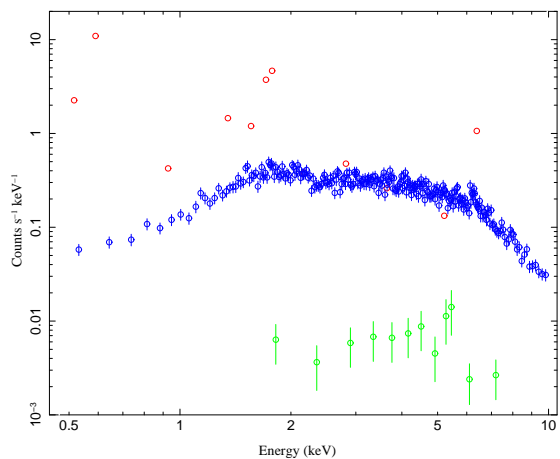}
\caption[Spectra of J044]{ \label{fig:J044_spec} The spectra of XMMSL1\,J044347.0+285822 from the Slew Survey 2003 (red), from an \xmm\ pointed observation 2007 (blue) and during the \swift\ follow-up observation in 2010 (green). Please note that the spectra are not relatively calibrated.}
\end{figure}

This narrow line Seyfert 1 galaxy is variable due to absorption. Figure \ref{fig:J044_spec} shows the spectra of the three observations made since 2003. 
The data points of the Slew Survey correspond to individual photons. However the number of photons at low energies is sufficient to exclude high absorption, while the two later spectra are heavily absorbed.\\

\citet{ricci2010} did a precise analysis of the \xmm\ pointed observation of 2007 and additional Integral data, where they found two thick partially covering absorbers ($N_\text{H}=1.4 \times 10^{22} \unit[]{cm^{-2}}$ with a covering fraction of 92\% and $N_\text{H}=4 \times 10^{22} \unit[]{cm^{-2}}$ with a smaller covering fraction of $\sim$60\%).\\

\begin{figure}[htbp]
\centering
\includegraphics[width=80mm]{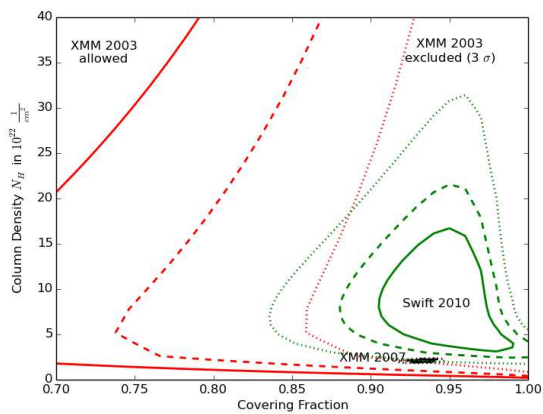}
\caption[Spectra of J044]{\label{fig:J044_contours} The allowed parameter space for one partially covering absorber in the three observations of XMMSL1\,J044347.0+285822. Please note that for the Slew observation nearly all values except high covering fractions are allowed.}
\end{figure}

To be able to compare this result with the additional less well constrained spectra, we simplify the model using one partially covering absorber only, which still leads to a reasonable fit for the \xmm\ pointed observation. We fix the photon index to 1.5 for all observations, which is the value found in \citet{ricci2010}. In Figure \ref{fig:J044_contours} we show the allowed parameter space of a partially covering absorber.\\
While the slew observation excludes high covering fractions and column densities, both the \xmm\ and the \swift\ spectra require those. The best fit of the \swift\ observation is at a higher column density of $\sim N_\text{H}=1\times10^{23} \unit[]{cm^{-2}}$ and a slightly higher covering fraction than in the \xmm\ observation. However at the 3$\sigma$ level the parameters of those two spectra are compatible, which matches the comparable X-ray fluxes of those two observations.\\
We hence conclude that this source is absorbed by a complex structure of neutral hydrogen, which changes with time. Probably at most times the observable flux is rather low, however in the slew observation in 2003 the covering fraction and/or column density was reduced considerably, such that the AGN appeared brighter by a factor of 3--4 compared to later observations. Very likely the even lower luminosity during the RASS in 1991 was also due to obscuration.\\

\subsection{XMMSL1\,J045740.0-503053}
For this unclassified galaxy no redshift is known. The source was not seen in the RASS, however in 2002 a bright source was detected in the Slew Survey. Three years later the source had faded at least by a factor of six and was fainter by a factor 18 in the {\it Swift} follow-up observation. The {\it Swift} spectrum has few counts, but can be fit with an unabsorbed power law with $\Gamma=1.6\pm0.6$. 

\subsection{XMMSL1\,J051935.5-323928}
The multi-epoch observations of this Seyfert 1.5 have recently been extensively 
analysed in \citep{agis14} and we perform no further analysis on our \swift\ spectrum here. \citet{agis14}
found that the high-quality spectra were well fit by a disc-reflection model  
absorbed by a neutral gas column. Strong flux variability was shown to be due to
variations in the depth of the absorber which ranged from $N_{\rm H}\sim5\times10^{21}$ cm$^{-2}$ to $N_{\rm H}\sim3-4\times10^{23}$ cm$^{-2}$. Simultaneous variability was seen
 in the UV confirming the presence of an extended absorber which was
 identified with the dusty, clumpy torus. 

\subsection{XMMSL1\,J064541.1-590851}
The redshift and AGN type are unknown for this source. The \xmm\ Slew survey observation in 2006 showed a large flux increase from a RASS upper limit. The \swift\ observation, taken 4 years later had a
similar flux to the slew observation and could be fit with a power law of $\Gamma = 1.9\pm0.2$
with no absorption above the Galactic value.

\subsection{XMMSL1\,J070841.3-493305}
\label{sec:J070}
This narrow-line Seyfert galaxy is a very well studied source with 1300\,ks of 
total observation time by \xmm. Instead of reanalysing the data we summarise the most relevant results here.\\
The spectra of this AGN have been interpreted as relativistically blurred
reflection off an accretion disc around a maximal spinning black hole. \citet{Dauser2012} analyse two \xmm\ spectra from 2008-02-02 and 2010-09-16 during which the source was relatively bright. They
find that both spectra are well fit with a steep power law, of photon index $\sim$2.85, and two reflectors.
In addition the fit is improved by adding a mildly relativistic, but highly ionised outflow, which changes its parameters between the two observations.\\
In January 2011 the source was observed to undergo an extremely X-ray weak phase, a factor 190 below the highest observed flux in the soft X-ray band. An \xmm\ spectrum taken on 2011-01-12 had a 
soft-band flux a factor 10 lower than the combined spectra from the year 2008. However in the hard band, which is dominated by the Fe line, a change by a factor of just two is observed.
\citet{fabian2012} analysed those spectra in detail finding that by 2011 the power law emission had disappeared. By examining the relativistically broadened Fe\,K$\alpha$ line, they are able to constrain the part of the disc at which the X-rays are scattered. While in 2008 around 50\% of the X-ray flux comes from within the inner disc, within one gravitational 
radius of the event horizon, 90\% of 
the radiation comes from this part of the disc in the January 2011 observation. They interpret this such that the corona collapsed to a compact source very close to the black hole. Due to light bending effects very few X-rays can escape and contribute to the power law continuum, while most of them are focussed on the disc, which increases the reflection features.\\
Even though this is an explanation of the temporary low luminosity state at the beginning of 2011, it is not clear whether the same mechanism is responsible for the large amplitude variability observed during more than one decade.\\
The relative X-ray luminosity of this source is, with $\alpha_\text{OX}=-2.5$, by far the lowest one. It deviates by more than a factor of 700 from the expected value. Since none of the other sources get close to this value, we conclude that this source is exceptional even in this sample of highly variable AGN.

\subsection{XMMSL1\,J082753.7+521800}
This source is a bright radio loud quasar at redshift 0.33 and a luminous 
infrared galaxy. The X-ray spectrum is well described by a simple power law, but a highly ionised absorber can not be ruled out and could cause a flux change by up to a factor of 3. The source is radio loud and thus jet activity might be a more likely explanation of the observed variability.

\subsection{XMMSL1\,J090421.2+170927}
This broad-line QSO at a redshift of 0.073 was not detected during the RASS 
or in two further {\it ROSAT} pointed observations. The detection in the Slew Survey is brighter by a factor of 28 compared to the lowest upper limit. Four years later in the {\it Swift} follow-up observation the source had faded by a factor of 3-4. The spectrum is rather flat. When fitted with a single power law, 
the photon index is as low as $\Gamma=0.9\pm0.5$. If the photon index 
is fixed to 1.7 and an ionised absorber added, the fitted column density of 
(1.6$^{+4.3}_{-1.6}$)$\times 10^{21}$ cm$^{-2}$ can explain a flux change 
sufficient to explain the brightest observation in the Slew Survey. However the spectrum has too few counts to verify that an absorber is present. We conclude therefore that variability due to absorption is a possible scenario for this source.

\subsection{XMMSL1\,J093922.5+370945}
\begin{figure}
\centering
\rotatebox{-90}{\includegraphics[height=80mm]{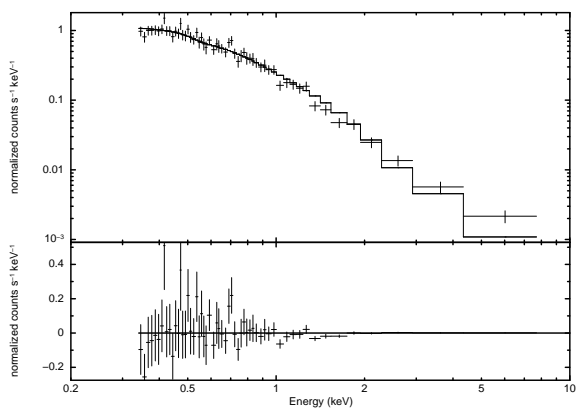}}
\caption[Spectrun of J093]{ \label{fig:j093_spec} A power law fit to the
\xmm\ pointed observation spectrum of XMMSL1\,J093922.5+370945 from 2006-11-01. The spectrum is
soft with $\Gamma\sim3$.} 
\end{figure}

The spectrum and flux of XMMSL1\,J093922.5+370945 varied little between an \xmm\ pointed observation of 2006 and \swift\ observations of 2007 and 2011 and can be well fit with a single power law ($\Gamma\sim3$) with no evidence of intrinsic absorption 
(Fig.~\ref{fig:j093_spec}). The steep 
power law slope is typical of a narrow-line Seyfert 1 galaxy \citep{esquej2007}.
The variability mechanism is unknown although a reasonable assumption
would be that it shares the same mechanism as the other NLS1 in our sample, XMMSL1\,J070841.3-493305.

\subsection{XMMSL1\,J100534.8+392856}
The Slew Survey and \swift\ observations of this galaxy
show a similar flux, a factor 9-15 above the RASS upper limit. The \swift\ spectra may both be fit with an unabsorbed power law and a common power law index of $\Gamma=1.9\pm 0.2$. The narrow [OIII]$\lambda$5007 line in the optical spectrum is
relatively weak, $L_{\rm{X}}/L_{[\rm{OIII}]}\sim1000$ well in excess of  
the ratio 1-100 usually seen in Seyfert galaxies \citep{Panessa06}. 
As $L_{[\rm{OIII}]}$ represents a measure of the integrated historical
bolometric luminosity this suggests that the intrinsic emission is currently at a high level.

\subsection{XMMSL1\,J104745.6-375932}
This AGN showed a clear Sy 1 profile at a redshift of 0.075, in the 6dF survey of the Anglo-Australian Telescope (AAT). It was not detected in the RASS but has been seen three times since 2003. The last, and faintest, detection with \swift\ revealed a power law spectrum of photon index 
$\Gamma=2\pm0.2$ with no additional absorption above the Galactic column. 

\subsection{XMMSL1\,J111527.3+180638}
This source has been reported as a low-redshift, candidate tidal 
disruption event by \citet{esquej2007} and \citet{esquej2008}.

\subsection{XMMSL1\,J112841.5+575017}
This Seyfert 2 was not detected during the RASS, but was found to be
bright during the Slew Survey and {\it Swift} follow-up observations.
The spectrum in 2013 may be fit adequately with a pure power law model
($\Gamma=1.7\pm0.1$). Since the only available spectrum corresponds to the brightest state we cannot exclude a varying column density of
the absorber as a variability mechanism.
In the brightest observation the X-ray luminosity nearly matches
the one expected from the UV flux while the {\it ROSAT} upper limit is X-ray weak.

\subsection{XMMSL1\,J113001.8+020007}
This is most likely a spurious detection. It was only detected in one \xmm\ slew observation
and the counterpart is a rather faint, g=21.4 galaxy, SDSS\,J113001.72+015956.6.
The significance of the slew detection is rather high, however a visual inspection shows that the photons lie in a line rather than coming from an obvious point source.

\subsection{XMMSL1\,J121335.0+325609}
This broad-line QSO at a redshift of $z=0.222$ has been observed to be variable for over 30 years. Nevertheless due to the short exposure times no good spectrum is available. The source was detected twice by {\it ROSAT}, the first time at an intermediate level, but shortly after at its currently lowest 
observed state, a factor of more than 20 dimmer. Since 2006 the source has been detected five times by {\it Swift} and during the Slew survey. In those observations the source was found to have a bright or intermediate luminosity and a change by a factor of seven has been observed within 17 months.

The UVOT flux was constant during the {\it Swift} observations.
However in an additional short observation, nine months after the last 
X-ray data the UV flux had increased by 30\%. 

The merged spectrum of the three \swift\ observations in 2006 with a flux five times below the Slew detection may not be explained by a thick absorber and the variability mechanism remains unclear.

\subsection{XMMSL1\,J132342.3+482701}
The variability in this non-active galaxy has been interpreted as a tidal 
disruption event (see \citet{esquej2007, esquej2008} for details).

\subsection{XMMSL1\,J162553.4+562735}
This QSO, at redshift 0.3, was not seen in the RASS but has been
detected in all observations since 2005 with a
slowly decreasing luminosity. The 2010-07-10 \swift\ spectrum has few counts and can be described by a pure power law with 
$\Gamma = 2.3\pm 0.5$ (see Table \ref{tab:models}).
A significant absorption column can be excluded. While the Slew 
observation matches the expected X-ray to UV ratio, the \rosat\ upper 
limit corresponds to a rather X-ray weak state. The quasar is bright 
in the infrared and can be classified as a LIRG. 

\subsection{XMMSL1\,J173738.2-595625}
\begin{figure}
\centering
\rotatebox{-90}{\includegraphics[height=80mm]{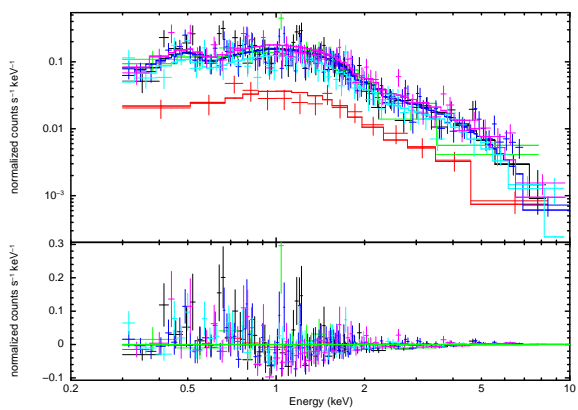}}
\caption[Spectra of J173]{ \label{fig:J173_sim} A simultaneous fit
to 8 \swift\ spectra from 
XMMSL1\,J173738.2-595625 of a power law, of fitted index 
$\Gamma=1.90\pm0.04$, absorbed by the Galactic column. 
The normalisation is fairly constant,
only varying significantly for the spectrum from 2008-06-03 (red), where it 
is lower than the brightest spectrum by a factor 5. This change in flux
rate is reflected as a dip in the X-ray light curve (see Figure \ref{fig:J173}). Additional absorption is not required to fit the spectra}.
\end{figure}

This is the Seyfert 2 galaxy, ESO 139-G12 which has been mooted as a 
possible site of cosmic-ray acceleration \citep{Terrano12}.
It was observed eight times by \swift\ in 2008 and 2013, where it varied 
by a factor 5. The \swift\ spectra can be fit simultaneously with
a power law of slope $\Gamma=1.90\pm0.04$ of variable normalisation, 
absorbed by the Galactic column (Figure \ref{fig:J173_sim}; 
C/dof = \cred\ = 1565/1641). There is no obvious evidence for
intrinsic absorption in the spectra. A small improvement in the fit can be 
achieved by adding a soft-excess component, represented
by a black body ($kT=151\pm28$ eV; \cred=1535/1633).

To check whether the dip in 2008-06-03 is due to absorption we fit this
spectrum simultaneously with the brighter spectrum from 2008-05-04.
A power law (index=$1.96\pm{0.10}$) with variable normalisation and Galactic
absorption gives a good fit (\cred=344/375). Fixing the normalisation between
the two spectra and adding a cold (warm) absorption component to the 2008-06-03
spectrum gives a poor fit of \cred=517/375 (431/374). Allowing the
normalisation to vary and adding a warm absorber to the fainter spectrum
gives no improvement (\cred=344/373). Therefore the
variability in this source is likely to be intrinsic.
The relative X-ray luminosity is somewhat low in this source, however,
even in the brightest state (see Table~\ref{tab:alphaox}) and so 
another possibility is that 
variability is due to partial covering of the central source by a 
compton-thick absorber.

\subsection{XMMSL1\,J183521.4+611942}
This flat-spectrum radio quasar, at a redshift of 2.2, is the most distant and 
luminous object in our sample,
with a peak luminosity of $L_\text{X}\sim6\times10^{46}$ erg s$^{-1}$ and  
$\text{M}_\text{BH}=10^{9.7}$ \msun\ derived from the 
k-band luminosity. The low count X-ray spectrum is not well fit by a simple power law but an absorber does not improve the fit. Therefore, variable jet activity is perhaps
the most likely cause.

\subsection{XMMSL1\,J193439.3+490922}
This unclassified galaxy with an unknown redshift was detected three
times in the Slew Survey from 2003 to 2006 but was not seen in the RASS nor during the {\it Swift} follow-up observation. 
By combining the 20 photons available from the three Slew survey detections we
can make a crude spectrum. A fit to this gives a typical AGN-like
power law slope of $\Gamma = 2.1\pm{0.5}$.

\begin{table*}[hp]
{\small
\hfill{}
\caption{Observations used in this study.}
\label{tab:data_table}    
\begin{center}
\begin{tabular}{l l c c c}       
\hline\hline               
XMMSL1 name			& Instrument	& obsID		& date\tablefootmark{a} 		& $T_\text{exp}$ (s) \tablefootmark{b}\\ 
\hline                       
J005953.1+314934 			& Einstein MPC	& 2619 		& 1980-01-10 	& 2211		\\
			& ROSAT 	& RASS 	& 1991-01-01 & 462			\\
			& ROSAT PSPC 	& WG701087P.N1 	& 1992-07-18 	& 4292.8		\\
			& XMMSL1 	& 9038500002 	& 2002-01-15 	& 8.8			\\
			& XMM EPIC pn 	& 0312190101 	& 2006-01-24 	& 11193			\\
			& Swift XRT 	& 00035243 	& 2006-05-29 -- 2006-05-31 	& 31181	\\
			& Suzaku 	& 704025010 	& 2010-01-06 -- 2010-01-07	& 91000	\\
			& Swift XRT 	& 00040300$^*$ 	& 2011-02-07 -- 2011-08-29 	& 3444	\\
[1ex] \\ [-2.5ex]
	
J015510.9-140028 	& ROSAT PSPC	& RASS 		& 1991-01-01  		& 468 		\\
	& ROSAT PSPC	& rp600005n00 	& 1992-01-15  		& 15850 		\\
			& XMMSL1	& 9075100004 	& 2004-01-15 		& 6.0 	\\
			& Swift XRT	& 00040302$^*$ 		& 2010-06-06 -- 2010-09-14 		& 1610 	\\
[1ex] \\ [-2.5ex]

J020303.1-074154  & ROSAT	& RASS		& 1991-01-01	&  400	\\
 			 & XMMSL1	& 9075000006	& 2004-01-14	& 4.7 \\
 			 & XMM EPIC pn	& 0411980201	& 2006-07-03	& 10828 \\
                         & Swift XRT	& 00035746001	& 2007-06-03	& 3270 	\\
                         & Swift XRT	& 00035746002	& 2008-03-02	& 3668 	\\
                         & Swift XRT	& 00040304001$^*$	& 2010-03-01	& 2422 	\\
[1ex] \\ [-2.5ex]

J024916.6-041244  & ROSAT PSPC	& RASS		& 1991-01-01	&  238		\\
			& XMMSL1	& 9084700002	& 2004-07-25	& 1.90 \\
			& XMM EPIC pn	& 0411980401	& 2006-07-14	& 9865 	\\
                          & Swift XRT	& 00035748002,3	& 2006-12-06 -- 2006-12-07	& 959 		\\
                          & Swift XRT	& 00035748004	& 2007-01-26	& 649 	\\
                          & Swift XRT	& 00035748005	& 2007-06-27	& 3054 	\\
                          & Swift XRT	& 00040306001,2,3$^*$	& 2010-06-13 -- 2011-03-08	& 2145 		\\
			& Swift XRT   & 00040306004,5$^*$	& 2015-03-03 -- 2015-03-06	& 2812	\\
[1ex] \\ [-2.5ex]

J034555.1-355959  & ROSAT	& RASS		& 1991-01-01	& 70 		 \\
		& ROSAT PSPC	& rp800300n00 	& 1992-08-27  	& 1217		 \\
		& ROSAT PSPC	& rp800300a01 	& 1993-01-26 	& 3600 		 \\
		& ROSAT PSPC	& rp190499n00	& 1997-02-06	& 2005 		 \\
			& XMMSL1	& 9156600004	& 2008-06-28	& 6.4 	\\
			& XMMSL1	& 9186200002	& 2010-02-07	& 5.2 	\\
			& Swift XRT	& 00040307001$^*$	& 2010-12-22 -- 2010-12-23	& 2361 	\\
[1ex] \\ [-2.5ex]

J044347.0+285822  & ROSAT       & RASS          & 1991-01-01    & 454                  \\
                        & XMMSL1        & 9058800004    & 2003-02-24    & 3.8          \\
                        & XMM EPIC pn   & 0401790101    & 2007-03-18    & 10042        \\
                        & Swift XRT     & 00040308001$^*$     & 2010-07-27    & 1806      \\
[1ex] \\ [-2.5ex]

J045740.0-503053  & ROSAT       & RASS          & 1991-01-01    & 400            \\
                        & XMMSL1        & 9049100003    & 2002-08-14    & 4.5       \\
                        & XMMSL1        & 9102600004    & 2005-07-16    & 8.3          \\
                        & XMMSL1        & 9188000003    & 2010-03-16    & 8.3      \\
                        & Swift XRT     & 00040309001$^*$  & 2010-11-06    & 2284     \\
                        & Swift XRT     & 00040309002,3,5$^*$    & 2013-04-05 -- 2013-05-11 & 5100   \\
[1ex] \\ [-2.5ex]

J051935.5-323928\tablefootmark{c}  & ROSAT   & RASS          & 1991-01-01    & 505           \\
& Swift campaign & 35234 & 2005-10-29 - 2005-11-26 &  \\
        & XMMSL1        & 0312190701 & 2006-01-28    & 8726  \\
          & XMMSL1      & 9124900007    & 2006-10-05    & 6.4   \\
          & Suzaku       & 703014010     & 2008-04-11    & 83410   \\
          & XMMSL1        & 9179700004  & 2009-10-01    & 5.66    \\
          & XMMSL1        & 9196400008    & 2010-08-31    & 7.41     \\
& Swift campaign & 31868& 2010-11-15--2011-01-21& \\
                        & Swift XRT     & 00040311001,2$^*$  & 2011-01-07 -- 2011-01-23       & 2309     \\
[1ex] \\ [-2.5ex]

J064541.1-590851  & ROSAT       & RASS          & 1991-01-01    & 645           \\
                         & XMMSL1       & 9126300002    & 2006-11-01    & 3.5      \\
                         & XMMSL1       & 9143600003    & 2007-10-13    & 6.2         \\
                         & XMMSL1       & 9159400002    & 2008-08-22    & 8.4          \\
                         & XMMSL1       & 9195100003    & 2010-08-04    & 7.3       \\
                         & Swift XRT    & 00040315001,2$^*$  & 2010-12-22 -- 2010-12-26       & 2455  \\
                         & XMMSL1       & 9241500003    & 2013-02-15    & 1.0   \\
                         & XMMSL1       & 9255200003    & 2013-11-15    & 6.1   \\
[1ex] \\ [-2.5ex]
\hline                                  
\end{tabular}
\end{center}
}
\end{table*}

\setcounter{table}{1}
\addtocounter{table}{-1} 

\begin{table*}[ht!]
{\small
\hfill{}
\caption{Used observations continued.} 
\begin{center}
\begin{tabular}{l l c c c}        
\hline\hline                 
XMMSL1 name			& Instrument	& obsID		& date\tablefootmark{a}            & $T_\text{exp}$ (s) \tablefootmark{b}\\
\hline     
J070841.3-493305\tablefootmark{c}  & Rosat   & RASS          & 1991-01-01    & 171           \\
                        & ROSAT         & rp180306n00   & 1998-12-09    & 9472      \\
			  & XMM EPIC pn	& 0110890201 & 2000-10-21	& 40700		\\
			  & XMM EPIC pn	& 0148010301 & 2002-10-13	& 78000	\\
                        & XMMSL1       & 9109200003    & 2005-11-25    & 2.9        \\
			  & XMM campaign & 0506200201-501 & 2007-05-16 -- 2007-07-06	& 	\\
                        & XMMSL1        & 9143400002    & 2007-10-09    & 9.3         \\
			  & XMM campaign & 0511580101-1201 & 2008-01-29 -- 2008-02-06	& \\
			  & XMM campaign & 0653510301-601 & 2010-09-13 -- 2010-09-19	& \\
                        & Swift campaign& 90393 & 2010-04-03 -- 2010-12-19 & \\
                        & Swift XRT     & 00040317001$^*$       & 2010-12-22    & 1569\\
			  & XMM EPIC pn	& 0554760801 & 2011-01-12 & 96000	\\
                        & Swift campaign& 90393 & 2010-12-23 -- 2011-01-25 & \\
 			& Swift XRT     & 00040317003$^*$       & 2011-01-27    & 476 \\
 			& Swift campaign& 90393 & 2011-01-28 -- 2011-03-29 & \\
                        & XMMSL1        & 9207000003    & 2011-03-30    & 6.9    \\
 			& Swift campaign& 91623 & 2013-05-20 -- 2013-06-19 & \\
 			& Swift campaign& 80720 & 2014-05-05 -- 2014-06-28 & \\
[1ex] \\ [-2.5ex]

J082753.7+521800  			& ROSAT	& RASS		& 1991-01-01	& 449 	\\
			& XMMSL1	& 9106500003	& 2005-10-02	& 8.4		\\
			& Swift XRT	& 00040320001$^*$	& 2010-03-11	& 2159 	\\
[1ex] \\ [-2.5ex]

J090421.2+170927  		& ROSAT		& RASS		& 1991-01-01	& 310 	\\
	& ROSAT PSPC	&  rs931424n00		& 1991-11-18  	& 4170 		\\
		& XMMSL1	& 9126300002	& 2006-11-01	& 4.6		\\
		& Swift XRT	& 00040322001$^*$	& 2010-06-12	& 1794 	\\
[1ex] \\ [-2.5ex]

J093922.5+370945   & ROSAT	& RASS		& 1991-01-01	&  470		 \\
	& XMMSL1	& 9081300003	& 2004-05-18	& 7.7	\\
	& XMM Epic pn	& 0411980301	& 2006-11-01	& 5038 	\\
	& Swift XRT	& 00035747001		& 2007-09-21	& 6554 \\
	& Swift XRT	& 00040325001,2$^*$	& 2011-03-28 -- 2011-05-11	& 1846 	\\
[1ex] \\ [-2.5ex]

J100534.8+392856   & ROSAT	& RASS		& 1991-01-01	& 481		\\
			  & XMMSL1	& 9107400003	& 2005-10-21	& 1.0	\\
			  & Swift XRT	& 00040326001$^*$		& 2011-01-31	& 2058 	\\
			  & Swift XRT	& 00040326004$^*$	& 2013-04-23 	&  	3654\\
[1ex] \\ [-2.5ex]

J104745.6-375932   & ROSAT	& RASS		& 1991-01-01	&  362		\\
			  & XMMSL1	& 9073500003	& 2003-12-14	& 7.6	\\
			  & XMMSL1	& 9173700002	& 2009-06-03	& 9.1	\\
			  & Swift XRT	& 00040328001,2$^*$	& 2010-04-11 -- 2010-04-16	& 2288 \\
[1ex] \\ [-2.5ex]

J111527.3+180638\tablefootmark{d}   & ROSAT	& RASS		& 1991-01-01	&  383		\\
			  & XMMSL1	& 9072400006	& 2003-11-22	& 8.0	\\
			  & XMM EPIC pn	& 0411980101	& 2006-06-23	& 5017	\\
			  & Swift XRT	& 00035745001		& 2006-12-01	& 5773 \\
  & XMM EPIC pn	& 0556090101	& 2008-12-02	& 41734	\\
			  & Swift XRT	& 00040332001,2$^*$		& 2010-10-23 -- 2010-11-16	& 2614 		\\
			  & Swift XRT	& 00084368001,2	& 2014-07-17 -- 2014-10-23 	& 5936	\\
[1ex] \\ [-2.5ex]

J112841.5+575017   		& ROSAT		& RASS		& 1991-01-01	& 519 	\\
		& XMMSL1	& 9107500006	& 2005-10-23	& 10.5		\\
		& XMMSL1	& 9116700004    & 2006-04-24	& 5.5		\\
		& Swift XRT	& 00040333001$^*$	& 2011-03-28	& 1872 	\\
		& Swift XRT	& 00040333002,3,4$^*$ & 2013-04-10 -- 2013-10-19	& 6196	\\
[1ex] \\ [-2.5ex]

J113001.8+020007   		& ROSAT		& RASS		& 1991-01-01	& 427	\\
		& XMMSL1	& 9109400004		& 2005-11-29	& 6.7	\\
		& XMMSL1	& 9165400002	& 2008-12-20	& 4.6		\\
		& XMMSL1	& 9183300003	& 2009-12-13	& 5.9		\\
		& Swift XRT	& 00040334001,2$^*$	& 2011-01-26 -- 2011-01-31	& 2041	\\
[1ex] \\ [-2.5ex]
\hline                                   
\end{tabular}
\end{center}
}
\end{table*}

\addtocounter{table}{-1}

\begin{table*}[ht!]
{\small
\hfill{}
\caption{Used observations continued.}
\begin{center}
\begin{tabular}{l l c c  c}        
\hline\hline                 
XMMSL1 name			& Instrument	& obsID		& date\tablefootmark{a}            & $T_\text{exp}$ (s) \tablefootmark{b}\\
\hline
J121335.0+325609      
	  & ROSAT	& RASS		& 1991-01-01	& 502 		\\
	  & ROSAT	& rp600130n00	& 1991-11-24	& 16310 \\
	  & XMMSL1	& 9072500005	& 2003-11-24	& 9.0	\\
	  & Swift campaign	& 55300		& 2006-06-02	& 2910\\
	  & Swift campaign	& 55300		& 2006-06-09 -- 2006-06-12	& 4369	\\
	  & Swift campaign	& 55300		& 2006-06-27	& 1814	\\
	  & XMMSL1	& 9155100004	& 2008-05-29	& 10.1	\\
	  & Swift XRT	& 00040335001$^*$	& 2010-10-19	& 1589 \\
	  & Swift XRT	& 00040335002$^*$	& 2011-07-06	& 398 \\
[1ex] \\ [-2.5ex]
J132342.3+482701   & ROSAT	& RASS		& 1991-01-01	&  560		\\
	& XMMSL1	& 9072900002	& 2003-12-01	& 8.6	\\
	& XMM EPIC pn	& 0411980501	& 2006-07-15	& 5018	\\
	& Swift XRT	& 00035749001	& 2007-01-11	& 3854	\\
	& Swift XRT	& 00035749002	& 2007-05-26	& 1922	\\
	& Swift XRT	& 00040336001$^*$	& 2011-02-04	& 1819	\\
[1ex] \\ [-2.5ex]
J162553.4+562735   & ROSAT	& RASS		& 1991-01-01	& 1035 	\\
			  & XMMSL1	& 9100800002	& 2005-06-10	& 5.8	\\
			  & XMMSL1	& 9131700004	& 2007-02-17	& 4.0	 \\
			  & Swift XRT	& 00040345001,2$^*$	& 2010-07-10 -- 2010-07-11	& 3753	\\
			  & Swift XRT	& 00040345004,5$^*$	& 2013-04-08 -- 2013-04-14 	& 3987	\\
[1ex] \\ [-2.5ex]
J173739.3-595625\tablefootmark{d}   & ROSAT	& RASS		& 1991-01-01	& 165 \\
			  & XMMSL1	& 9123400007	& 2006-09-05	& 2.9 \\
			  & XMMSL1	& 9151100003	& 2008-03-10	& 5.5	\\
			  & Swift XRT	& 00037391001	& 2008-05-04	& 1958	\\
			  & Swift XRT	& 00037705001	& 2008-06-03	& 2671	\\
			  & Swift XRT	& 00037705002	& 2008-07-10	& 553	\\
			  & Swift XRT	& 00037391002	& 2008-07-21	& 5001	\\
			  & Swift XRT	& 00037391003	& 2008-07-29	& 3359	\\
			  & Swift XRT	& 00037705004	& 2008-11-02	& 2078	 \\
			  & Swift XRT	& 00040349001$^*$	& 2013-03-18	& 125\\
			  & Swift XRT	& 00040349002,3$^*$	& 2013-10-12 -- 2013-10-18 	& 1747	\\
			  [1ex] \\ [-2.5ex]
J183521.4+611942   & ROSAT	& RASS		& 1991-01-01	& 2270 	\\
		& XMMSL1	& 9100300002	& 2005-05-31	& 1.0	\\
		& XMMSL1	& 9149600002	& 2008-02-09	& 8.4	\\
		& XMMSL1	& 9173900002	& 2009-06-08	& 10.1	\\
		& Swift XRT	& 00040352001$^*$	& 2011-02-04	& 1914	\\
		& Swift XRT	& 00040352003,4,5,6$^*$	& 2013-06-16 -- 2013-07-24 	& 2970	\\
[1ex] \\ [-2.5ex]
J193439.3+490922   & ROSAT	& RASS		& 1991-01-01	& 823 	\\
& XMMSL1      & 9072800002 & 2003-11-30    &9.2   \\
			  & XMMSL1	& 9100700002	& 2005-06-08	& 8.2	\\
			  & XMMSL1	& 9127200003	& 2006-11-19	& 4.8	\\
			  & Swift XRT	& 00040353002,3,4,5$^*$		& 2011-11-24 -- 2012-09-15	& 2237	 \\
\hline                                   
\end{tabular}
\end{center}
}
\tablefoot{
{\it Swift} observations taken specifically for this programme are marked with $^*$.\\
\tablefoottext{a}{For the {\it ROSAT} All Sky Survey, 1991-01-01 is assumed as an approximate observation date for all observations.}\\
\tablefoottext{b}{The exposure time for the {\it XMM-Newton} Slew observations is 
given for the total energy band. There is a small difference, due to the 
energy-dependent vignetting, when considering the soft or hard band only.}\\
\tablefoottext{c}{Numerous observations from other campaigns are available, many are 
shown in the light curve. Here we only list {\it ROSAT}, {\it XMM} and 
our {\it Swift} observations.}\\
\tablefoottext{d}{In addition there are {\it Chandra} observations which are not considered here.}
}
\hfill{}
\end{table*}

\begin{sidewaystable*}
{\small
\hfill{}
\caption{Summary of proposed variability mechanism.}
\label{tab:summary}      
\begin{center}
\begin{tabular}{l l l l}
\hline \hline
XMMSL1 name			& Type		& $z$ 		& variability causes and comments \\
\hline                     
\multicolumn{4}{l}{Sources displaying factor $>10$ variability between ROSAT and XMM, and XMM and Swift epochs} \\ \hline
J015510.9-140028 	&- 		& - & spurious Slew detection \\
J024916.6-041244     & Sy 1.9 	& 0.0186 & soft, absorbed black body spectrum, variability likely intrinsic in nature due to constant spectral shape \\ 	
J034555.1-355959  	& - 		& - 	   & inconclusive \\
J045740.0-503053  	& - 		& - 	   & inconclusive \\	
J051935.5-323928  	& Sy 1.5 	& 0.0125   & other studies suggest variable absorption \\
J070841.3-493305  	& NLS 1 	& 0.0406   & intrinsic power law changes, studies suggest variable reflection off disc \\
J111527.3+180638  	& liner	        & 0.00278  & tidal disruption event \\	
J113001.8+020007          & -	        & - & spurious Slew detection \\
J132342.3+482701  	& non-active	& 0.0875 & tidal disruption event \\	
J193439.3+490922 		& - 		& - 	& unclear   \\ 
\hline
\multicolumn{4}{l}{Sources displaying factor $>10$ variability between ROSAT and XMM epochs} \\ \hline
J005953.1+314934 &  Sy 1.2 	& 0.0149 		& neutral and ionised partially covering absorbers \\		
J020303.1-074154 & 		Sy 1 		& 0.0615 & spectral shape unaltered since 2004 indicating intrinsic flux change, little variability since 2006 \\	
J044347.0+285822  	& Sy 1 		& 0.0217 	& neutral partial covering absorbers \\
J064541.1-590851 & - 	& - 			& inconclusive; no evidence for significant variability since 2006, bright state spectra do not require absorption \\
J082753.7+521800 		& QSO, RL		& 0.3378 &  no absorption required in bright state, likely jet variability \\
J090421.2+170927  		& QSO 		& 0.0733 & possibly ionised absorption \\	
J093922.5+370945  		& NLS 1 	& 0.1861 & inconclusive, bright state spectra do not require absorption \\
J100534.8+392856  		 & Sy 1 	& 0.1409 & bright state spectrum does not require absorption. Possibly increased intrinsic emission. \\	
J104745.6-375932  		 & Sy 1 	& 0.0755 & absorption-only excluded, faintest state does not require absorption \\	
J112841.5+575017 	        & Sy 2 	        & 0.0509 & inconclusive, brightest state does not require absorption \\	
J121335.0+325609  	& QSO 		& 0.222 	& inconclusive \\
J162553.4+562735  	& QSO 		& 0.307 	& soft spectrum, absorption-only excluded \\
J173739.3-595625  	& Sy 2 		& 0.0170 	& intrinsic; constant spectral shape from 2008-2013 \\
J183521.4+611942 	        & blazar, RL	& 2.274 & inconclusive, likely jet contribution \\
\hline                                  
\end{tabular}
\end{center}
}
\end{sidewaystable*}

\end{appendix}

\end{document}